\documentclass[aps,prd,superscriptaddress,amsmath,amssymb,twocolumn,fleqn,epsf,epsfig,showpacs,nofootinbib]{revtex4-1} 

\usepackage{dcolumn}
\usepackage{bm}
\usepackage{amsfonts}

\usepackage{color}

\usepackage{graphicx}
\usepackage{colordvi}
\usepackage{mathtools}
\usepackage{empheq}

\def\be{\begin{equation}}
\def\ee{\end{equation}}
\def\ba{\begin{eqnarray}}
\def\ea{\end{eqnarray}}
\newcommand{\tk}{\tilde{k}}

\DeclarePairedDelimiter{\abs}{\lvert}{\rvert}

\begin{document}


\title{CMB anisotropies generated by a stochastic background of primordial magnetic fields 
with non-zero helicity} 

\author{Mario Ballardini}\email{ballardini@iasfbo.inaf.it}
\affiliation{DIFA, Dipartimento di Fisica e Astronomia,\\
Alma Mater Studiorum, Universit\`a degli Studi di Bologna,\\
viale Berti Pichat 6/2, I-40127 Bologna, Italy}
\affiliation{INAF/IASF-BO, Istituto di Astrofisica Spaziale e Fisica Cosmica di Bologna, \\
via Gobetti 101, I-40129 Bologna - Italy}
\affiliation{INFN, Sezione di Bologna, \\
via Irnerio 46, I-40126 Bologna, Italy}
\author{Fabio Finelli}\email{finelli@iasfbo.inaf.it}
\affiliation{INAF/IASF-BO, Istituto di Astrofisica Spaziale e Fisica Cosmica di Bologna, \\
via Gobetti 101, I-40129 Bologna - Italy}
\affiliation{INFN, Sezione di Bologna, \\
via Irnerio 46, I-40126 Bologna, Italy}
\author{Daniela Paoletti}\email{paoletti@iasfbo.inaf.it}
\affiliation{INAF/IASF-BO, Istituto di Astrofisica Spaziale e Fisica Cosmica di Bologna, \\
via Gobetti 101, I-40129 Bologna - Italy}
\affiliation{INFN, Sezione di Bologna, \\
via Irnerio 46, I-40126 Bologna, Italy}

\date{\today}
\begin{abstract}
We consider the impact of a stochastic background of primordial magnetic fields 
with non-vanishing helicity on CMB anisotropies in temperature and polarization. 
We compute the exact expressions for the scalar, vector and tensor part of 
the energy-momentum tensor including the helical contribution, by assuming a power-law 
dependence for the spectra and a comoving cutoff which mimics the damping due to 
viscosity. We also compute the parity-odd correlator between the helical and 
non-helical contribution which generate the $TB$ and $EB$ cross-correlation in 
the CMB pattern. We finally show the impact of including the helical term on
the power spectra of CMB anisotropies up to multipoles with 
$\ell \sim \mathcal{O}(10^3)$.
\end{abstract}

\pacs{Valid PACS appear here}
\maketitle
\section{Introduction}

A stochastic background of primordial magnetic fields (PMF) generated 
prior to recombination can leave several footprints on the anisotropy 
pattern of the cosmic microwave background (see e.g. Ref.~\cite{Durrer:2013pga} for a review). 
A stochastic background of PMF generates compensated scalar 
\cite{stochastic background,Giovannini:2004,KR,FPP,Kunze:2010ys,Yamazaki:2014xna}, vector 
and tensor perturbations \cite{Durrer:1999bk,MKK,lewis,Giovannini:2006,PFP}
whose contribution to the cosmic microwave background (CMB) 
anisotropies in temperature and polarization is not suppressed by the 
Silk damping. The dominant vector contribution to temperature anisotropies 
from PMF at high multipoles which drives the current CMB constraints \cite{PF} 
needs therefore to be disentangled from the foreground residuals and secondary 
anisotropies \cite{PF_WMAP7SPT,Planck2013:Parameters}. 

The $\chi^2$ statistics of a stochastic background of PMF \cite{Brown:2005kr} 
makes the contribution to CMB anisotropies fully non-Gaussian. The CMB 
bispectrum was therefore targeted as a probe for PMF which is independent from the
constraints based on the CMB power spectrum \cite{SS,CFPR,Planck2013:fNL}. 
Subsequent works have been dedicated to refine the predictions for the CMB 
bispectrum for compensated and passive initial conditions 
\cite{Cai:2010uw,Trivedi:2010gi} and to compute the CMB trispectrum predictions 
\cite{Trivedi,Trivedi:2013wqa}. 

A stochastic background of PMF has also distinctive predictions for 
the CMB polarization pattern. Vector perturbations sourced by PMF lead to a 
B-mode power spectrum with a broad maximum at high multipole as 
$\ell \sim {\cal O} (10^3)$. Such spectrum is not degenerate with the one 
produced by tensor perturbations, either these were originated during inflation 
\cite{Giovannini:2000,Durrer:2010mq} or passively sourced when neutrinos free 
stream after the stochastic background of PMF was generated \cite{lewis,shawlewis}. 
A stochastic background of PMF can also modify the CMB polarization pattern by 
the Faraday effect with the characteristic frequency dependence 
$\propto 1/\nu^4$ \cite{Kosowsky:2004zh}.

In this paper we study in detail another interesting aspect of the interplay 
between PMF and CMB anisotropies in temperature and polarization. A stochastic 
background of PMF is characterized in general both by a symmetric and antisymmetric 
power spectrum and its {\em helicity}. Helicity measures the complexity 
of the topology of the magnetic field. Being helicity a P and CP odd-function, 
its search in the CMB pattern is of primary importance for the understanding 
of the generation mechanism of PMF. As examples for generation mechanisms, 
helicity can be produced by a coupling to a primordial pseudo-scalar field
\cite{GFC,Anber:2006xt,Caprini:2014mja,Atmjeet:2014cxa} and be affected
by the presence of chiral anomaly in the early Universe \cite{BFR}.

The helical contribution in a stochastic background of PMF has also been subject of 
previous investigations \cite{PVW,CDK,Kahniashvili:2005xe,Kunze_helical,Kahniashvili:2014dfa}. 
If the stochastic background of PMF has non-vanishing helicity, its 
contribution to CMB parity {\em even} correlators such as $TT$, $EE$, $BB$, $TE$, 
is modified. In addition,
CMB parity {\em odd} correlators such as $TB$ and $EB$
are also generated. Parity {\em odd} cross-correlators, 
since are generated only by helical components,  may be used to 
break the intrinsic degeneracy between the helical and non-helical 
contributions of PMF to CMB parity even correlators. 
Helicity turns on terms in the bispectrum which would vanish 
in the non-helical case \cite{Shiraishi:2012sn}. 
In the general case of non-vanishing helicity, Faraday rotation could be 
useful in breaking the degeneracy between non-helical and helical components of a stochastic background, 
since it does not generate odd-correlators \cite{Campanelli:2004pm} 
\footnote{Whereas Faraday rotation from a stochastic
background of PMF can generate only $BB$, a homogeneous
magnetic field can generate $BB$, $TB$ and $EB$ by Faraday rotation
(in this latter case it is the configuration
of the magnetic field which breaks the parity symmetry)}.

The goal of this paper is to present an original study of PMF including the helical part which 
covers from the analytic computations of the Fourier components of the energy-momentum tensor 
to the predictions for CMB anisotropies in temperature and polarization.
We give for the first time the exact expressions for the Fourier power spectra of the EMT tensor, 
by extending the exact integration scheme for a sharp cut-off used for the non-helical case 
\cite{FPP,PFP}. By implementing these original results for the EMT tensor, 
we present the numerical results for the CMB power spectra in temperature and polarization by a 
modifed version of CAMB \cite{CAMB}.

Our paper is organized as follows. In Sec.~\ref{sec:two} we present the energy-momentum tensor 
(EMT) of PMF in the general case of non-vanishing helicity. In Secs.~\ref{sec:three}, 
\ref{sec:four}, \ref{sec:five} we compute 
the helical contribution to the scalar, vector and tensor parts of the 
EMT of PMF in Fourier space, respectively. For the vector 
and tensor parts we also compute the parity-odd correlators in Fourier space.
In Sec.~\ref{sec:six} we discuss the impact onto CMB anisotropies including the 
power spectra of the parity-odd cross-correlations $TB$ and $EB$. In Sec.~\ref{sec:conclusion} 
we draw our conclusions. In the appendices we describe the methodology to compute the 
convolutions following the integration scheme of Ref.~\cite{FPP} and present 
the corresponding exact formul{\ae} for specific spectral indices.

\section{Stochastic background of primordial magnetic fields with non-zero helicity}
\label{sec:two}

Following Ref.~\cite{PVW}, the most general two-point correlation function for a stochastic 
background, which preserve homogeneity and isotropy, is:
\begin{align}
\label{eqn:ansatz}
\langle B_i(\mathbf{k})B_j^*(\mathbf{h}) \rangle = \frac{(2\pi)^3}{2}\delta^{(3)}(\mathbf{k}-\mathbf{h})
\Bigl[& P_{ij}(k) P_B (k) \notag\\
&+ \imath \epsilon_{ijl} \hat{k}_l P_H(k) \Bigr]\,,
\end{align}
where $P_{ij}(k) = \delta_{ij} - \hat{k}_i \hat{k}_j$, $P_B$ and $P_H$   
are the non-helical and helical part of the spectrum of the stochastic background, 
respectively. The symmetric part of the power spectrum represents the averaged magnetic 
field energy density whereas the antisymmetric part is related to 
the absolute value of the averaged helicity:
\begin{align}
\label{symm}
\langle B_i(\mathbf{k})B_i^*(\mathbf{h}) \rangle
&= (2\pi)^3\delta^{(3)}(\mathbf{k}-\mathbf{h})P_B(k)\,,\\
\label{antisymm}
\langle (\widehat{\mathbf{\nabla}\times\mathbf{B}})_i(\mathbf{k})B_i^*(\mathbf{h}) \rangle 
&= (2\pi)^3\delta^{(3)}(\mathbf{k}-\mathbf{h})P_H(k)\,.
\end{align}
Note that $P_B(k) \propto \langle \big| B \big|^2 \rangle$ so it is defined positive, whereas 
the averaged magnetic helicity can be of either sign and its value is limited by 
combining Eqs.~\eqref{symm} and \eqref{antisymm} with the Schwarz's inequality:
\be
\label{eqn:Schwarz}
\lim_{\mathbf{h} \to \mathbf{k}} \langle (\widehat{\mathbf{k}\times\mathbf{B}})_i(\mathbf{k})B_i^*(\mathbf{h}) \rangle \le \lim_{\mathbf{h} \to \mathbf{k}} \langle B_i(\mathbf{k})B_i^*(\mathbf{h}) \rangle
\ee
implying:
\be
\label{eqn:diseq}
\big| P_H(k) \big| \le P_B(k)
\ee
as detailed discussed in \cite{Durrer:2003ja, CDK}.\\
We model both non-helical and helical terms of the PMF power spectrum with a power law:
\begin{align}
\label{eqn:powerS}
&P_B(k)=A_B\left(\frac{k}{k_*}\right)^{n_B}\,, \\
\label{eqn:powerA}
&P_H(k)=A_H\left(\frac{k}{k_*}\right)^{n_H}\,,
\end{align}
where $A_{B,\,H}$ are the amplitudes, $n_{B,\,H}$ the spectral indices of the 
non-helical and helical parts respectively and $k_*$ is a pivot scale. The 
Eq.~\eqref{eqn:diseq} begins:
\be
\big| A_H \big| \le A_B \left( \frac{k}{k_*} \right)^{n_B-n_H}
\ee
and we can derive as limit condition of {\em maximal helicity} $A_B = A_H$ and 
$n_B = n_H$, valid for small $k$.

We introduce a sharp cutoff at the damping scale, $k_D$, to mimic the 
damping of the PMF on small angular scales \cite{Jedamzik:1996wp,stochastic background}: 
as in previous works we assume that Eqs.~\eqref{eqn:powerS} 
and \eqref{eqn:powerA} hold up to $k \leq k_D$ and $P_{B,\, H}=0$ for $k>k_D$.

We can express the amplitudes $A_B$ and $A_H$ in terms of mean-square values of 
the magnetic field and of the absolute value of the helicity as:
\begin{gather}
\langle B^2 \rangle=\int_\Omega\frac{d^3k}{(2\pi)^3}P_B(k)=\frac{A_B}{2\pi^2}\frac{k_D^{n_B+3}}{k_*^{n_B} \left( n_B+3 \right)}\,,\\
\langle \mathcal{B}^2 \rangle=\frac{1}{k_D}\int_\Omega\frac{d^3k}{(2\pi)^3}k|P_H(k)|
=\frac{|A_H|}{2\pi^2}\frac{k_D^{n_H+3}}{k_*^{n_H} \left( n_H+4 \right) }\,.
\end{gather}
An alternative convention is to parametrize the fields through a convolution with a
3D-Gaussian window function, smoothed over a sphere of comoving radius 
$\lambda$. In order to calculate these quantities, we convolve the 
magnetic field and its helicity with a Gaussian filter function:
\begin{align}
\langle B^2_\lambda \rangle &= \int_\Omega\frac{d^3k}{(2\pi)^3}P_B(k)\,e^{-\lambda^2k^2} \notag\\
&=\,\frac{A_B}{(2\pi)^2}\frac{1}{k_*^{n_B} \lambda^{n_B+3}}\Gamma\Bigl(\frac{n_B+3}{2}\Bigr)\,,\\
\langle \mathcal{B}^2_\lambda \rangle &= \,\lambda\int_\Omega\frac{d^3k}{(2\pi)^3}k|P_H(k)|\,e^{-\lambda^2k^2} \notag\\
&=\,\frac{|A_H|}{(2\pi)^2}\frac{1}{k_*^{n_H} \lambda^{n_H+3}}\Gamma\Bigl(\frac{n_H+4}{2}\Bigr)\,,
\end{align}
where we consider $n_B>-3$ and $n_H>-4$ in order to ensure 
the convergence of the integrals above without introducing infrared cut-offs.

The definition of helicity in Eq.~\eqref{antisymm} is called {\em kinetic helicity}, is 
gauge-invariant and gives a measure of the turbulence developed by the stochastic
magnetic field \cite{MB}. An alternative definition is the {\em magnetic helicity} $\mathcal{H}$, 
defined as ${\bf A} \cdot {\bf B}$, with ${\bf B} = \nabla \times {\bf A}$ where ${\bf A}$ 
is the gauge field, which measures the complexity of the topology of the magnetic field and is 
gauge invariant only under particular boundary condition on the field 
\cite{Kunze_helical,Kahniashvili:2014dfa}:
\begin{align}
\label{eqn:MagHel}
\langle \mathcal{H} \rangle &= \frac{1}{4\pi}\int_\Omega\frac{d^3k}{(2\pi)^3}\frac{1}{k}|P_H(k)| \notag\\
&=\frac{|A_H|}{8\pi^3}\frac{k_D^{n_H+3}}{k_*^{n_H} \left( n_H+2 \right) }\,, 
\end{align}
where the factor $1/(4\pi)$ has been introduced to recover the definition of magnetic helicity 
density used in \cite{Kahniashvili:2014dfa}.
Note that $n_H>-2$ is required in order to have integrability at small $k$
for the integrated magnetic helicity $\mathcal{H}$ \cite{Kunze_helical,Kahniashvili:2014dfa}, differently from $\mathcal{B}$. 

The PMF described have an impact on cosmological perturbations. In particural the PMF source all 
types of metric perturbations: scalar, vector and tensor and induce a Lorentz force on baryons. 
The EMT scalar, vector and tensor components are:
\begin{align}
\tau^{0\,\,\text{PMF}}_{\,\,\,0}(\mathbf{x})=&-\frac{1}{8\pi a^4}|\mathbf{B}(\mathbf{x})|^2\,, \\
\tau^{0\,\,\text{PMF}}_{\,\,\,i}(\mathbf{x})=&\,0\,,\\
\tau^{i\,\,\text{PMF}}_{\,\,\,j}(\mathbf{x})=&\,\frac{1}{4\pi a^4}\left[\delta^i_j
\frac{|\mathbf{B}(\mathbf{x})|^2}{2}-B^i(\mathbf{x})B_j(\mathbf{x})\right] \,,
\end{align}
where, due to the high conductivity in the primordial plasma, $\sigma \gg 1$, we have omitted 
terms $\propto \mathbf{E}\cdot\mathbf{B}$ and $E^2$ which 
are suppressed by $1/\sigma$ and $1/\sigma^2$, respectively.
The spatial part of magnetic field EMT in Fourier space is given by:
\begin{align}
\label{eqn:SpatialEMT}
\tau_{ij}^{\text{PMF}}(\mathbf{k})=\frac{1}{32\pi^4}
\int_\Omega d^3p\Bigl[&B_i(\mathbf{p})B_j(\mathbf{k}-\mathbf{p}) \notag\\
&-\frac{\delta_{ij}}{2}B_l(\mathbf{p})B_l(\mathbf{k}-\mathbf{p})\Bigr]\,.
\end{align}

\begin{widetext}
The two-point correlation tensor related to Eq.~\eqref{eqn:SpatialEMT} takes the form:
\be
\langle \tau_{ab}(\mathbf{k})\tau^*_{cd}(\mathbf{h}) \rangle =
\frac{1}{1024\pi^8} 
\int_\Omega d^3p\int_\Omega d^3q
\langle B_a(\mathbf{p})B_b(\mathbf{k}-\mathbf{p})
B_c(\mathbf{-q})B_d(\mathbf{q}-\mathbf{h}) \rangle +\dots\delta_{ab}+\dots\delta_{cd}+\dots\delta_{ab}\delta_{cd}\,,
\ee
and after a little algebra results:
\ba
\label{eqn:stresstensor}
\langle \tau_{ab}(\mathbf{k})\tau^{*}_{cd}(\mathbf{h}) \rangle 
&=& \frac{1}{4(4\pi)^2}\delta^{(3)}(\mathbf{k}-\mathbf{h})\int_\Omega d^3p
\biggl\{\biggl[
P_B(p)P_B(|\mathbf{k}-\mathbf{p}|) \Bigl(P_{ac}(p)P_{bd}(|\mathbf{k}-\mathbf{p}|)+P_{ad}(p)P_{bc}(|\mathbf{k}-\mathbf{p}|)\Bigr) 
\nonumber \\
&& -\, P_H(p)P_H(|\mathbf{k}-\mathbf{p}|)
\bigl(\epsilon_{aci}\epsilon_{bdj}\hat{p}_i(\widehat{\mathbf{k}-\mathbf{p}})_j
+\epsilon_{adi}\epsilon_{bcj}\hat{p}_i(\widehat{\mathbf{k}-\mathbf{p}})_j\bigr) \nonumber \\
&& +\, \imath P_B(p)P_H(|\mathbf{k}-\mathbf{p}|)
\Bigl(P_{ac}(p)\epsilon_{bdi}(\widehat{\mathbf{k}-\mathbf{p}})_i 
+P_{ad}(p)\epsilon_{bci}(\widehat{\mathbf{k}-\mathbf{p}})_i\Bigr) \nonumber \\
&& +\, \imath P_B(p)P_H(|\mathbf{k}-\mathbf{p}|)\Bigl(\epsilon_{aci}P_{bd}(|\mathbf{k}-\mathbf{p}|)\hat{p}_i + 
\epsilon_{adi}P_{bc}(|\mathbf{k}-\mathbf{p}|)\hat{p}_i\Bigr)\biggr] \notag\\
&& +\dots\delta_{ab}+\dots\delta_{cd}+\dots\delta_{ab}\delta_{cd} \biggr\}\,.
\ea
\end{widetext}
In the following three sections we will present the scalar, vector, tensor contributions to the PMF 
EMT, respectively.

Following the integration scheme used in Refs.~\cite{FPP,PFP} and reviewed in 
\appendixname~\ref{appendix1}, we will perform the integration in the 
convolutions for the various contributions. We will report the exact results for the 
contributions to the EMT for given spectral indices in \appendixname~\ref{appendix2}.

\section{The scalar contribution}
\label{sec:three}

Scalar magnetized perturbations are sourced by the energy density, the scalar part of the 
Lorentz force and the scalar part of the anisotropic stress of the stochastic 
background of PMF. Due to the inhomogeneous nature of the stochastic background, the 
conservation law for the EMT of PMF implies that only two of the above quantities 
are independent and the following relation held:
\be
\sigma^{\text{PMF}} = \frac{\rho^{\text{PMF}}}{3} + L^{\text{PMF}}\,.
\ee
We will omit for simplicity the label {\em PMF} in the equations which follow.

\subsection{The energy density}

\begin{figure*}[!]
\begin{minipage}[c]{8.5cm}
\centering
\includegraphics[width=8cm]{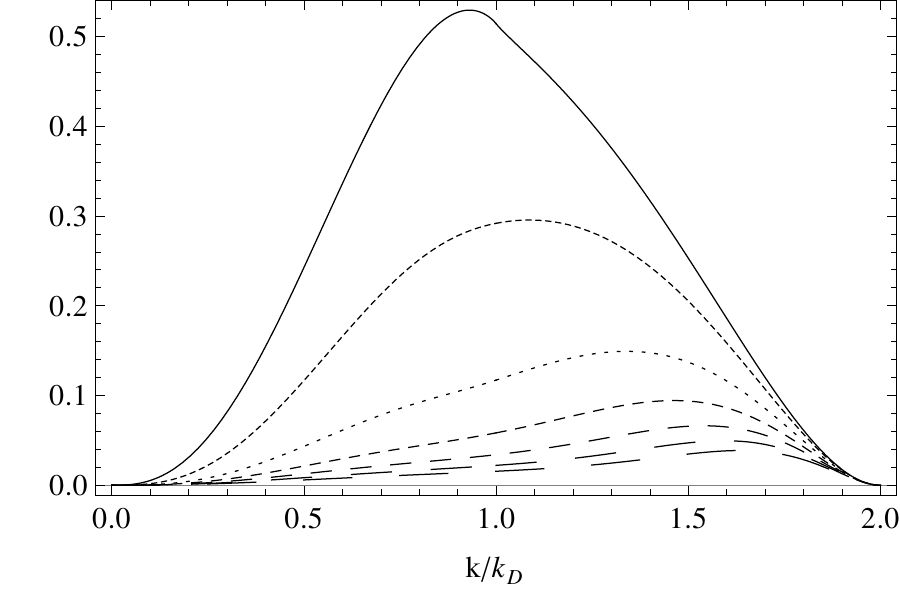} 
\includegraphics[width=8cm]{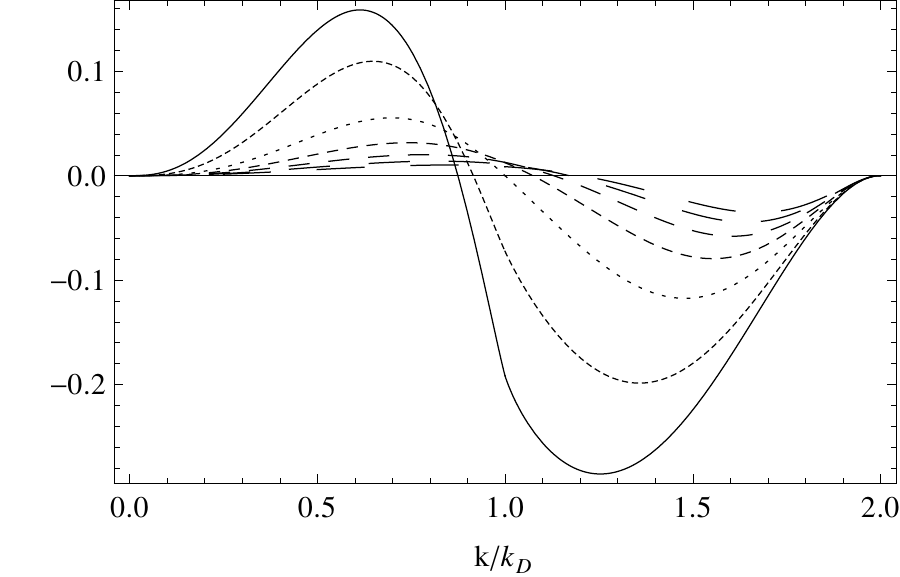}
\end{minipage}
\hspace{2mm}
\begin{minipage}[c]{8.5cm}
\centering
\includegraphics[width=8cm]{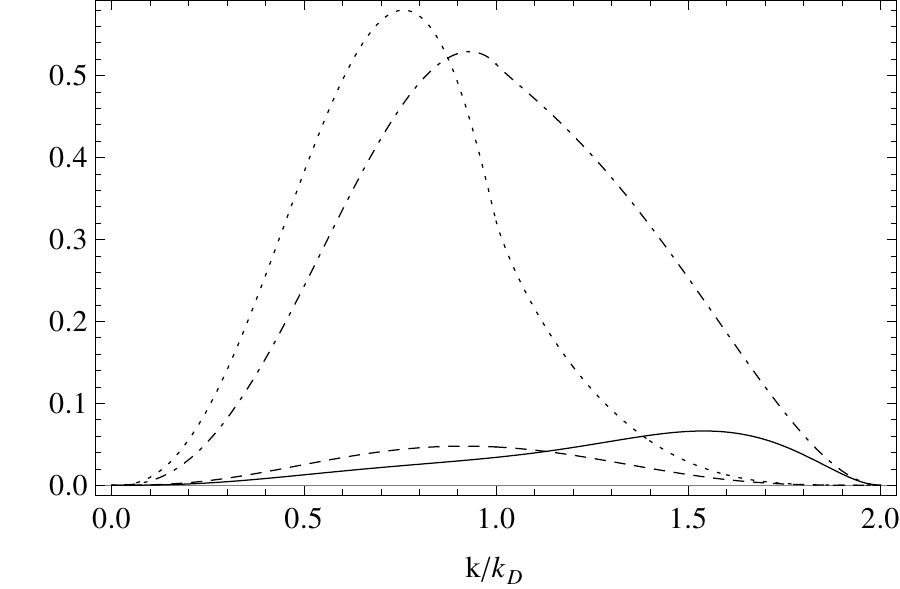}
\includegraphics[width=8cm]{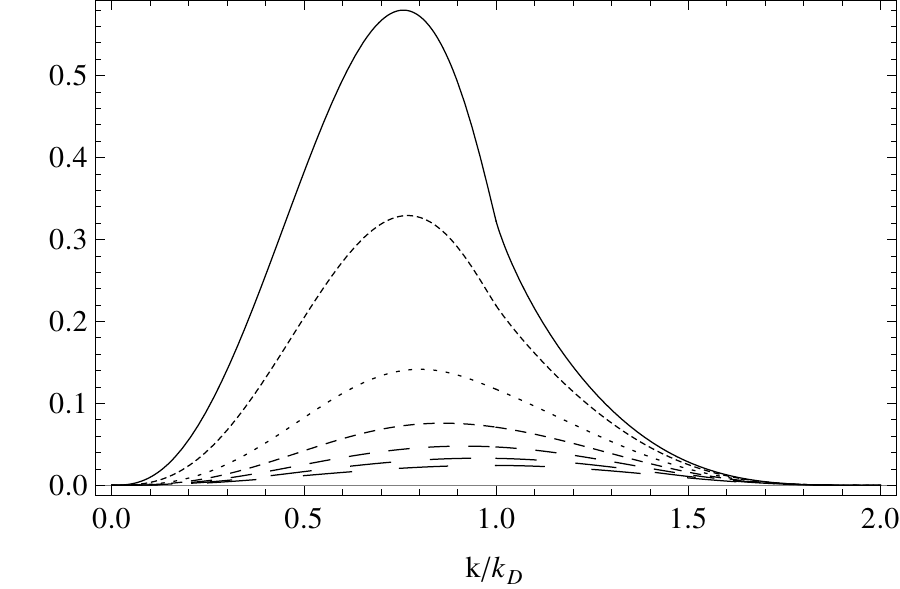}
\end{minipage}
\caption{Non-helical (helical) contribution to $k^3\big|\rho(k)\big|^2$ in units of 
$\langle B^2\rangle^2/(4\pi)^4$ ($\langle \mathcal{B}^2\rangle^2/(4\pi)^4$) versus 
$k/k_D$ is plotted in the upper left (bottom left) panel. The total contribution is 
displayed in the bottom right panel for $\langle B^2\rangle^2 = \langle \mathcal{B}^2\rangle^2$ 
and $n_B = n_H$. The different lines are for 
$n_B$ ($n_H$) = -3/2, -1, 0, 1, 2, 3, 4 
ranging from the solid to the longest dashed. The panel in the upper right display the 
comparison between the non-helical case and the maximal helical case for $n_{B,\,H}=1$ 
(solid vs dashed) and $n_{B,\,H}=-3/2$ (dot-dashed vs dotted).}
\label{fig:DensFig}
\end{figure*}

In this section we will describe the relevant terms of the scalar sector. The two-point 
correlation function of the energy density can be written in the Fourier space as:
\begin{align}
\label{eqn:ScalarCorrelator}
\langle\rho(\mathbf{k})\rho^*(\mathbf{h})\rangle\equiv&\,(2\pi)^3\delta^{(3)}(\mathbf{k}-\mathbf{h})\bigl|\rho(k)\bigr|^2\,, \notag\\
=&\,\delta_{ab}\delta_{cd}\langle\tau_{ab}(\mathbf{k})\tau^*_{cd}(\mathbf{h})\rangle\,.
\end{align}
Only the first two terms from Eq.~\eqref{eqn:stresstensor}, and their permutations, 
will contribute to this term and the energy density spectrum is therefore:
\begin{align}
\label{eqn:ScalarSpectrum}
\bigl|\rho(k)\bigr|^2 &\equiv \bigl|\rho_B(k)\bigr|^2 -2 \bigl|\rho_H(k)\bigr|^2 \notag \\
&=\int_\Omega \frac{d^3p}{(4\pi)^5}\Bigl[P_B(p)P_B(|\mathbf{k}-\mathbf{p}|)(1+\mu^2) \notag \\
&\quad \quad \quad \quad \quad \quad -2P_H(p)P_H(|\mathbf{k}-\mathbf{p}|)\mu\Bigr] \,,
\end{align}
where $\mu \equiv \hat{p}\cdot\widehat{(\mathbf{k}-\mathbf{p})}=\frac{k\gamma-p}{\sqrt{k^2-2kp\gamma+p^2}}$ and $\gamma \equiv \hat{k} \cdot \hat{p}$. 

For $k\ll k_D$ and $n_{B,\,H} > -3/2$ the energy density spectrum is:
\begin{align}
\bigl|\rho(k)\bigr|^2 \simeq\, &\frac{A_B^2\,k_D^{2n_B+3}}{128\pi^4 k_*^{2n_B}(2n_B+3)} \notag\\
&+ \frac{A_H^2\,k_D^{2n_H+3}}{128\pi^4 k_*^{2n_H}(2n_H+3)}\,.
\end{align}
For $n_{B,\,H}=-3/2$ we have a removable parametric divergence which is replaced by a 
logarithmic divergence in $k$, see the exact results in \appendixname~\ref{appendix2}.

See the panels in the left in Fig.~\ref{fig:DensFig} for the shape of $\bigl|\rho_B(k)\bigr|^2$ and 
$-2\bigl|\rho_H(k)\bigr|^2$ for different spectral indices. See the panel in the bottom right of Fig.~\ref{fig:DensFig} 
for the total contribution $\bigl|\rho(k)\bigr|^2$ in the maximal helical case, $A_B = A_H$, 
and $n_B = n_H$. The panel in the upper right of Fig.~\ref{fig:DensFig} displays the comparison 
of $\bigl|\rho(k)\bigr|^2$ in the non-helical case, $A_H = 0$, and in the maximal helical case.

\subsection{The scalar part of the Lorentz force}

\begin{figure*}[!]
\begin{minipage}[c]{8.5cm}
\centering
\includegraphics[width=8cm]{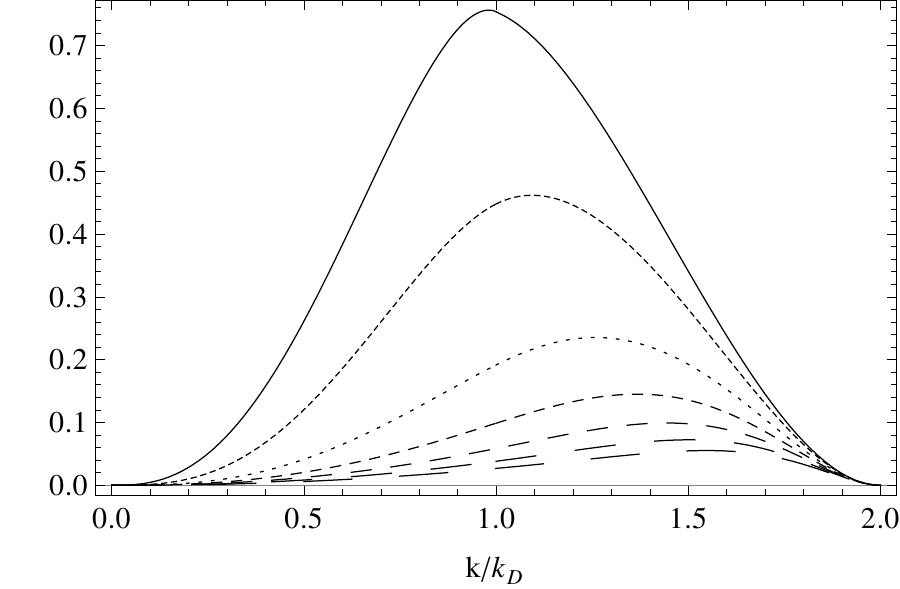}
\includegraphics[width=8cm]{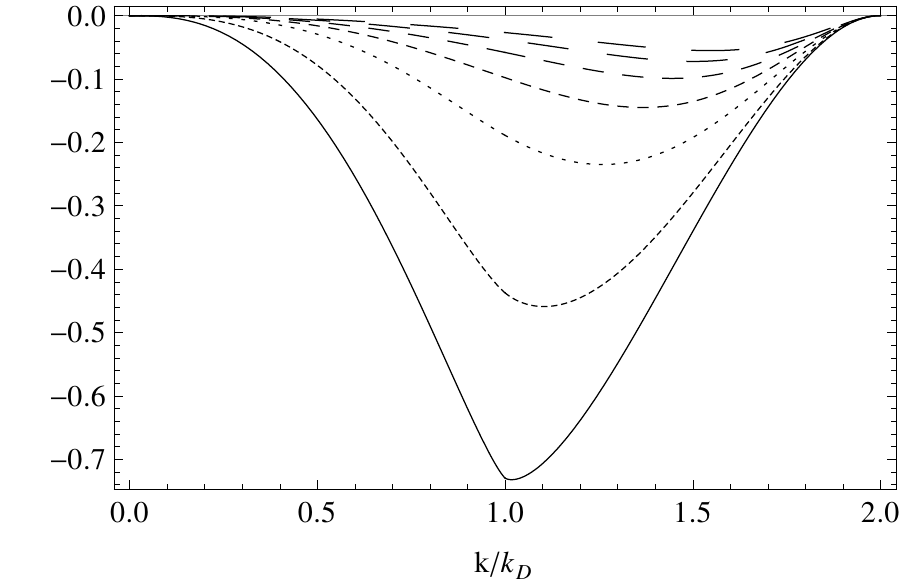}
\end{minipage}
\hspace{2mm}
\begin{minipage}[c]{8.5cm}
\centering
\includegraphics[width=8cm]{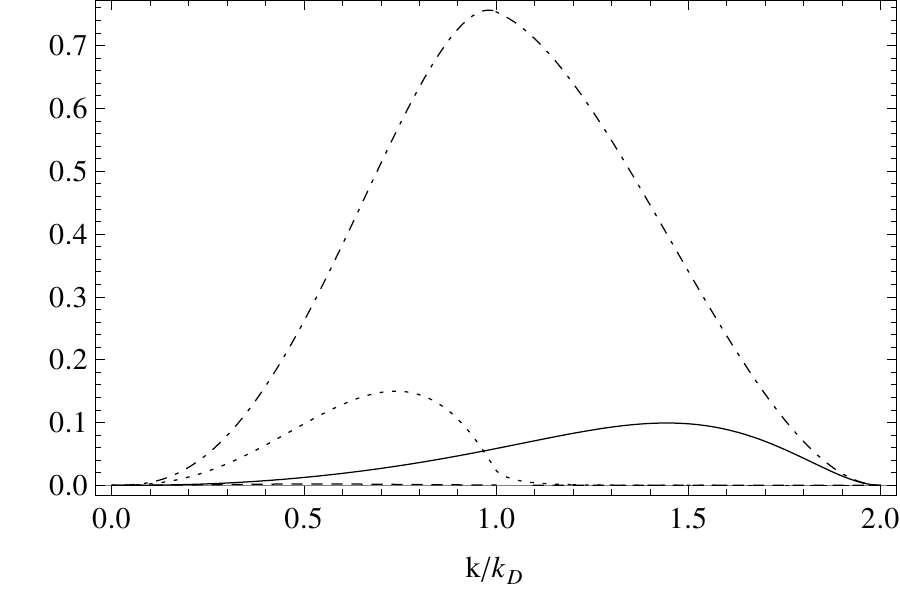}
\includegraphics[width=8cm]{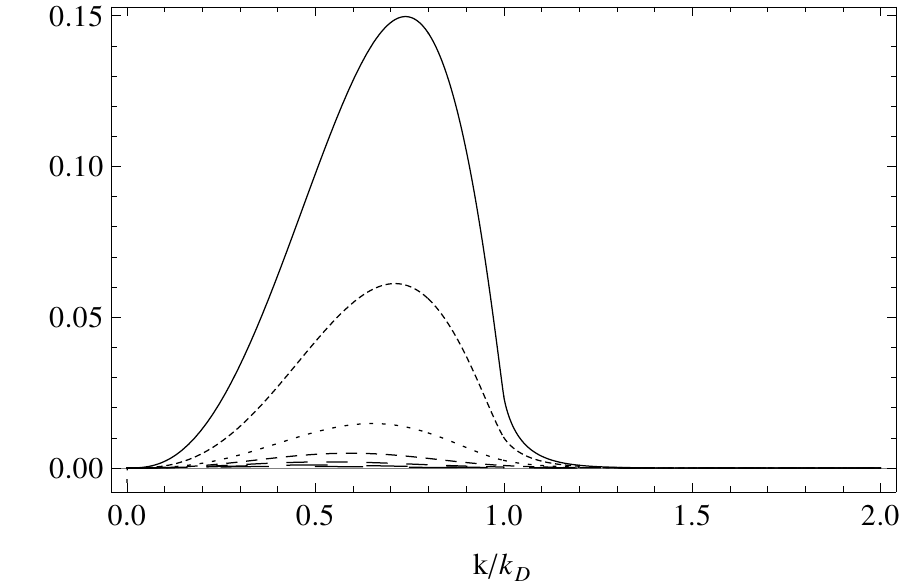}
\end{minipage}
\caption{Non-helical (helical) contribution to $k^3\big|L(k)\big|^2$ in units of 
$\langle B^2\rangle^2/(4\pi)^4$ ($\langle \mathcal{B}^2\rangle^2/(4\pi)^4$) versus 
$k/k_D$ is plotted in the upper left (bottom left) panel. The different lines are for 
$n_B$ ($n_H$) = -3/2, -1, 0, 1, 2, 3, 4 
ranging from the solid to the longest dashed. The panel in the upper right display the 
comparison between the non-helical case and the maximal helical case for $n_{B,\,H}=1$ 
(solid vs dashed) and $n_{B,\,H}=-3/2$ (dot-dashed vs dotted). The total contribution is 
displayed in the panel in the bottom right for $\langle B^2\rangle^2 = \langle \mathcal{B}^2\rangle^2$ 
and $n_B = n_H$.}
\label{fig:LorFig}
\end{figure*}

In order to compute the scalar contribution of a stochastic background of PMF to the 
cosmological perturbations, the convolution for the scalar part of the Lorentz force power 
spectrum is also necessary. In the MHD approximation, the Lorentz force is:
\be
L(\mathbf{x})=-\frac{1}{4\pi}\Bigl[\mathbf{B}(\mathbf{x})\times\bigl(\nabla\times\mathbf{B}(\mathbf{x})\bigr)\Bigr]\,,
\ee
and so the two-point correlation function in Fourier space is:
\begin{align}
\langle L(\mathbf{k})L^*(\mathbf{h})\rangle\equiv&\,(2\pi)^3\delta^{(3)}(\mathbf{k}-\mathbf{h})\big|L(k)\big|^2 \notag\\
=&\,\hat{k}_a\hat{k}_b\hat{k}_c\hat{k}_d\langle\tau_{ab}(\mathbf{k})\tau^*_{cd}(\mathbf{h})\rangle\,.
\end{align}
\begin{widetext}
The spectrum of the Lorentz force is:
\begin{align}
\label{eqn:LorentzSpectrum}
\big|L(k)\big|^2 &\equiv \big|L_B(k)\big|^2 + \big|L_H(k)\big|^2 \notag \\
&= \int_\Omega \frac{d^3p}{(4\,\pi)^5}\Bigl[P_B(p)P_B(|\mathbf{k}-\mathbf{p}|)\bigl(1+\mu^2+4\gamma^2\beta^2-4\gamma\beta\mu\bigr)+\,P_H(p)P_H(|\mathbf{k}-\mathbf{p}|)\bigl(2\mu-4\gamma\beta\bigr)\Bigr]\,,
\end{align}
\end{widetext}
where $\beta \equiv \hat{k}\cdot\widehat{(\mathbf{k}-\mathbf{p})}=\frac{k-p\gamma}{\sqrt{k^2-2kp\gamma+p^2}}$.

Also in this case the spectrum in the infrared limit, for $k\ll k_D$ and $n_{B,\,H} > -3/2$, behaves as:
\begin{align}
\big|L(k)\big|^2 \simeq\, &\frac{11A_B^2\,k_D^{2n_B+3}}{1920\pi^4 k_*^{2n_B}(2n_B+3)} \notag\\
&- \frac{A_H^2\,k_D^{2n_H+3}}{384\pi^4 k_*^{2n_H}(2n_H+3)}\,.
\end{align}

See the panels on the left in Fig.~\ref{fig:LorFig} for the shape of $\bigl|L_B(k)\bigr|^2$ and
$\bigl|L_H(k)\bigr|^2$ for different spectral indices. See the panel in the bottom right of 
Fig.~\ref{fig:LorFig} for the total contribution $\bigl|L(k)\bigr|^2$ in the maximal 
helical case, $A_B = A_H$, and $n_B = n_H$. Note from the panel in the upper right of 
Fig.~\ref{fig:LorFig} how the Lorentz force is decreased in the maximal helical case.
\begin{widetext}
The expression for the density-Lorentz force cross correlation \cite{shawlewis,PF}, 
including the helicity contribution, looks:
\be
\langle \rho(k)L^*(k)\rangle=\int_\Omega \frac{d^3p}{(4\,\pi)^5}\left[P_B(p)P_B(|\mathbf{k}-\mathbf{p}|)\bigl(1-\mu^2-2\gamma^2-2\beta^2-2\gamma\beta\mu\bigr)-\,P_H(p)P_H(|\mathbf{k}-\mathbf{p}|)\left(\frac{2}{3}\mu-2\gamma\beta\right)\right]\,.
\ee
\end{widetext}

\subsection{The scalar part of the anisotropic stress}

\begin{figure*}[!]
\begin{minipage}[c]{8.5cm}
\centering
\includegraphics[width=8cm]{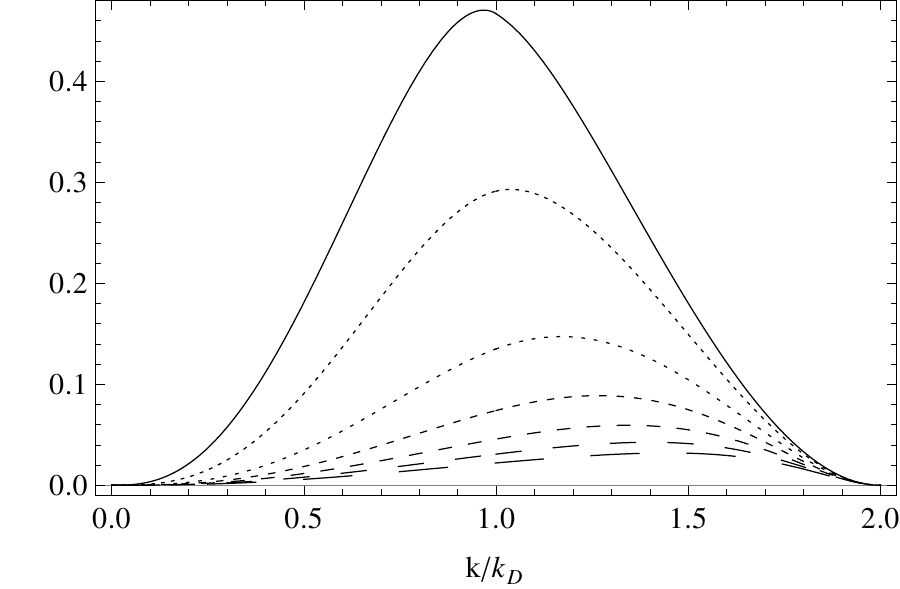}
\includegraphics[width=8cm]{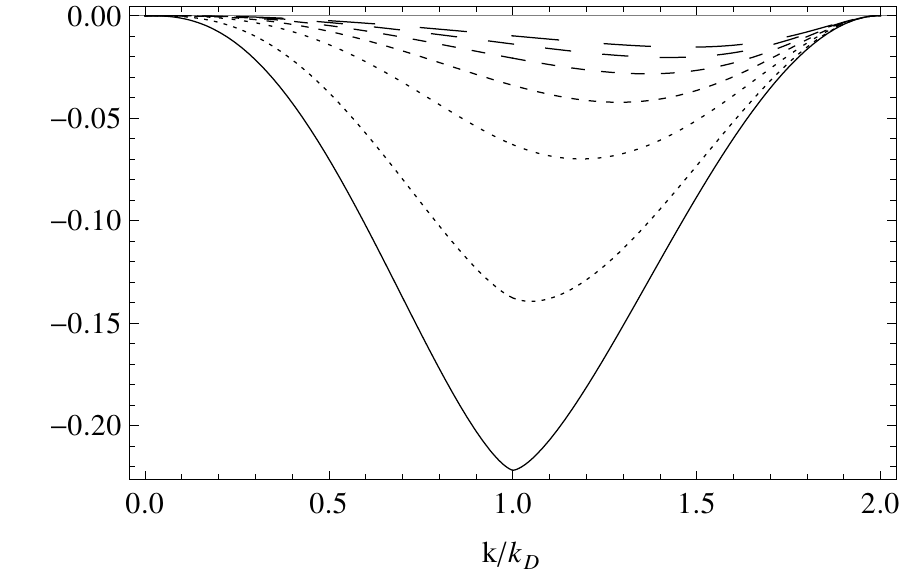}
\end{minipage}
\hspace{2mm}
\begin{minipage}[c]{8.5cm}
\centering
\includegraphics[width=8cm]{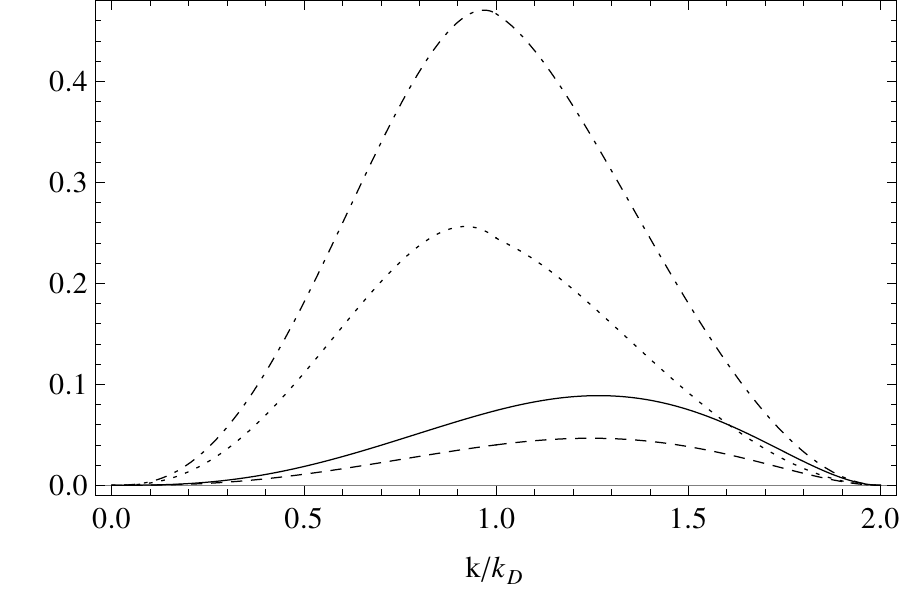}
\includegraphics[width=8cm]{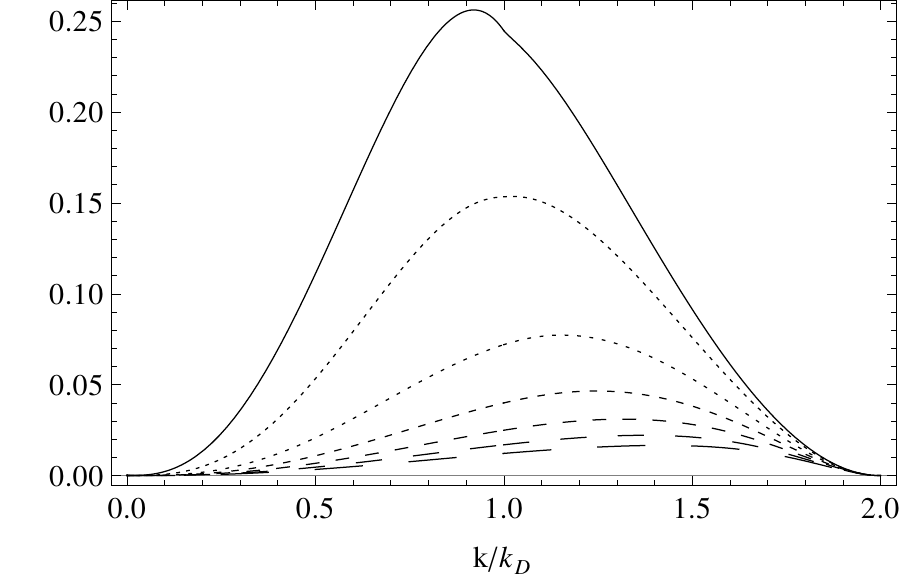}
\end{minipage}
\caption{Non-helical (helical) contribution to $k^3\big|\sigma(k)\big|^2$ in units of 
$\langle B^2\rangle^2/(4\pi)^4$ ($\langle \mathcal{B}^2\rangle^2/(4\pi)^4$) versus 
$k/k_D$ is plotted in the upper left (bottom left) panel. The different lines are for 
$n_B$ ($n_H$) = -3/2, -1, 0, 1, 2, 3, 4 
ranging from the solid to the longest dashed. The panel in the upper right display the 
comparison between the non-helical case and the maximal helical case for $n_{B,\,H}=1$ 
(solid vs dashed) and $n_{B,\,H}=-3/2$ (dot-dashed vs dotted). The total contribution is 
displayed in the bottom right panel for $\langle B^2\rangle^2 = \langle \mathcal{B}^2\rangle^2$ 
and $n_B = n_H$.}
\label{fig:StrFig}
\end{figure*}

For completeness we also write the scalar part of the anisotropic stress in function 
of the shear stress like \cite{Ma}:
\be
\sigma\equiv-\Bigl(\widehat{k}_i\widehat{k}_j-\frac{1}{3}\delta_{ij}\Bigr)\Sigma_{ij} \,,
\ee
where the stress shear is defined in our convention by:
\be
\Sigma_{ij}=\tau_{ij}-\frac{1}{3}\delta_{ij}\tau_{ll}\,. 
\ee
We are interested in the power spectrum of this quantity that is derived from the 
two-point correlator function as:
\be
\langle\Pi^{(S)}_{ij}(\mathbf{k})\Pi^{(S)*}_{lm}(\mathbf{h})\rangle\equiv(2\pi)^3\delta^{(3)}(\mathbf{k}-\mathbf{h})\big|\sigma(k)\big|^2\,.
\ee
\begin{widetext}
After a little algebra the spectrum reads:
\begin{align}
\label{eqn:StressSpectrum}
\big|\sigma(k)\big|^2 &\equiv \big|\sigma_B(k)\big|^2 + \big|\sigma_H(k)\big|^2 \notag \\
&= \int_\Omega \frac{d^3p}{(4\pi)^5}\left\{P_B(p)P_B(|\mathbf{k}-\mathbf{p}|)\left[\frac{4}{9}(4+\mu^2-3\gamma^2-3\beta^2+9\gamma^2\beta^2-6\gamma\beta\mu)\right]+P_H(p)P_H(|\mathbf{k}-\mathbf{p}|)\left(\frac{16}{9}\mu-\frac{8}{3}\gamma\beta\right)\right\}\,.
\end{align}
\end{widetext}

See the panels on the right in Fig.~\ref{fig:StrFig} for the shape of $\bigl|\sigma_B(k)\bigr|^2$ and 
$\bigl|\sigma_H(k)\bigr|^2$ for different spectral indices. See the panel in the bottom right of 
Fig.~\ref{fig:StrFig} for the total contribution $\bigl|\sigma(k)\bigr|^2$ in the maximal 
helical case, $A_B = A_H$, and $n_B = n_H$.

\section{The vector contribution}
\label{sec:four}

\begin{figure*}[!]
\begin{minipage}[c]{8.5cm}
\centering
\includegraphics[width=8cm]{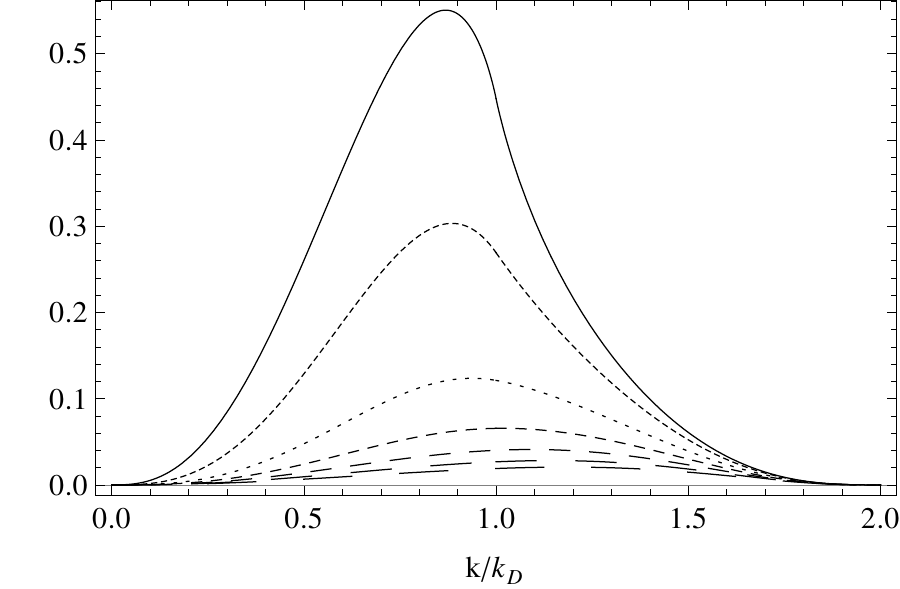} 
\includegraphics[width=8cm]{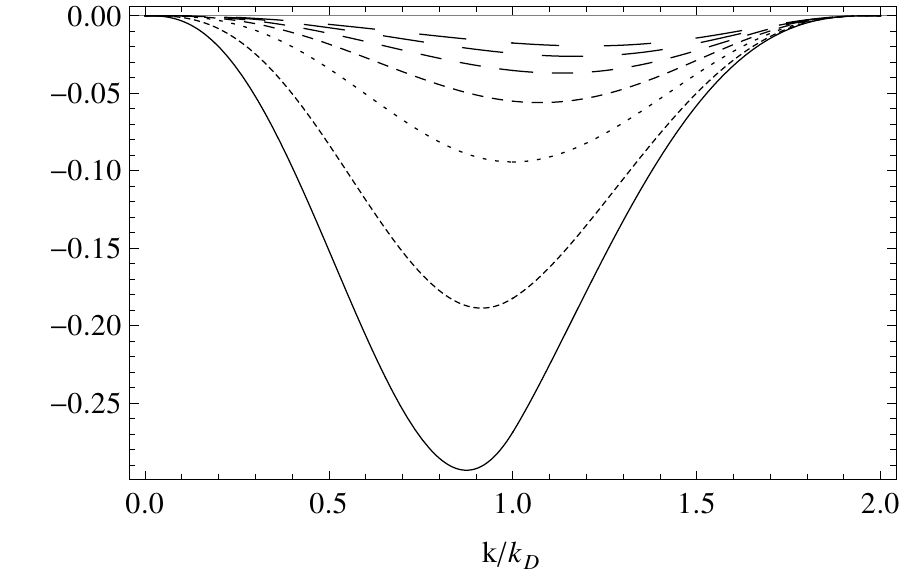}
\end{minipage}
\hspace{2mm}
\begin{minipage}[c]{8.5cm}
\centering
\includegraphics[width=8cm]{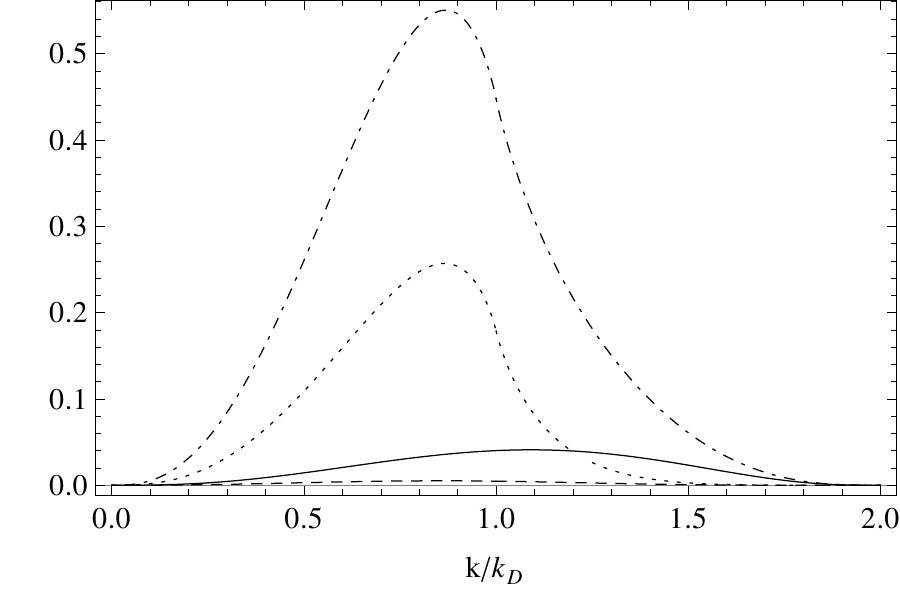}
\includegraphics[width=8cm]{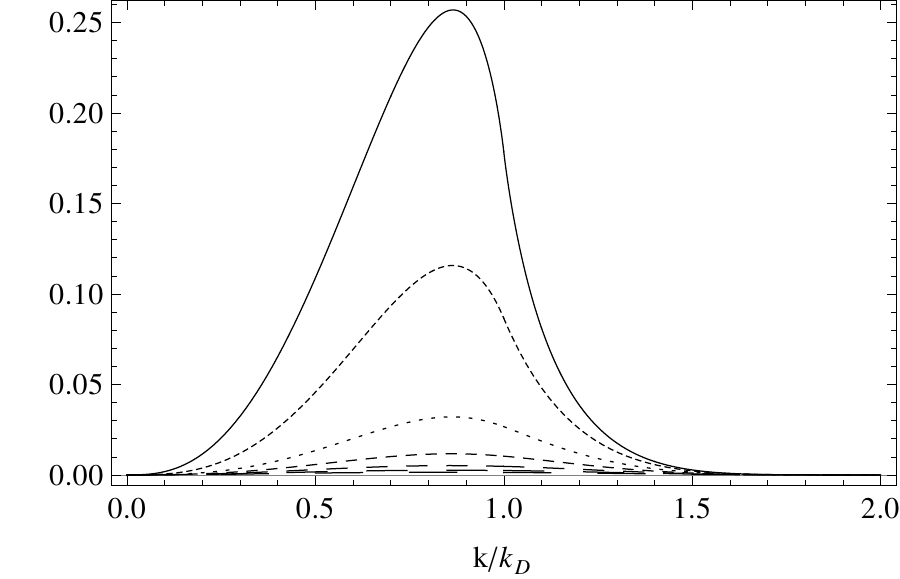}
\end{minipage}
\caption{Non-helical (helical) contribution to $k^3\big|\Pi^{(V)}(k)\big|^2$ in units of 
$\langle B^2\rangle^2/(4\pi)^4$ ($\langle \mathcal{B}^2\rangle^2/(4\pi)^4$) versus 
$k/k_D$ is plotted in the upper left (bottom left) panel. The different lines are for 
$n_B$ ($n_H$) = -3/2, -1, 0, 1, 2, 3, 4 
ranging from the solid to the longest dashed. The panel in the upper right display the 
comparison between the non-helical case and the maximal helical case for $n_{B,\,H}=1$ 
(solid vs dashed) and $n_{B,\,H}=-3/2$ (dot-dashed vs dotted). The total contribution is 
displayed in the bottom left panel for $\langle B^2\rangle^2 = \langle \mathcal{B}^2\rangle^2$ 
and $n_B = n_H$.}
\label{fig:VecFig}
\end{figure*}

In the standard $\Lambda$CDM model vector modes decay with the expansion of the Universe 
and have no observational signature at any significant level. However the associated 
temperature fluctuations, once generated, do not decay but in this case they have to be 
sourced by some shear, \cite{lewis:vector}.

PMF carrying vector anisotropic stress generate a fully magnetized vector mode that is 
the dominant PMF compensated contribution to the CMB angular power spectra on small angular 
scales. On these scales the primary CMB is suppressed by Silk damping therefore magnetic 
vector mode dominates over CMB angular power spectrum as shown in \cite{PF_WMAP7SPT}.
The vector contribution to $\tau_{ab}$ is given by:
\be
\Pi_i^{(V)}\equiv\hat{k}_a P_{ib}(k) \tau_{ab}\,.
\ee
We introduce the two-point correlation function for the vector source in the Fourier 
space, which can be parametrized as:
\ba
&\langle \Pi^{(V)}_i(\mathbf{k})\Pi^{(V)*}_j(\mathbf{h}) \rangle
\equiv\frac{(2\pi)^3}{2}\delta^{(3)}(\mathbf{k}-\mathbf{h}) \times \notag\\
&\qquad\qquad\times\Bigl[P_{ij}(k) \big|\Pi^{(V)}(k)\big|^2 +\imath\epsilon_{ijl}\hat{k}_l X^{(V)}(k)\Bigr]\,.
\ea
Differently from the scalar case, the two-point correlation function for the vector 
source include an antisymmetric component. It is easy to separate the symmetric and 
the antisymmetric parts of the source spectra:
\begin{widetext}
\begin{align}
\label{eqn:SymmVec}
(2\pi)^3\delta^{(3)}(\mathbf{k}-\mathbf{h}) \big|\Pi^{(V)}(k)\big|^2 \equiv&\,
\frac{1}{2}\Bigl[P_{ai}(k)\hat{k}_b P_{ci}(h)\hat{h}_d+P_{bi}(k)\hat{k}_aP_{di}(h)\hat{h}_c\Bigr] 
\langle \tau_{ab}(\mathbf{k})\tau^*_{cd}(\mathbf{h})\rangle\,, \\
\label{eqn:AntiVec}
(2\pi)^3\delta^{(3)}(\mathbf{k}-\mathbf{h}) X^{(V)}(k)\equiv&
-\frac{\imath}{2}\hat{k}_i \Bigl[\epsilon_{bdi}\hat{k}_a\hat{h}_c+\epsilon_{aci}\hat{k}_b\hat{h}_d\Bigr] 
\langle \tau_{ab}(\mathbf{k})\tau^*_{cd}(\mathbf{h})\rangle\,.
\end{align}
We obtain:
\begin{align}
\label{eqn:VectorSpectrumS}
\big|\Pi^{(V)}(k)\big|^2 &\equiv \big|\Pi^{(V)}_B(k)\big|^2 - \big|\Pi^{(V)}_H(k)\big|^2 \notag \\
&= 2\int_\Omega \frac{d^3p}{(4\pi)^5}\Bigl\{P_B(p)P_B(|\mathbf{k}-\mathbf{p}|)
\bigl[(1+\beta^2)(1-\gamma^2)+\gamma\beta(\mu-\gamma\beta)\bigr] -P_H(p)P_H(|\mathbf{k}-\mathbf{p}|)(\gamma\beta-\mu)\Bigr\}\,, \\
\label{eqn:VectorSpectrumA}
X^{(V)}(k)
&=\int_\Omega \frac{d^3p}{(4\pi)^5}\biggl\{P_B(p)P_H(|\mathbf{k}-\mathbf{p}|)
\bigl[\beta(1-\gamma^2)-(\gamma\beta-\mu)\gamma\bigr] +P_H(p)P_B(|\mathbf{k}-\mathbf{p}|)\bigl[\gamma(1-\beta^2)-(\gamma\beta-\mu)\beta\bigr]\biggr\}\,.
\end{align}
\end{widetext}

The behaviour of $\big|\Pi^{(V)}(k)\big|^2$ for $k\ll k_D$ and $n_{B,\,H} > -3/2$ 
has a white noise behaviour:
\begin{align}
\label{eqn:VecPole}
\big|\Pi^{(V)}(k)\big|^2 \simeq\, &\frac{7A_B^2\,k_D^{2n_B+3}}{960\pi^4 k_*^{2n_B}(2n_B+3)} \notag\\
&- \frac{A_H^2\,k_D^{2n_H+3}}{192\pi^4 k_*^{2n_H}(2n_H+3)}\,.
\end{align}
with a logarithmic divergence at $n_{B,\,H}=-3/2$. 
The antisymmetric spectrum has a different slope and is linear in $k$ for large wavelengths 
$k\ll k_D$ and for $n_B+n_H > -2$:
\be
X^{(V)}(k) \simeq \frac{A_B\,A_H\,k_D^{n_B+n_H+2}}{960\pi^4 k_*^{n_B+n_H}(n_B+n_H+2)}k \,.
\ee
The numerical coefficients obtained with 
semi-analytical approximation of the angular integral for the vector spectra in 
\cite{Kahniashvili:2005xe} need to be multiplied, in our conventions, to 14/15 for 
$|\Pi^{(V)}_B(k)\big|^2$, as pointed in \cite{PFP}; the numerical coefficient for 
$|\Pi^{(V)}_H(k)\big|^2$ is in agreement with Ref.~\cite{Kahniashvili:2005xe}. A larger 
numerical coefficient $1/5$ is needed for previous calculations  
which neglected the angular integration to match our result $X^{(V)}(k)$ \cite{Kahniashvili:2005xe}.\\
The pole at $n_B+n_H=-2$ in Eq.~\eqref{eqn:VecPole} is removable and we find for this choice 
of parameters:
\be
X^{(V)}(k) \simeq -\frac{A_B\,A_H\,k_*^2}{5760\pi^4}k\log (k/k_D)\,.
\ee
For $n_{B,\,H}=-3/2$ we obtain:
\be
X^{(V)}(k) \simeq \frac{A_B\,A_H\,k_*^3}{1536\pi^3} \,.
\ee
Note that the convolution integral for $X^{(V)}(k)$ in Eq.~\eqref{eqn:VectorSpectrumA} in the 
maximal helical case does not require infrared cut-offs for $n_{B}+n_{H}>-3$.

As for the scalar parts, Fig.~\ref{fig:VecFig} displays on the left column the non-helical and 
helical part of the vector anisotropies $\big|\Pi^{(V)}(k)\big|^2$ when the spectral index is varied. 
The panel in the upper right displays the comparison between the non-helical and the helical 
case for the symmetric vector spectrum. The panel in the bottom right displays the total 
$\big|\Pi^{(V)}(k)\big|^2$.

\section{The tensor contribution}
\label{sec:five}

\begin{figure*}[!]
\begin{minipage}[c]{8.5cm}
\centering
\includegraphics[width=8cm]{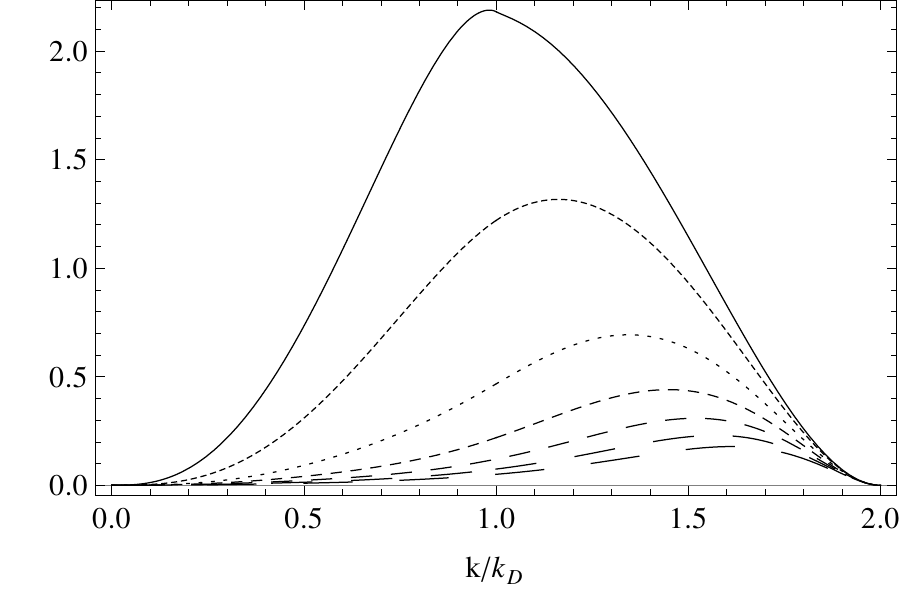} 
\includegraphics[width=8cm]{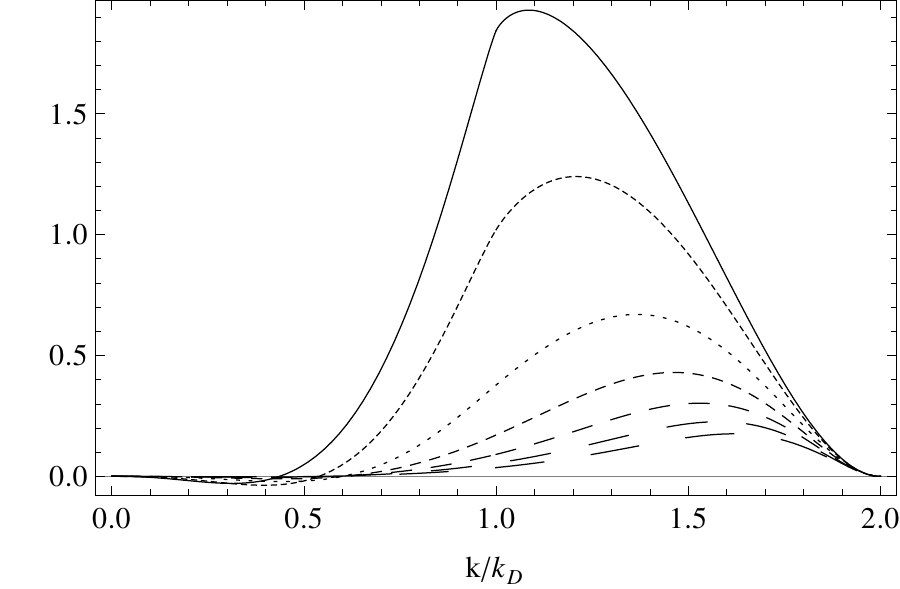}
\end{minipage}
\hspace{2mm}
\begin{minipage}[c]{8.5cm}
\centering
\includegraphics[width=8cm]{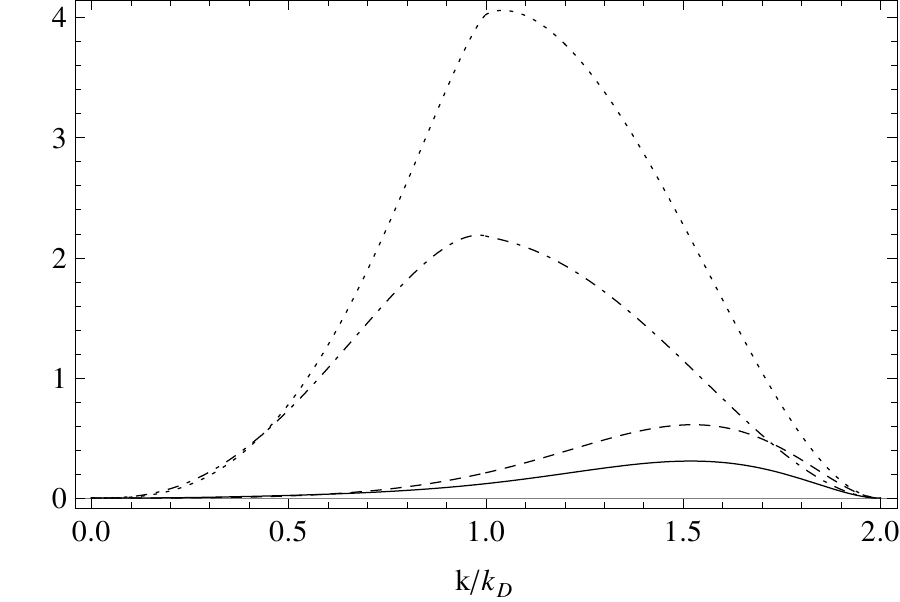}
\includegraphics[width=8cm]{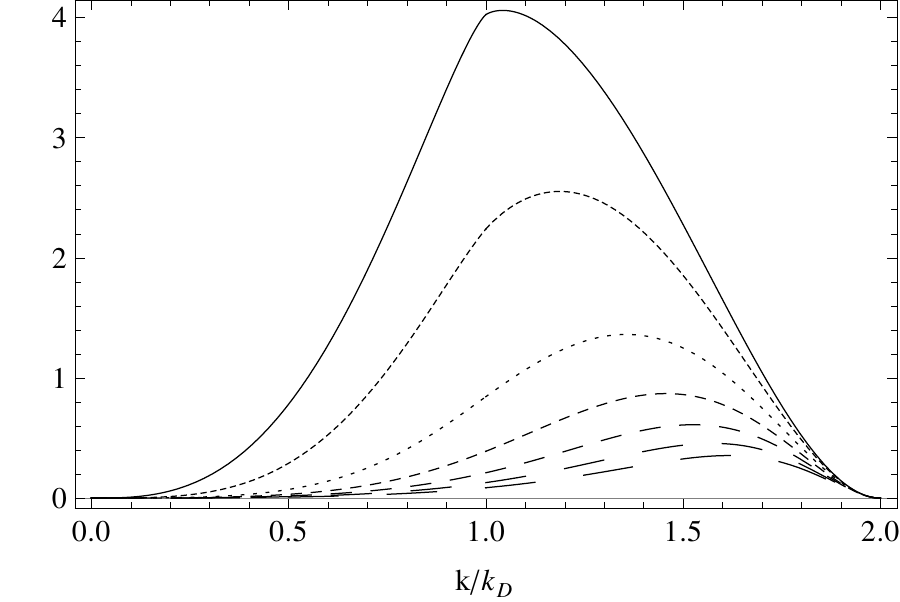}
\end{minipage}
\caption{Non-helical (helical) contribution to $k^3\big|\Pi^{(T)}(k)\big|^2$ in units of 
$\langle B^2\rangle^2/(4\pi)^4$ ($\langle \mathcal{B}^2\rangle^2/(4\pi)^4$) versus 
$k/k_D$ is plotted in the upper left (bottom left) panel. The different lines are for 
$n_B$ ($n_H$) = -3/2, -1, 0, 1, 2, 3, 4 
ranging from the solid to the longest dashed. The panel in the upper right display the 
comparison between the non-helical case and the maximal helical case for $n_{B,\,H}=1$ 
(solid vs dashed) and $n_{B,\,H}=-3/2$ (dot-dashed vs dotted). The total contribution is 
displayed in the bottom right panel for $\langle B^2\rangle^2 = \langle \mathcal{B}^2\rangle^2$ 
and $n_B = n_H$.}
\label{fig:TenFig}
\end{figure*}

PMF source tensor modes from tensor anisotropic pressure. The tensor part of the magnetic 
field EMT is given by:
\begin{equation}
\Pi_{ij}^{(T)}(\mathbf{k})\equiv\mathcal{P}_{ijab}(k)\tau_{ab}(\mathbf{k})\,,
\end{equation}
with the tensor projector $\mathcal{P}_{ijab}$ as:
\be
\mathcal{P}_{ijab}(k)=P_{ia}(k)P_{jb}(k)-\frac{1}{2}P_{ij}(k)P_{ab}(k)\,.
\ee
We define the tensor projector to apply on 
$\langle \tau_{ab}(\mathbf{k})\tau^*_{cd}(\mathbf{h})\rangle$ as:
\begin{equation}
\label{eqn:tensorproj}
\mathcal{P}^{abcd}_{ijlm}(k,h)\equiv\mathcal{P}_{ijab}(k)\mathcal{P}_{lmcd}(h)\,.
\end{equation}
As for the vector case, we introduce the two-point correlation function for the tensor source as:
\ba
\label{eqn:tensorsource}
&\langle \Pi^{(T)}_{ij}(\mathbf{k})\Pi^{(T)*}_{lm}(\mathbf{h})\rangle\equiv\frac{(2\pi)^3}{4}\delta^{(3)}(\mathbf{k}-\mathbf{h}) \times \notag\\
&\qquad\qquad\times\Bigl[\mathcal{M}_{ijlm}\big|\Pi^{(T)}(k)\big|^2+\imath\mathcal{A}_{ijlm} X^{(T)}(k)\Bigr]\,,
\ea
where the tensors $\mathcal{M}_{ijlm}$ and $\mathcal{A}_{ijlm}$ are given by:
\begin{align}
\mathcal{M}_{ijlm}\equiv P_{il}&P_{jm}+P_{im}P_{jl}-P_{ij}P_{lm}\,,\\
\mathcal{A}_{ijlm}\equiv
\frac{\hat{k}_t}{2}\bigl(&P_{il}\epsilon_{jmt}+P_{im}\epsilon_{jlt}  \notag\\
&+P_{jl}\epsilon_{imt}+P_{jm}\epsilon_{ilt}\bigr)\,.
\end{align}
Both $\mathcal{M}_{ijlm}$ and $\mathcal{A}_{ijlm}$ are symmetric under permutations 
$(i\leftrightarrow j)$ and $(l\leftrightarrow m)$; $\mathcal{M}_{ijlm}$ is also 
symmetric under the exchange of $(ij)\leftrightarrow(lm)$, whereas $\mathcal{A}_{ijlm}$ 
is antisymmetric under this permutation. We can summarize the previous rules with 
the properties:
\begin{align}
\label{eqn:propertyM}
&\mathcal{M}_{ijij}=4\,,\quad\mathcal{M}_{iilm}=\mathcal{M}_{ijll}=0\,,\\
\label{eqn:propertyA}
&\mathcal{A}_{ijij}=\mathcal{A}_{iilm}=\mathcal{A}_{ijll}=0\,,\\
&|\mathcal{M}|^2=|\mathcal{A}|^2=8\,,\\
&\mathcal{M}_{ijlm}\mathcal{A}_{ijlm}=0\,.
\end{align}
The source terms for the tensor parts are:
\begin{align}
(2\pi)^3\delta^{(3)}(\mathbf{k}-\mathbf{h}) \big|\Pi^{(T)}(k)\big|^2 &\equiv \frac{1}{2}\mathcal{M}_{abcd}\langle \tau_{ab}(\mathbf{k})\tau^*_{cd}(\mathbf{h})\rangle\,,\\
(2\pi)^3\delta^{(3)}(\mathbf{k}-\mathbf{h}) X^{(T)}(k) &\equiv -\frac{\imath}{2}\mathcal{A}_{abcd}\langle \tau_{ab}(\mathbf{k})\tau^*_{cd}(\mathbf{h}) \rangle\,.
\end{align}
\begin{widetext}
We find for the source spectra:
\begin{align}
\label{eqn:TensorSpectrumS}
\big|\Pi^{(T)}(k)\big|^2 &\equiv \big|\Pi^{(T)}_B(k)\big|^2 + 4\big|\Pi^{(T)}_H(k)\big|^2
= 2\int_\Omega \frac{d^3p}{(4\pi)^5} \Bigl[ 
P_B(p)P_B(|\mathbf{k}-\mathbf{p}|)(1+\gamma^2)(1+\beta^2)  
 +4 P_H(p)P_H(|\mathbf{k}-\mathbf{p}|)\gamma\beta\Bigr]\,, \\
\label{eqn:TensorSpectrumA}
X^{(T)}(k)
&=4\int_\Omega \frac{d^3p}{(4\pi)^5}\biggl[P_B(p)P_H(|\mathbf{k}-\mathbf{p}|)(1+\gamma^2)\beta 
+P_H(p)P_B(|\mathbf{k}-\mathbf{p}|)\gamma(1+\beta^2)\biggr]\,.
\end{align}
\end{widetext}
As for the vector sector we obtain an antisymmetric power spectrum. The tensor anisotropic 
stress spectra is similar to the vector ones for $n_{B,\,H}>-3/2$:
\begin{align}
\big|\Pi^{(T)}(k)\big|^2 \simeq\,& \frac{7A_B^2\,k_D^{2n_B+3}}{480\pi^4 k_*^{2n_B}(2n_B+3)} \notag\\
&- \frac{A_H^2\,k_D^{2n_H+3}}{96\pi^4 k_*^{2n_H}(2n_H+3)}\,, \notag\\
X^{(T)}(k) \simeq\,& \frac{A_B\,A_H\,k_D^{n_B+n_H+2}}{480\pi^4 k_*^{n_B+n_H}(n_B+n_H+2)}k\,.
\end{align}
In this case the numerical coefficients obtained with semi-analytical approch in 
\cite{CDK} differ from the exact result of a factor 28/15 for $|\Pi^{(T)}_B(k)\big|^2$ and 
1/2 for $|\Pi^{(T)}_H(k)\big|^2$. Moreover the relation between the vector and tensor 
anisotropic stresses is different: we found that for the white noise spectra is still valid 
the relation $|\Pi^{(T)}(k)\big|^2\simeq2|\Pi^{(V)}(k)\big|^2$ taking into account these new 
contributions to the even correlators. $X^{(T)}(k)$ is different by a factor 1/5.

The pole at $n_B+n_H=-2$ is removable and we find for the antisymmetric part:
\be
X^{(T)}(k) \simeq -\frac{A_B\,A_H\,k_*^2}{120\pi^4}k\log (k/k_D)\,.
\ee
For $n_{B,\,H}=-3/2$ we obtain:
\be
X^{(T)}(k) \simeq \frac{7A_B\,A_H\,k_*^3}{768\pi^3 }\,.
\ee
Note that the convolution integral for $X^{(T)}(k)$ in Eq.~\eqref{eqn:VectorSpectrumA} in 
the maximal helical case does not require infrared cut-offs for $n_{B}+n_{H}>-3$.

Fig.~\ref{fig:TenFig} displays on the left column the non-helical, 
$\big|\Pi_B^{(T)}(k)\big|^2$, and helical part, $4\big|\Pi_H^{(T)}(k)\big|^2$, 
of the tensor anisotropies $\big|\Pi^{(T)}(k)\big|^2$ when the spectral index is varied. 
The panel in the bottom right displays the total $\big|\Pi^{(T)}(k)\big|^2$ for the maximal 
helical case when $n_B = n_H$ is varying.

\begin{figure}[!htp]
\centering
\includegraphics[height=5cm]{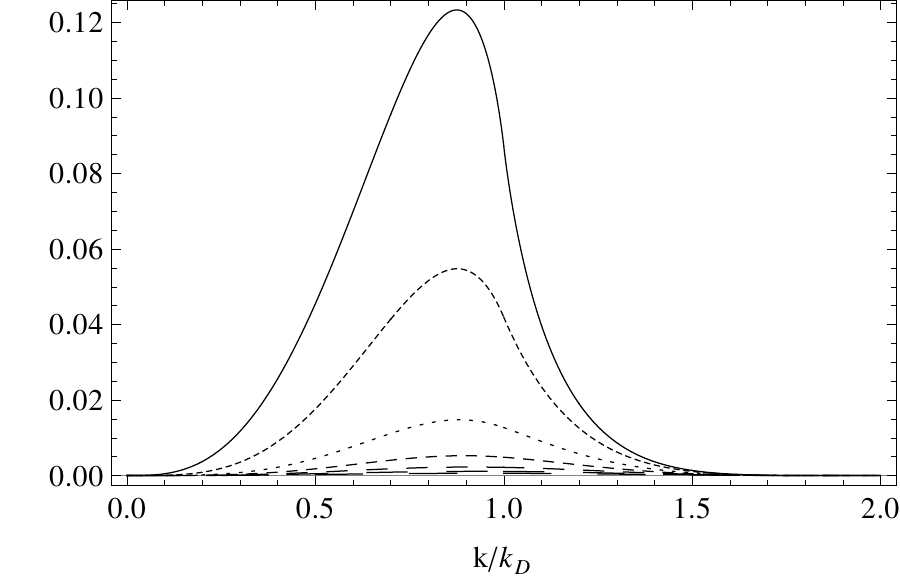}
\includegraphics[height=5cm]{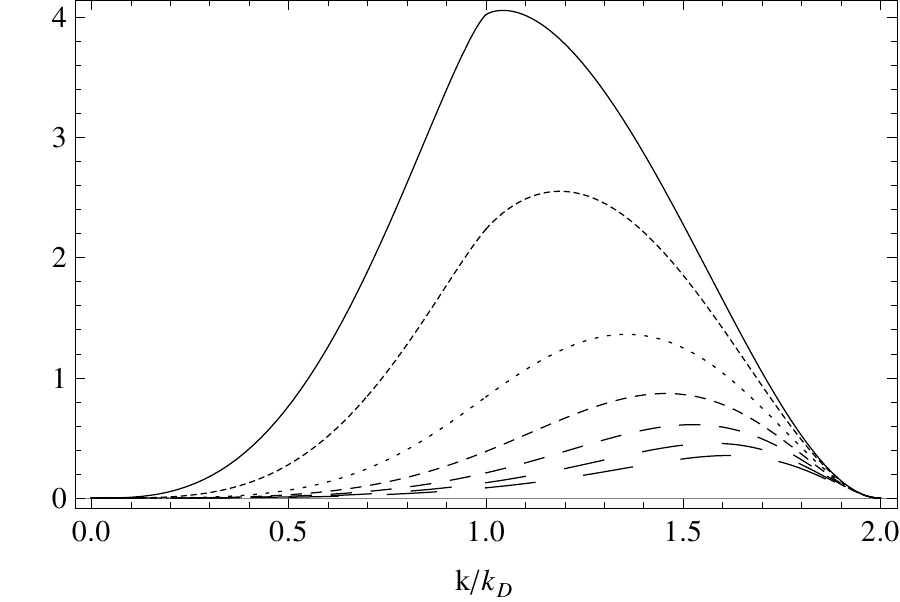}
\caption{Comparison of antisymmetric correlators in 
units of $\langle B^2\rangle \langle \mathcal{B}^2\rangle/(4\pi)^4$, 
the different lines are for $n_B = n_H = -3/2, -1, 0, 1, 2, 3, 4$ 
ranging from the solid to the longest dashed. The vector one, 
$k^3 X^{(V)}(k)$, in the upper panel and the tensor one, $k^3 X^{(T)}(k)$, in the bottom panel.}
\label{fig:AntiFig}
\end{figure}

The left panel of Fig.~\ref{fig:AntiFig} displays the antisymmetric $X^{(V)}(k)$ 
when varying $n_B = n_H$. The right panel correspond to the tensor one $X^{(T)}(k)$.

\section{CMB anisotropies}
\label{sec:six}

We now investigate how helicity changes the PMF contribution to CMB power spectrum anisotropies 
in temperature and polarization. We included the helical contribution of the PMF EMT in our 
modified version of the public Einstein-Boltzmann code CAMB \cite{CAMB} which was used  based 
on the already existent one from \cite{FPP,PFP} to derive the angular power spectra.

\subsection{The scalar contribution to CMB anisotropies} 

The scalar contribution is the sum of the helical and non-helical terms in the density, 
Lorentz and corresponding cross-correlations.

\begin{figure}[htp]
\includegraphics[width=8.5cm]{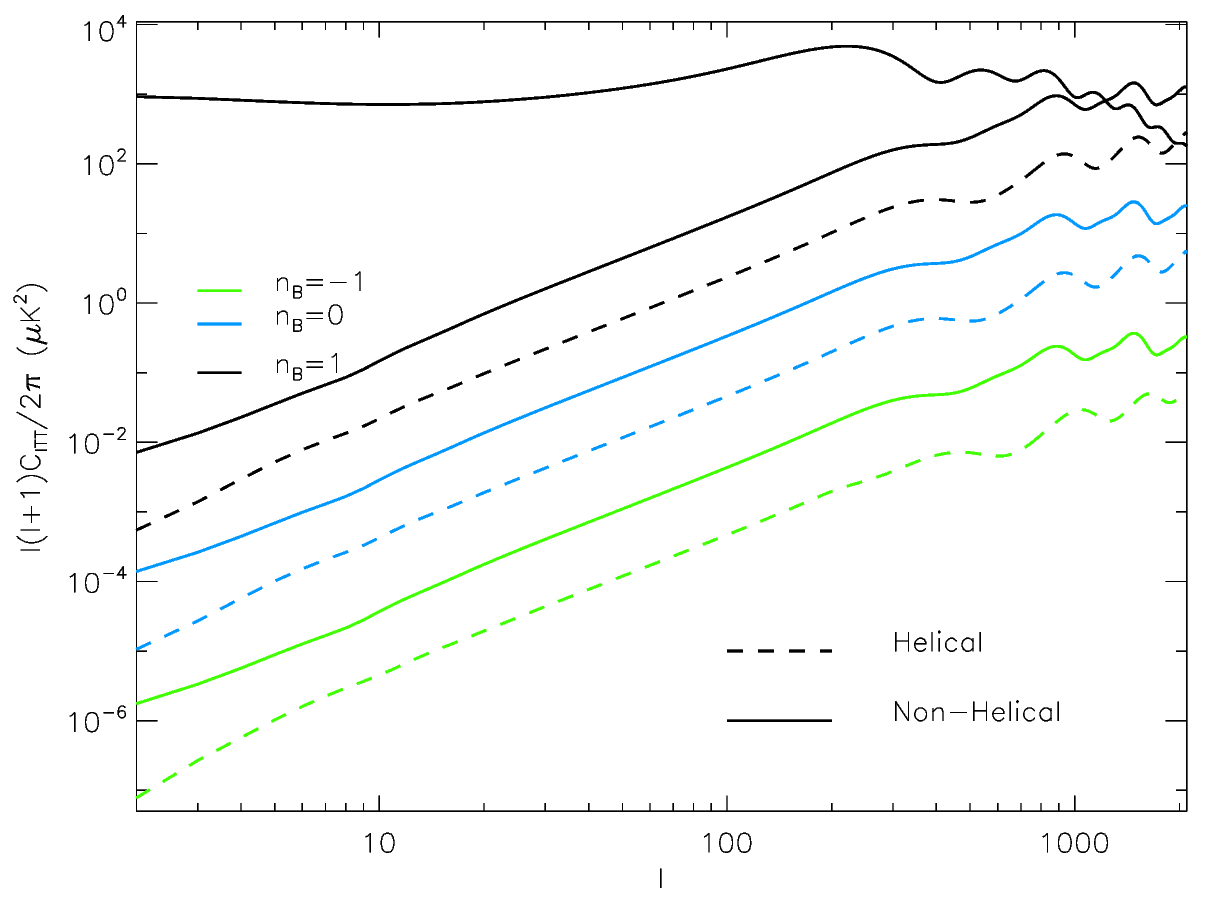} 
\caption{We show the scalar power spectrum with the cross-correlation between $\rho_B$ and $L_B$. The solid line is the adiabatic scalar contribution in comparison with the scalar contributions of a stochastic background of PMF for $\sqrt{\langle B^2_\lambda \rangle}=3.5\,nG$.}
\label{fig:ScalarCl}
\end{figure}

In Fig.~\ref{fig:ScalarCl} we show the contributions to the total CMB temperature angular 
power spectra from the scalar pure magnetic mode for different fixed spectral indices and its 
comparison with the adiabatic mode.

\subsection{The vector contribution to CMB anisotropies}

\begin{figure}[htp]
\centering
\includegraphics[width=8.5cm]{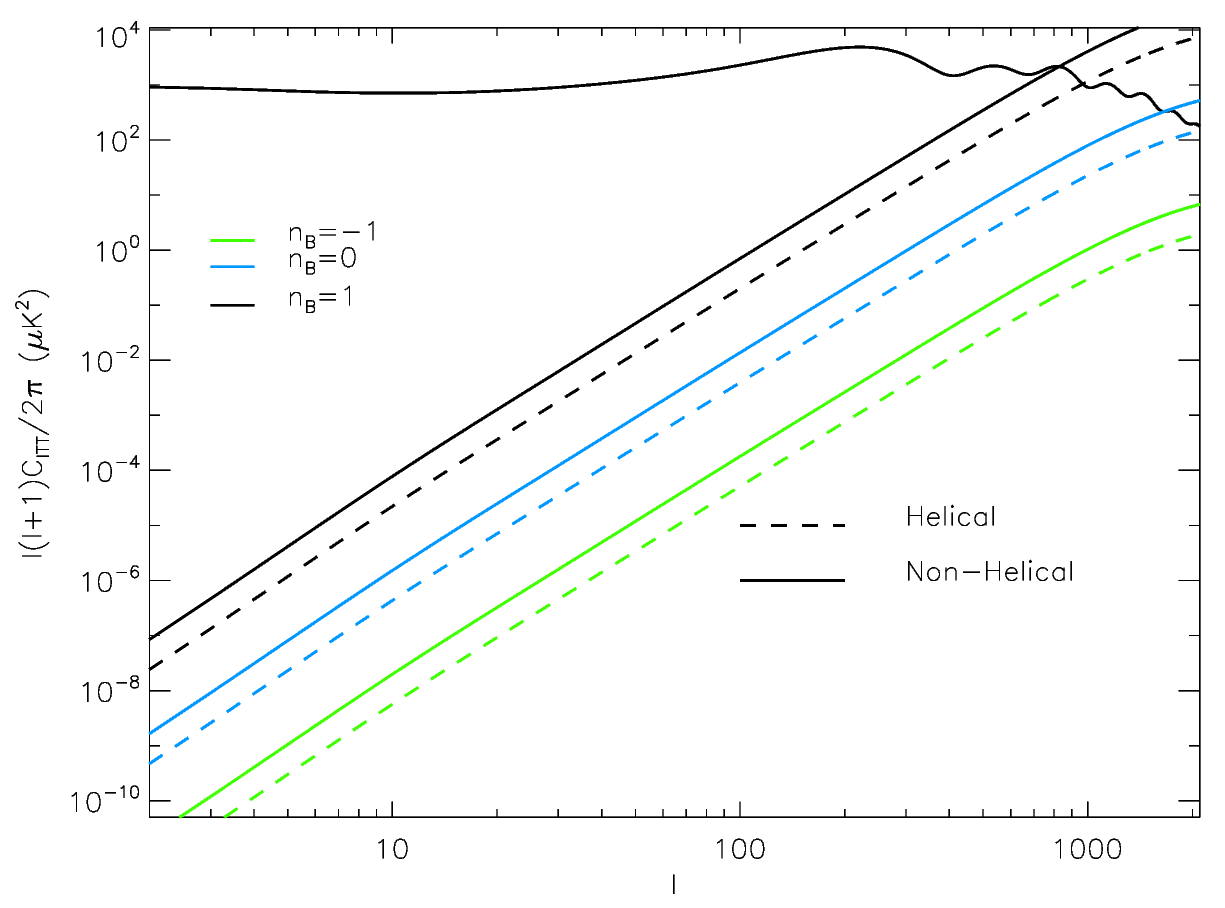}
\caption{CMB anisotropies angular power spectrum for temperature. 
The solid line is the adiabatic scalar contribution in comparison with the vector contributions of a stochastic background of PMF for $\sqrt{\langle B^2_\lambda\rangle}=3.5\ nG$.}
\label{fig:VecCl}
\end{figure}

To understand how the antisymmetric component of the vector source term 
in Eq.~\eqref{eqn:VectorSpectrumA} afflict the CMB power spectrum anisotropies it is 
useful to rewrite the spectrum in a polarization orthonormal base that for the helical 
case will be:
\be
\mathbf{e}^\pm(\mathbf{k})=-\frac{\imath}{\sqrt{2}}(\mathbf{e}^+\pm\imath\mathbf{e}^-)\,,
\ee
with the following properties:
\begin{align}
&\mathbf{e}^\pm\cdot\mathbf{e}^\mp=-1\,,\\
&\mathbf{e}^\pm\cdot\mathbf{e}^\pm=0\,,\\
&\mathbf{e}^\pm(\mathbf{k})=\mathbf{e}^\mp(-\mathbf{k})\,.
\end{align}
With this choice we obtain the decomposition:
\be
\Pi^{(V)}_i(\mathbf{k})=e_i^+\Pi^+_V(\mathbf{k})+e_i^-\Pi^-_V(\mathbf{k})
\ee
that allow us to rewrite \eqref{eqn:SymmVec} and \eqref{eqn:AntiVec} into:
\begin{align}
&(2\pi)^3\delta^{(3)}(\mathbf{k}-\mathbf{h})\big|\Pi^{(V)}_B(k)\big|^2= \notag\\
&\qquad\langle\Pi^+_V(\mathbf{k})\Pi^{+*}_V(\mathbf{-h})+\Pi^-_V(\mathbf{k})\Pi^{-*}_V(\mathbf{-h})\rangle\,, \\
&(2\pi)^3\delta^{(3)}(\mathbf{k}-\mathbf{h})X^{(V)}(k)= \notag\\
&\qquad-\langle\Pi^+_V(\mathbf{k})\Pi^{+*}_V(\mathbf{-h})-\Pi^-_V(\mathbf{k})\Pi^{-*}_V(\mathbf{-h})\rangle\,.
\label{eqn:AntiVecP}
\end{align}
In conclusion for the vector sector we will have two independent metric perturbation 
modes which  are sourced by combinations of $\big|\Pi^{(V)}\big|^2$ and $X^{(V)}$:
\be
\dot{h}^\pm_V+2\mathcal{H}h^\pm_V=-16\pi Ga^2\frac{\Pi_\nu^{(V)}+\Pi_\gamma^{(V)}+\Pi_V^\pm}{k}\,.
\ee
We note that the angular power spectrum peaks around $l\sim2000$ according to \cite{PFP,lewis}. 
The peak is in the region where primary CMB is suppressed by Silk damping, therefore magnetized 
vector anisotropies are the dominant compensated contribution on small scales. The vector 
part of the Lorentz force induced on baryons modifies the baryon vector velocity equation:
\be
\dot{v}_b+\mathcal{H}v_b=-\frac{\rho_\gamma}{\rho_b}\Bigl[\frac{4}{3}n_ea\sigma_T(v_b-v_\gamma)-\frac{L^V}{\rho_\gamma}\Bigr]\,.
\ee
Considering Eq.~\eqref{eqn:LorentzSpectrum} we will have a slightly deviation from the non-helical case.
Fig.~\ref{fig:VecCl} shows the vector contribution to the $TT$ spectrum and its dependence 
from the spectral indices.

Due to the helical contribution the parity odd CMB power spectra are non-zero. In 
particular their presence is due to the antisymmetric source Eq.~\eqref{eqn:VectorSpectrumA} 
which emphasizes the difference between the two polarizations $+$ and $-$. As shown in 
\cite{PVW} these antisymmetric sources generate the parity odd spectra 
$C_l^{TB},\,C_l^{EB}$, since they are given by momentum integrals of $X^{(V)}(k)$:
\begin{align}
&C_l^{TB}=\frac{2}{\pi} \int_0^\infty k^2dk \,X^{(V)}(k)\Delta_l(k)B_l(k) \,, \\
&C_l^{EB}=\frac{2}{\pi} \int_0^\infty k^2dk \,X^{(V)}(k)E_l(k)B_l(k) \,,
\end{align}
where $\Delta_l(k)$, $E_l(k)$ and $B_l(k)$ contain all the information about the CMB transfer 
functions.

\begin{figure}[htp]
\centering
\includegraphics[width=8.5cm]{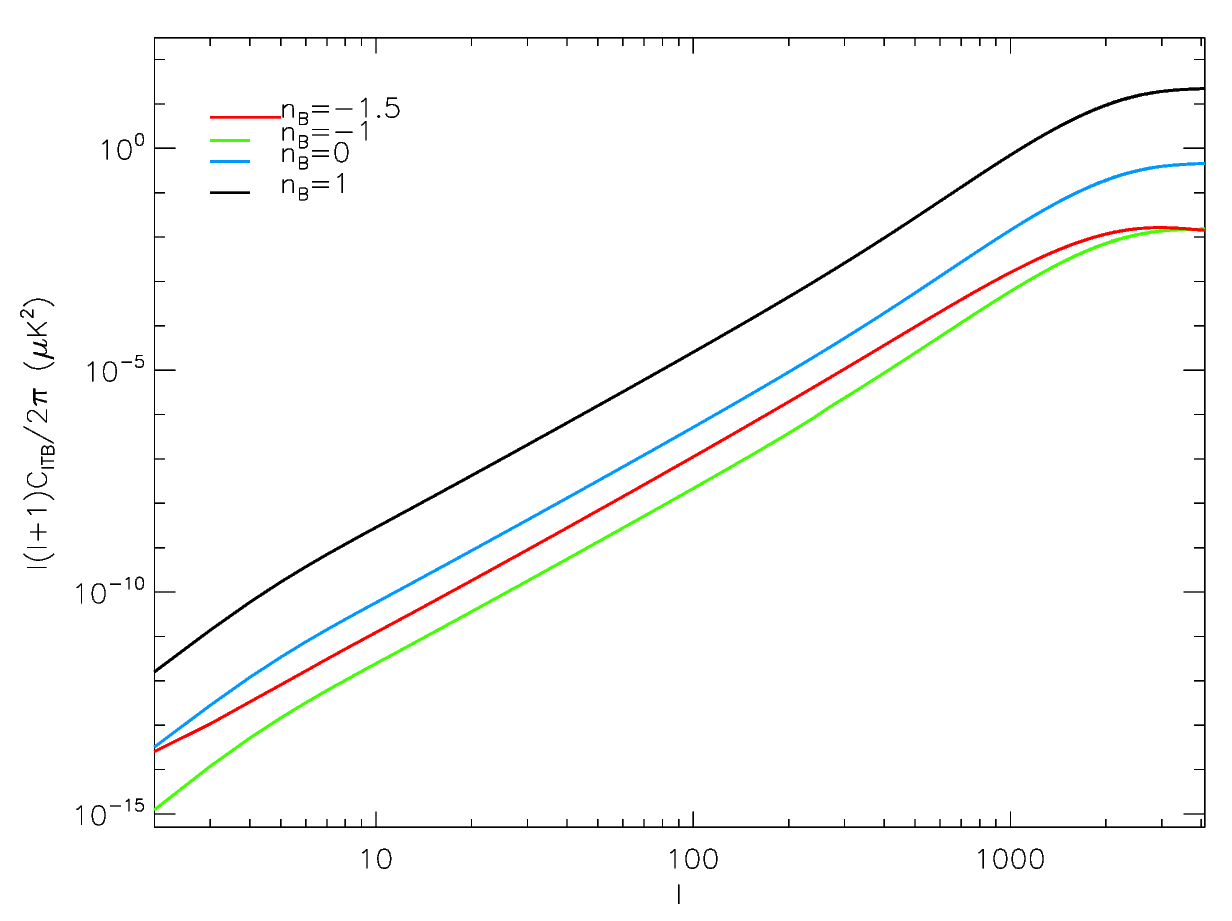}
\includegraphics[width=8.5cm]{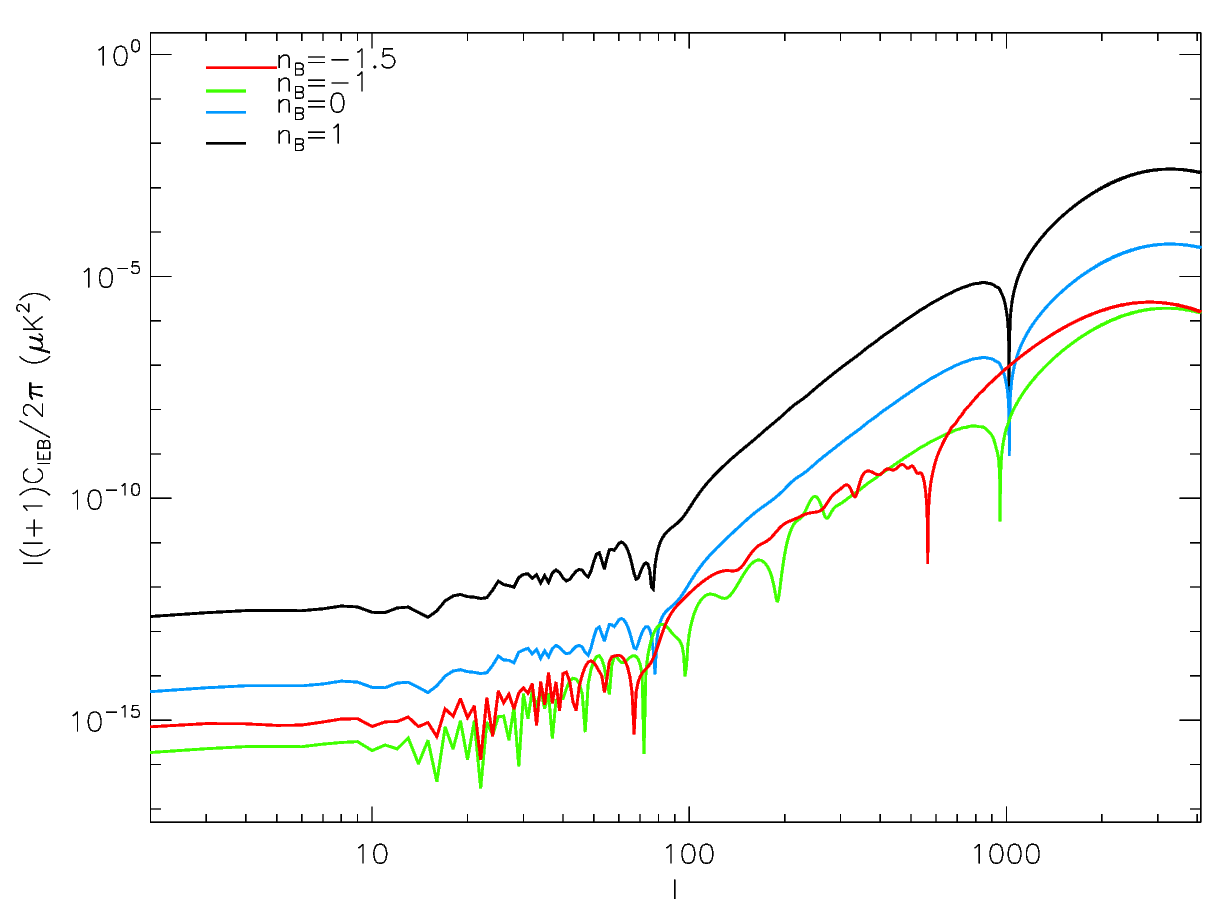}
\caption{On the top panel we show the parity-odd vector power spectrum $TB$ and in the bottom panel 
the parity-odd correlator $EB$, with a magnetic field $\sqrt{\langle B^2_\lambda\rangle}=3.5\ nG$.}
\label{figure:VecOdd}
\end{figure}

From Fig.~\ref{figure:VecOdd}, we can see that the resulting $\ell (\ell+1) C_l^{TB}/(2 \pi)$ is of 
the order of ${\cal O} (10^{-1})$ $\mu$K$^2$ for $n_B=n_H=0$ at $\ell \sim 10^3$ for the maximal 
helical case with $\sqrt{\langle B_\lambda^2\rangle} = 3.5\ nG$. 
For comparison, the vector contribution $\ell (\ell+1) C_l^{TT}/(2 \pi)$ 
to the temperature anisotropies for a non-helical stochastic background is larger than 
${\cal O} (10^2)$ $\mu$K$^2$ for $\sqrt{\langle B_\lambda^2\rangle} = 3.5\ nG$ and $n_B=0$ and 
is roughly ${\cal O} (10^2)$ $\mu$K$^2$ in the maximal helical case at $\ell = 10^3$. 
These values need to be compared with a typical value for the 
$\Lambda$CDM best-fit model of the order of ${\cal O} (10^3)$ $\mu$K$^2$ at $\ell = 10^3$.

In a recent paper \cite{Kahniashvili:2014dfa}, WMAP 9 yr $TB$ data have been used to constrain 
the helical odd-parity vector contribution of a stochastic background of primordial magnetic fields.\\
In \cite{Kahniashvili:2014dfa} the  basic assumptions in terms of simple power spectra for the 
non-helical and helical contributions with a sharp cut-off at $k=k_D$ are the same as in this 
paper, however, there is a strong difference in the treatment of the maximum helical condition. 
They use the integrated measure ${\cal H}$ in 
Eq.~\eqref{eqn:MagHel} and therefore allows the range $n_H > -2$ 
without the use of an integrated cut-off; this results in a bound of ${\cal H} < 10\ nG^2\ Gpc$ 
as a 95\% CL for $\sqrt{\langle B^2\rangle} = 3\ nG$ and $n_B = n_H - 1 = -2.9$ from 
WMAP 9 yr $TB$ data.

We first show that the bound quoted in Kahniashvili et al. \cite{Kahniashvili:2014dfa} 
is much larger than what admitted by the Schwarz's inequality for amplitudes of the non-helical 
part constrained by current CMB data.
We obtain the maximum value for $A_H^{\mathrm max} = A_B (k_*/k_D)$ by imposing the inequality in 
Eq.~\eqref{eqn:diseq} to be valid at all $k \le k_D$ for $n_B =n_H - 1 = -2.9$. As a maximum 
value for ${\cal H}$, we therefore obtain for the same values of the two spectral indices:
\be
{\cal H}^{\mathrm max} = \frac{\langle B^2 \rangle }{16 \pi k_D} \,.
\label{eqn:Hmax}
\ee
The bounds coming from Eq.~\eqref{eqn:Hmax} for a typical value of the damping scale according to 
Refs.~\cite{Jedamzik:1996wp,stochastic background}, 
i.e. in the range of $10^2\ Mpc^{-1}$ is about seven orders of magnitude smaller 
than the 95\% bound $10\ nG^2\ Gpc$. In order to respect the maximum helical condition imposed 
by Eq.~\eqref{eqn:Hmax} it would be necessary to consider a damping scale of the order of 
$k_D \sim 2 \times 10^{-2}\ Gpc^{-1}$ which would suppress all the contributions of primordial 
magnetic fields apart from the very large angular scales, namely only the very first multipoles 
of the CMB anisotropy angular power spectra.

\begin{figure}[!htp]
\centering
\includegraphics[height=5cm]{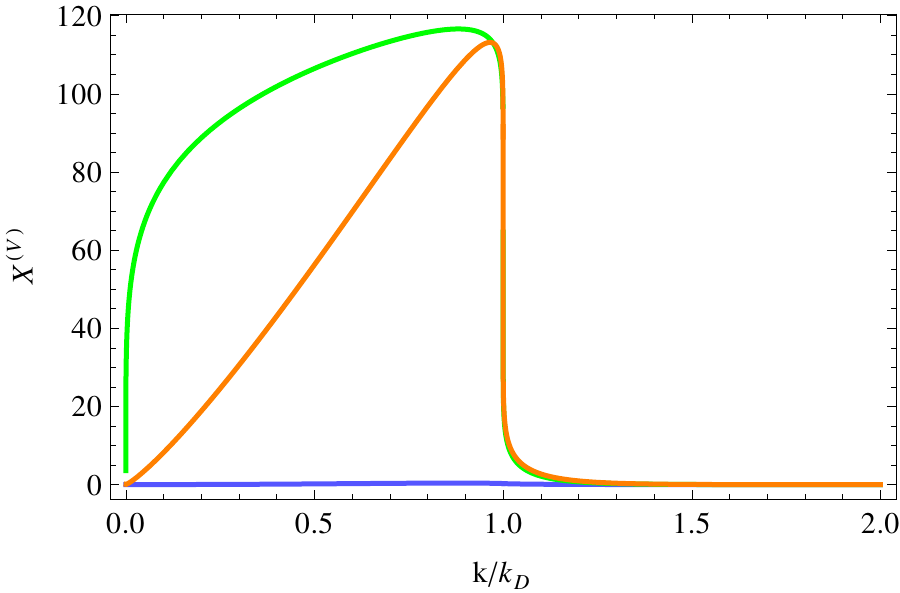}
\caption{Comparison of $k^3 X^{(V)} (k)$
(in units of $\langle B^2\rangle \langle \mathcal{B}^2\rangle/(4\pi)^4$), 
for $n_B = n_H = -1.9$ (blue line), $n_B = n_H - 1= -2.9$ (orange line) and $n_B = n_H = -2.9$ (green line).}
\label{offdiag}
\end{figure}

In addition, there are values of parameters which are excluded 
by considering ${\cal H}$ instead of ${\cal B}$ for which $C_\ell^{TB}$ could be larger.
Our treatment allows to compute the parity-odd $X^{(V)} (k)$ for spectral indices $n_B \ne n_H$.
In Fig.~\ref{offdiag}, $X^{(V)} (k)$ with $n_B = n_H - 1 = -2.9$ is compared with the two 
maximal helical cases $n_B = n_H = -2.9$ and $n_B = n_H = -1.9$. As expected, $X^{(V)} (k)$ 
for the $n_B =n_H - 1 = -2.9$ lies between the two maximal helical cases with $n_B = n_H = -2.9$ 
and $n_B = n_H = -1.9$. For the wavenumbers relevant for CMB anisotropies, i.e. 
$k \ll k_D$, the maximal helical nearly scale-invariant case with $n_B = n_H = -2.9$ 
is larger than the $n_B =n_H - 1 = -2.9$.

The results of this analysis show how the currently publicly available WMAP 9 yr $TB$ data 
are hardly sensitive to constrain the helical odd-parity {\em vector} contribution at values 
comparable with those obtained by the inequality in Eq.~\eqref{eqn:diseq} for amplitudes 
of the non-helical part at the level of $nG$ and values of the spectral indices as 
$n_B =n_H - 1 = -2.9$ \footnote{Note that Ref.~\cite{Kahniashvili:2014dfa} mentions both the 
inequality in the Fourier space in Eq.~\ref{eqn:diseq}, but also a realizability 
condition in an integral form ${\cal H} \le \xi_M \langle B^2 \rangle/(4 \pi)$ with a 
correlation length $\xi_M = 2 \pi/k_D (n_B+3)/(n_B+2)$. 
This latter realizability condition is ill defined even for values of the non-helical spectral 
index $-3 < n_B < -2$, and we stress again that $X^{(V)}(k)$ in Eq.~\eqref{eqn:VectorSpectrumA} 
is infrared finite for any value $n_B + n_H > -6$.}. Different considerations would hold for 
the tensor contribution.


\subsection{The tensor contribution to CMB anisotropies}

\begin{figure}[htp]
\centering
\includegraphics[width=8.5cm]{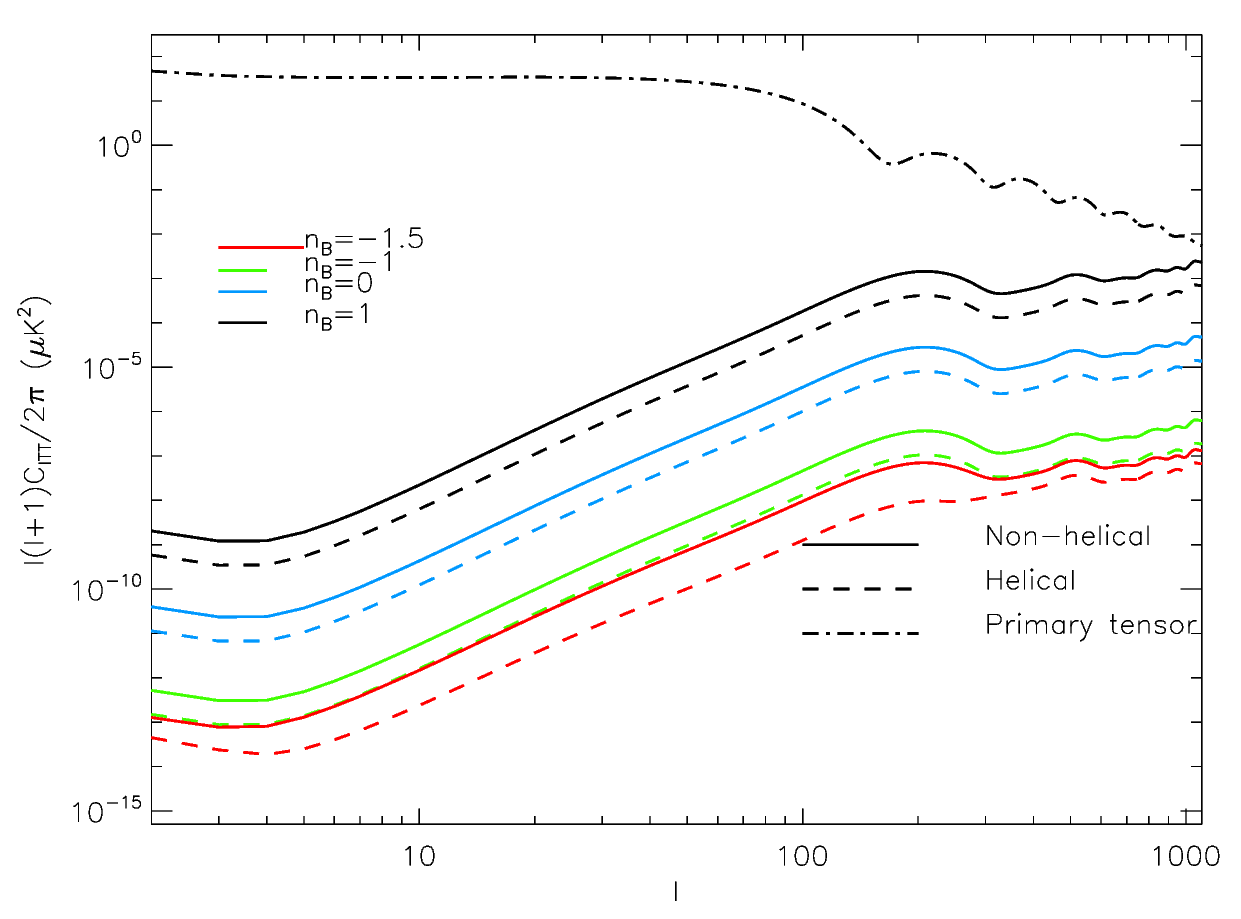}
\includegraphics[width=8.5cm]{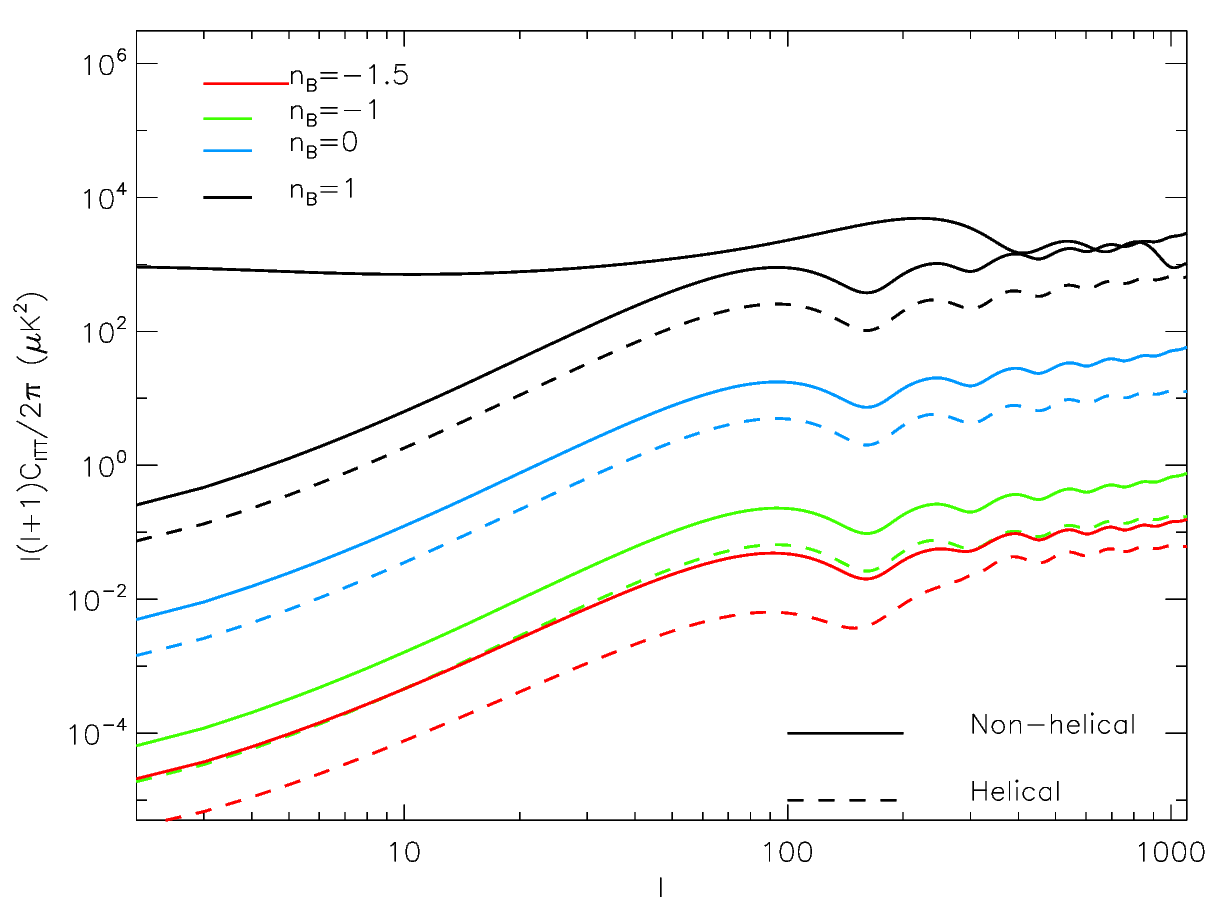}
\caption{CMB anisotropies angular power spectrum for temperature. We include the tensor primary contribution from adiabatic inflation in comparison with the tensor contributions 
of a stochastic background of PMF for $\sqrt{\langle B^2_\lambda\rangle}=3.5\ nG$. In the top panel we show the compensated mode and in the bottom one the passive mode.}
\label{figure:TensCl}
\end{figure}

The evolution of tensor metric perturbations is described by Einstein equations where 
PMF contribution is again an additional source term, given by PMF stress tensor:
\be
\label{eqn:EinsteinTens}
\ddot{h}_{ij}+2\mathcal{H}\dot{h}_{ij}+k^2h_{ij}=16\pi Ga^2\bigl(\rho_\nu\pi^\nu_{ij}+\Pi^{(T)}_{ij}\bigr)\,.
\ee
As in the vector case we can use a consistent tensor orthonormal polarization 
base to divide the metric solution respect to the two independent sources. Consider:
\be
e_{ij}^\pm=-\sqrt{\frac{3}{8}}(\mathbf{e}_1\pm\imath\mathbf{e}_2)_i\times(\mathbf{e}_1\pm\imath\mathbf{e}_2)_j\,,
\ee
with the following properties:
\begin{align}
&e_{ij}^\pm e_{ij}^\pm=0\,,\\
&e_{ij}^\pm e_{ij}^\mp=\frac{3}{2}\,,\\
&(e_{ij}^\pm)^*=e_{ij}^\mp\,.
\end{align}
In this basis the tensor part of the anisotropic stress is expressed as:
\be
\Pi_{ij}^{(T)}(\mathbf{k})=e_{ij}^+\Pi^+_T(\mathbf{k})+e_{ij}^-\Pi^-_T(\mathbf{k})\,.
\ee
Now, we can rewrite the EMT source in terms of the component $\Pi^\pm_T$ and viceversa as:
\begin{align}
&(2\pi)^3\delta^{(3)}(\mathbf{k}-\mathbf{h}) \big|\Pi^{(T)}_B(k)\big|^2= \notag\\
&\qquad\frac{3}{2}\langle\Pi^-_T(\mathbf{k})\Pi^{-\,*}_T(\mathbf{h})+\Pi^+_T(\mathbf{k})\Pi^{+\,*}_T(\mathbf{h})\rangle\,, \\
&(2\pi)^3\delta^{(3)}(\mathbf{k}-\mathbf{h}) X^{(T)}(k)= \notag\\
&\qquad-\frac{3}{2}\langle-\Pi^-_T(\mathbf{k})\Pi^{-\,*}_T(\mathbf{h})+\Pi^+_T(\mathbf{k})\Pi^{+\,*}_T(\mathbf{h})\rangle\,, 
\end{align}
and so we can split the Eq.~\eqref{eqn:EinsteinTens} in two polarization states $+$ and $-$, as we previously made for the vector case.

\begin{figure}[htp]
\centering
\includegraphics[width=8.5cm]{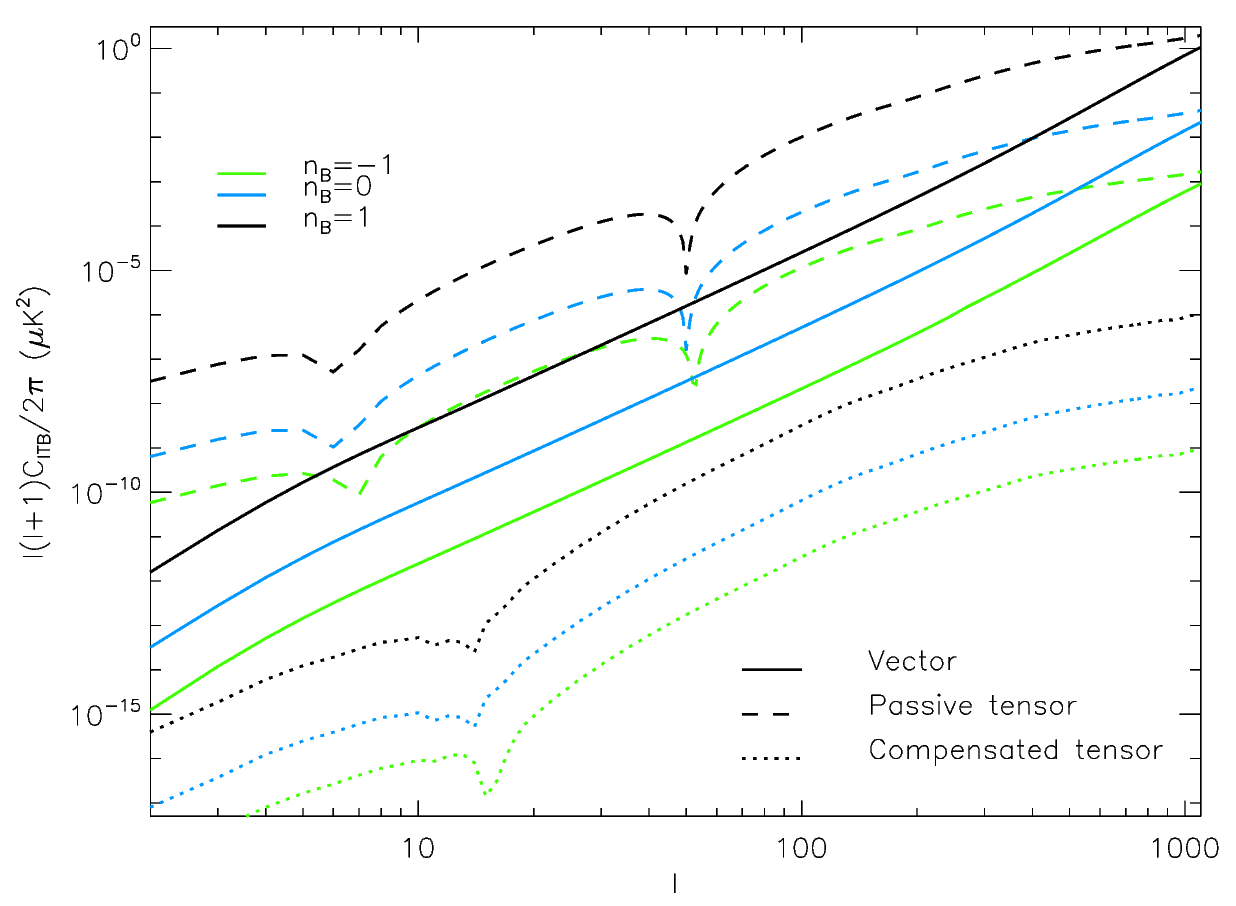}
\caption{Comparison between the vector, compensated tensor and passive tensor $C_l^{TB}$ spectrum.}
\label{figure:TensComparison}
\end{figure}

As for the vector case, tensor magnetic source spectrum has an helical contribution that gives 
non-vanishing odd CMB power spectra. In Figs.~\ref{figure:TensCl} and \ref{figure:TensComparison} 
are shown the angular power spectra of the temperature polarization CMB's anisotropies due to 
the tensor modes for compensated and passive initial condition.


\section{Conclusions}
\label{sec:conclusion}

We have studied the helical contribution to the EMT of a 
stochastic primordial background of PMF extending the previous treatment 
in the non-helical case \cite{FPP,PFP}. 

Under the assumption of a sharp cutoff for the damping scale, we gave the exact 
expressions of the Fourier convolutions of the EMT for the values of the 
selected spectral index $n_H$. The helical contribution to the EMT components 
is of a similar order of magnitude of the non-helical case.
As for the non-helical case, the integration of the angular part 
leads to different numerical coefficient with respect to 
the previous results \cite{CDK,Kahniashvili:2005xe}.

We have then computed the CMB anisotropy power spectra in temperature and polarization of the 
stochastic background for $\ell < 3000$. Such numerical 
computation for the power spectra to high $\ell$ allows the 
comparison of theoretical predictions with observations in the regime where the 
PMF contribution is higher.

There are two main effects when taking into account a possible 
helical contribution. The first effect is the modification of the parity even contribution to 
$C_\ell^{TT} \,, C_\ell^{EE} \,, C_\ell^{BB} \,, C_\ell^{TE}$. 
This contribution in the case of maximal helicity 
is negative for scalar, vector and tensor and decrease the $C_\ell$. 
Since the helical and non-helical parity-even contributions have a similar 
asymptotic dependence on $k$ for $k \ll k_D$, a maximal helical contribution is nearly 
degenerate to the non-helical one with smaller amplitude. The EMT Fourier spectra and the CMB predictions derived here are used in Ref.~\cite{Adam:2015rua} to derive the {\em Planck} 2015 
constraints for the maximal helical case.

The second effect is the generation of the parity odd cross-correlation $C_\ell^{TB}$ and $C_\ell^{EB}$, 
which would otherwise vanish in absence of helicity. Current \cite{QUAD,BICEP,WMAP9} 
and future {\it Planck} data will be useful to help breaking this degeneracy.


\vspace{1cm}

{\bf Acknowledgements.}
We acknowledge discussions and suggestions by Chiara Caprini.
We acknowledge support by PRIN MIUR 2009 grant n. 2009XZ54H2 and ASI
through ASI/INAF Agreement I/072/09/0 for the Planck LFI Activity of Phase
E2.

\setcounter{section}{0}

\section{Appendix}
\label{appendix1}

As for the non-helical EMT components studied in \cite{FPP,PFP,PF}, 
our computations include a careful integration of the angular part, often neglected 
\cite{MKK,KR,CDK} previous to Ref.~\cite{FPP}.

We use the convolutions for the PMF EMT  spectra
with the parametrization for the magnetic field PS given in Eqs.~\eqref{eqn:powerS} and 
\eqref{eqn:powerA}.
Since $P_B (k) =0$ and $P_H (k) =0$ for $k > k_D$,
two conditions need to be taken into account: $p<k_D$ and
$\big| {\bf k}-{\bf p} \big| <k_D$. 

The second condition introduces
a $k$-dependence on the angular integration domain and the two
allow the energy power spectrum to be non zero only for $0<k<2k_D$.
Such conditions split the double integral (over $\gamma$ and over $p$)
in three parts
depending on the $\gamma$ and $p$ lower and upper limit of integration.
A sketch of the integration is thus the following:

\begin{widetext}
\begin{eqnarray}
1)&&  0<k<k_D \nonumber\\
&&\int_{0}^{k_D-k}dp
\int_{-1}^{1}d\gamma\,\dots + \int_{1-k}^{1}dp
\int_{\frac{k^2+p^2-k_D^2}{2kp}}^{1}d\gamma\,\dots \equiv \int_{0}^{k_D-k}dp I_a (p,k) + \int_{k_D-k}^{k_D}dp I_S (p,k)
\nonumber\\
2) && k_D<k<2k_D \nonumber\\
&& \int_{k-k_D}^{k_D}dp
\int_{\frac{k^2+p^2-k_D^2}{2kp}}^{1} d\gamma\,\dots \equiv
\int_{k-k_D}^{k_D}dp I_c (p,k)
\label{intscheme}
\end{eqnarray}

Particular care must be used in the radial integrals.
In particular, the presence of the term $|k-p|^{n+2}$ in both integrands,
needs a further splitting of the integral domain for odd $n$:\\
\begin{displaymath}
\int_0^{(k_D-k)} dp\rightarrow
\left\{\begin{array}{ll}
  k<k_D/2
\left\{\begin{array}{ll}
\int_0^k dp...
\quad
& {\rm with}  \quad p < k  \\
\int_k^{(k_D-k)}dp...
& {\rm with}\quad p > k \\
\end{array} \right.\\
 k>k_D/2 \quad\int_0^{(1-k)}dp...\quad
 {\rm with}\quad p < k \\
\end{array} \right.
\end{displaymath}

\begin{displaymath}
\int_{(k_D-k)}^1 dp\rightarrow
\left\{\begin{array}{ll}
 k<k_D/2 \quad\int_{(1-k)}^1dp... \qquad
 {\rm with}\quad p > k \\
 k>k_D/2
\left\{\begin{array}{ll}
\int_{(k_D-k)}^k dp...
\quad
& {\rm with}  \quad p < k  \\
\int_k^1 dp...
& {\rm with}\quad p > k \\
\end{array} \right.
 \end{array} \right.
\end{displaymath}

\begin{displaymath}
\int_{k-k_D}^1 dp\rightarrow
\left\{\begin{array}{ll}
 1<k<2 \quad\int_{k-k_D}^1dp... \quad
 {\rm with}\quad p < k \\
\end{array}  \right.
\end{displaymath}

\section{Appendix}
\label{appendix2}

Following the scheme in \appendixname~\ref{appendix1} we can now perform the integration over $\mathbf{p}$ for the selected correlators 
in Eqs.~\eqref{eqn:ScalarSpectrum}, \eqref{eqn:VectorSpectrumS}, \eqref{eqn:VectorSpectrumA}, \eqref{eqn:TensorSpectrumS}, \eqref{eqn:TensorSpectrumA}, \eqref{eqn:LorentzSpectrum} and \eqref{eqn:StressSpectrum}.\\

{\bf Correlators for scalar perturbations}\\
Our exact results for $|\rho_B(k)|^2$ and $|\rho_H(k)|^2$ are given for particular values of $n_B$ and $n_H$.
\setcounter{subsubsection}{0}


\subsubsection{$n_B,\,n_H=4$}
\be
|\rho_B(k)|^2=\frac{A_B^2\,k_D^{11}}{512\,\pi^4\,k_*^8}\biggl[\frac{4}{11}-\tk+\frac{4 \tk^2}{3}-\tk^3+\frac{8 \tk^4}{21}-\frac{\tk^5}{24}-\frac{\tk^7}{192}+\frac{\tk^{11}}{9856}\biggl] \nonumber\,,
\label{scalarNE_n4}
\ee
\begin{displaymath}
|\rho_H(k)|^2=\frac{A_H^2\,k_D^{11}}{512\,\pi^4\,k_*^8}
\left\{\begin{array}{ll}
-\frac{2}{11}+\frac{\tk}{2}-\frac{2 \tk^2}{3}+\frac{\tk^3}{2}-\frac{6 \tk^4}{35}+\frac{2 \tk^6}{175}-\frac{\tk^{11}}{7700} \quad
& {\rm for}  \quad 0 \le \tk \le 1 \\
\frac{2}{11}-\frac{2}{35 \tk}-\frac{23 \tk}{50}+\frac{2 \tk^2}{3}-\frac{\tk^3}{2}+\frac{6 \tk^4}{35}-\frac{2 \tk^6}{175}+\frac{\tk^{11}}{23100}
& {\rm for}  \quad 1 \le \tk \le 2 
\end{array} 
\label{scalarE_n4}
\right. \,,
\end{displaymath}
\be
|L_B(k)|^2=\frac{A_B^2\,k_D^{11}}{512\,\pi^4\,k_*^8}\biggl[\frac{4}{15}-\frac{2 \tk}{3}+\frac{44 \tk^2}{45}-\frac{5 \tk^3}{6}+\frac{8 \tk^4}{21}-\frac{17 \tk^5}{240}-\frac{\tk^7}{960}+\frac{\tk^{11}}{16128}\biggl] \nonumber\,,
\label{scalarNE_n4}
\ee
\begin{displaymath}
|L_H(k)|^2=\frac{A_H^2\,k_D^{11}}{512\,\pi^4\,k_*^8}
\left\{\begin{array}{ll}
-\frac{4}{33}+\frac{4 \tk^2}{15}-\frac{\tk^3}{3}+\frac{36 \tk^4}{245}-\frac{4 \tk^6}{315}+\frac{\tk^{11}}{5390} \quad
& {\rm for}  \quad 0 \le \tk \le 1 \\
\frac{4}{33}+\frac{16}{2205 \tk^3}-\frac{4}{35 \tk}-\frac{4 \tk^2}{15}+\frac{\tk^3}{3}-\frac{36 \tk^4}{245}+\frac{4 \tk^6}{315}-\frac{\tk^{11}}{16170}
& {\rm for}  \quad 1 \le \tk \le 2 
\end{array} 
\label{scalarE_n4}
\right. \,,
\end{displaymath}
\be
|\sigma_B(k)|^2=\frac{A_B^2\,k_D^{11}}{1152\,\pi^4\,k_*^8}\biggl[\frac{28}{55}-\tk+\frac{52 \tk^2}{45}-\frac{7 \tk^3}{8}+\frac{8 \tk^4}{21}-\frac{17 \tk^5}{240}-\frac{\tk^7}{1920}+\frac{37 \tk^{11}}{709632}\biggl] \nonumber\,,
\label{scalarNE_n4}
\ee
\begin{displaymath}
|\sigma_H(k)|^2=\frac{A_H^2\,k_D^{11}}{512\,\pi^4\,k_*^8}
\left\{\begin{array}{ll}
-\frac{4}{99}+\frac{\tk}{18}-\frac{4 \tk^2}{135}+\frac{4 \tk^4}{735}-\frac{4 \tk^6}{4725}+\frac{2 \tk^{11}}{121275} \quad
& {\rm for}  \quad 0 \le \tk \le 1 \\
\frac{4}{99}+\frac{8}{6615 \tk^3}-\frac{8}{315 \tk}-\frac{23 \tk}{450}+\frac{4 \tk^2}{135}-\frac{4 \tk^4}{735}+\frac{4 \tk^6}{4725}-\frac{2 \tk^{11}}{363825}
& {\rm for}  \quad 1 \le \tk \le 2 
\end{array} 
\label{scalarE_n4}
\right. \,.
\end{displaymath}

\subsubsection{$n_B,\,n_H=3$}
\begin{displaymath}
|\rho_B(k)|^2=\frac{A_B^2\,k_D^9}{512\,\pi^4\,k_*^6}
\left\{\begin{array}{ll}
\frac{4}{9}-\tk+\frac{20 \tk^2}{21}-\frac{5 \tk^3}{12}+\frac{4 \tk^4}{75}+\frac{4 \tk^6}{315}-\frac{\tk^9}{525} \quad
& {\rm for}  \quad 0 \le \tk \le 1 \\
-\frac{4}{9}+\frac{88}{525 \tk}+\frac{13 \tk}{15}-\frac{20 \tk^2}{21}+\frac{17 \tk^3}{36}-\frac{4 \tk^4}{75}-\frac{4 \tk^6}{315}+\frac{\tk^9}{1575}
& {\rm for}
\quad 1 \le \tk  \le 2
\end{array} 
\label{scalarNE_n3}
\right. \,,
\end{displaymath}
\be
|\rho_H(k)|^2=\frac{A_H^2\,k_D^9}{512\,\pi^4\,k_*^6}\biggl[-\frac{2}{9}+\frac{\tk}{2}-\frac{10 \tk^2}{21}+\frac{5 \tk^3}{24}-\frac{\tk^5}{48}+\frac{\tk^9}{4032}\biggr] \nonumber\,,
\label{scalarE_n3}
\ee
\begin{displaymath}
|L_B(k)|^2=\frac{A_H^2\,k_D^{9}}{512\,\pi^4\,k_*^6}
\left\{\begin{array}{ll}
\frac{44}{135}-\frac{2 \tk}{3}+\frac{556 \tk^2}{735}-\frac{4 \tk^3}{9}+\frac{164 \tk^4}{1575}+\frac{4 \tk^6}{2079}-\frac{11 \tk^9}{11025} \quad
& {\rm for}  \quad 0 \le \tk \le 1 \\
-\frac{44}{135}+\frac{64}{24255 \tk^5}-\frac{16}{945 \tk^3}+\frac{88}{525 \tk}+\frac{2 \tk}{3}-\frac{556 \tk^2}{735} \\
+\frac{4 \tk^3}{9}-\frac{164 \tk^4}{1575}-\frac{4 \tk^6}{2079}+\frac{11 \tk^9}{33075}
& {\rm for}  \quad 1 \le \tk \le 2 
\end{array} 
\label{scalarE_n4}
\right. \,,
\end{displaymath}
\be
|L_H(k)|^2=\frac{A_B^2\,k_D^{9}}{512\,\pi^4\,k_*^6}\biggl[-\frac{4}{27}+\frac{4 \tk^2}{21}-\frac{5 \tk^3}{36}+\frac{\tk^5}{48}-\frac{\tk^9}{3024}\biggl] \nonumber\,,
\label{scalarNE_n4}
\ee
\begin{displaymath}
|\sigma_B(k)|^2=\frac{A_H^2\,k_D^{9}}{1152\,\pi^4\,k_*^6}
\left\{\begin{array}{ll}
\frac{28}{45}-\tk+\frac{628 \tk^2}{735}-\frac{7 \tk^3}{16}+\frac{52 \tk^4}{525}+\frac{4 \tk^6}{3465}-\frac{\tk^9}{1225} \quad
& {\rm for}  \quad 0 \le \tk \le 1 \\
-\frac{28}{45}+\frac{16}{2695 \tk^5}-\frac{16}{315 \tk^3}+\frac{232}{525 \tk}+\frac{13 \tk}{15}-\frac{628 \tk^2}{735} \\
+\frac{65 \tk^3}{144}-\frac{52 \tk^4}{525}-\frac{4 \tk^6}{3465}+\frac{\tk^9}{3675}
& {\rm for}  \quad 1 \le \tk \le 2 
\end{array} 
\label{scalarE_n4}
\right. \,,
\end{displaymath}
\be
|\sigma_H(k)|^2=\frac{A_B^2\,k_D^{9}}{512\,\pi^4\,k_*^6}\biggl[-\frac{4}{81}+\frac{\tk}{18}-\frac{4 \tk^2}{189}+\frac{\tk^5}{864}-\frac{\tk^9}{36288}\biggl] \nonumber\,.
\label{scalarNE_n4}
\ee

\subsubsection{$n_B,\,n_H=2$}
\be
|\rho_B(k)|^2=\frac{A_B^2\,k_D^7}{512\,\pi^4\,k_*^4}\biggl[\frac{4}{7}-\tk+\frac{8 \tk^2}{15}-\frac{\tk^5}{24}+\frac{11 \tk^7}{2240}\biggl] \nonumber\,,
\label{scalarNE_n2}
\ee
\begin{displaymath}
|\rho_H(k)|^2=\frac{A_H^2\,k_D^7}{512\,\pi^4\,k_*^4}
\left\{\begin{array}{ll}
-\frac{2}{7}+\frac{\tk}{2}-\frac{4 \tk^2}{15}+\frac{2 \tk^4}{45}-\frac{\tk^7}{210} \quad
& {\rm for}  \quad 0 \le \tk \le 1 \\
\frac{2}{7}-\frac{2}{15 \tk}-\frac{7 \tk}{18}+\frac{4 \tk^2}{15}-\frac{2 \tk^4}{45}+\frac{\tk^7}{630}
& {\rm for}  \quad 1 \le \tk \le 2 
\end{array} 
\label{scalarE_n2}
\right. \,,
\end{displaymath}
\be
|L_B(k)|^2=\frac{A_B^2\,k_D^{7}}{512\,\pi^4\,k_*^4}\biggl[\frac{44}{105}-\frac{2 \tk}{3}+\frac{8 \tk^2}{15}-\frac{\tk^3}{6}-\frac{\tk^5}{240}+\frac{13 \tk^7}{6720}\biggl] \nonumber\,,
\label{scalarNE_n4}
\ee
\begin{displaymath}
|L_H(k)|^2=\frac{A_H^2\,k_D^{7}}{512\,\pi^4\,k_*^4}
\left\{\begin{array}{ll}
-\frac{4}{21}+\frac{8 \tk^2}{75}-\frac{4 \tk^4}{105}+\frac{\tk^7}{175} \quad
& {\rm for}  \quad 0 \le \tk \le 1 \\
\frac{4}{21}+\frac{16}{525 \tk^3}-\frac{4}{15 \tk}-\frac{8 \tk^2}{75}+\frac{4 \tk^4}{105}-\frac{\tk^7}{525}
& {\rm for}  \quad 1 \le \tk \le 2 
\end{array} 
\label{scalarE_n4}
\right. \,,
\end{displaymath}
\be
|\sigma_B(k)|^2=\frac{A_B^2\,k_D^{7}}{1152\,\pi^4\,k_*^4}\biggl[\frac{4}{5}-\tk+\frac{8 \tk^2}{15}-\frac{\tk^3}{8}-\frac{\tk^5}{240}+\frac{\tk^7}{640}\biggl] \nonumber\,,
\label{scalarNE_n4}
\ee
\begin{displaymath}
|\sigma_H(k)|^2=\frac{A_H^2\,k_D^{7}}{512\,\pi^4\,k_*^4}
\left\{\begin{array}{ll}
-\frac{4}{63}+\frac{\tk}{18}-\frac{8 \tk^2}{675}-\frac{4 \tk^4}{2835}+\frac{2 \tk^7}{4725} \quad
& {\rm for}  \quad 0 \le \tk \le 1 \\
\frac{4}{63}+\frac{8}{1575 \tk^3}-\frac{8}{135 \tk}-\frac{7 \tk}{162}+\frac{8 \tk^2}{675}+\frac{4 \tk^4}{2835}-\frac{2 \tk^7}{14175}
& {\rm for}  \quad 1 \le \tk \le 2 
\end{array} 
\label{scalarE_n4}
\right. \,.
\end{displaymath}

\subsubsection{$n_B,\,n_H=1$}
\begin{displaymath}
|\rho_B(k)|^2=\frac{A_B^2\,k_D^5}{512\,\pi^4\,k_*^2}
\left\{\begin{array}{ll}
\frac{4}{5}-\tk+\frac{\tk^3}{4}+\frac{4 \tk^4}{15}-\frac{\tk^5}{5} \quad
& {\rm for}  \quad 0 \le \tk \le 1 \\
-\frac{4}{5}+\frac{8}{15 \tk}+\frac{\tk}{3}+\frac{\tk^3}{4}-\frac{4 \tk^4}{15}+\frac{\tk^5}{15}
& {\rm for}  \quad 1 \le \tk \le 2 \\	
\end{array} 
\label{scalarNE_n1}
\right. \,,
\end{displaymath}
\be
|\rho_H(k)|^2=\frac{A_H^2\,k_D^5}{512\,\pi^4\,k_*^2}\biggl[-\frac{2}{5}+\frac{\tk}{2}-\frac{\tk^3}{8}+\frac{\tk^5}{80}\biggl] \nonumber\,,
\label{scalarE_n1}
\ee
\begin{displaymath}
|L_B(k)|^2=\frac{A_H^2\,k_D^{5}}{512\,\pi^4\,k_*^2}
\left\{\begin{array}{ll}
\frac{44}{75}-\frac{2 \tk}{3}+\frac{32 \tk^2}{105}+\frac{4 \tk^4}{315}-\frac{\tk^5}{25} \quad
& {\rm for}  \quad 0 \le \tk \le 1 \\
-\frac{44}{75}+\frac{64}{1575 \tk^5}-\frac{16}{105 \tk^3}+\frac{8}{15 \tk}+\frac{2 \tk}{3}-\frac{32 \tk^2}{105}-\frac{4 \tk^4}{315}+\frac{\tk^5}{75}
& {\rm for}  \quad 1 \le \tk \le 2 
\end{array} 
\label{scalarE_n4}
\right. \,,
\end{displaymath}
\be
|L_H(k)|^2=\frac{A_B^2\,k_D^{5}}{512\,\pi^4\,k_*^2}\biggl[-\frac{4}{15}+\frac{\tk^3}{12}-\frac{\tk^5}{80}\biggl] \nonumber\,,
\label{scalarNE_n4}
\ee
\begin{displaymath}
|\sigma_B(k)|^2=\frac{A_H^2\,k_D^{5}}{1152\,\pi^4\,k_*^2}
\left\{\begin{array}{ll}
\frac{28}{25}-\tk+\frac{16 \tk^2}{105}+\frac{\tk^3}{16}+\frac{4 \tk^4}{105}-\frac{\tk^5}{25} \quad
& {\rm for}  \quad 0 \le \tk \le 1 \\
-\frac{28}{25}+\frac{16}{175 \tk^5}-\frac{16}{35 \tk^3}+\frac{8}{5 \tk}+\frac{\tk}{3}-\frac{16 \tk^2}{105}+\frac{\tk^3}{16}-\frac{4 \tk^4}{105}+\frac{\tk^5}{75}
& {\rm for}  \quad 1 \le \tk \le 2 
\end{array} 
\label{scalarE_n4}
\right. \,,
\end{displaymath}
\be
|\sigma_H(k)|^2=\frac{A_B^2\,k_D^{5}}{512\,\pi^4\,k_*^2}\biggl[-\frac{4}{45}+\frac{\tk}{18}-\frac{\tk^5}{1440}\biggl] \nonumber\,.
\label{scalarNE_n4}
\ee

\subsubsection{$n_B,\,n_H=0$}
\begin{displaymath}
|\rho_B(k)|^2=\frac{A_B^2\,k_D^3}{512\,\pi^4}
\left\{\begin{array}{ll}
\frac{29}{24}-\frac{17 \tk}{16}-\frac{7 \tk^2}{8}+\frac{53 \tk^3}{96}+\frac{\tk^3 \pi ^2}{24}-\frac{\text{Log}[1-\tk]}{8 \tk}+\frac{1}{2} \tk \text{Log}[1-\tk] \notag\\
-\frac{3}{8} \tk^3 \text{Log}[1-\tk]+\frac{1}{2} \tk^3 \text{Log}[1-\tk] \text{Log}[\tk]-\frac{1}{2} \tk^3 \text{PolyLog}\left[2,\frac{-1+\tk}{\tk}\right] \quad
& {\rm for}  \quad 0 \le \tk \le 1 \\	
\frac{29}{24}-\frac{17 \tk}{16}-\frac{7 \tk^2}{8}+\frac{53 \tk^3}{96}-\frac{\text{Log}[-1+\tk]}{8 \tk}+\frac{1}{2} \tk \text{Log}[-1+\tk] \notag\\
-\frac{3}{8} \tk^3 \text{Log}[-1+\tk]+\frac{1}{4} \tk^3 \text{Log}[-1+\tk] \text{Log}[\tk]+\frac{1}{4} \tk^3 \text{PolyLog}\left[2,\frac{1}{\tk}\right] \notag\\
-\frac{1}{4} \tk^3 \text{PolyLog}\left[2,\frac{-1+\tk}{\tk}\right]
& {\rm for}
\quad 1 \le \tk  \le 2
\end{array} 
\label{scalarNE_n0}
\right. \,,
\end{displaymath}
\begin{displaymath}
|\rho_H(k)|^2=\frac{A_H^2\,k_D^3}{512\,\pi^4}
\left\{\begin{array}{ll}
-\frac{2}{3}+\frac{\tk}{2}+\frac{2 \tk^2}{3}-\frac{\tk^3}{2} \quad
& {\rm for}  \quad 0 \le \tk \le 1 \\	
\frac{2}{3}-\frac{2}{3 \tk}+\frac{\tk}{2}-\frac{2 \tk^2}{3}+\frac{\tk^3}{6}
& {\rm for}
\quad 1 \le \tk  \le 2
\end{array} 
\label{scalarE_n0}
\right. \,,
\end{displaymath}
\begin{align}
|L_B(k)|^2=\frac{A_H^2\,k_D^{3}}{512\,\pi^4}\biggl[&	\frac{43}{48}-\frac{1}{16 \tk^4}-\frac{1}{32 \tk^3}+\frac{7}{48 \tk^2}+\frac{13}{192 \tk}-\frac{67 \tk}{96}+\frac{\tk^2}{48}+\frac{17 \tk^3}{384}-\frac{\text{Log}[\abs{1-\tk}]}{16 \tk^5} \notag\\
&+\frac{\text{Log}[\abs{1-\tk}]}{6 \tk^3}-\frac{\text{Log}[\abs{1-\tk}]}{8 \tk}+\frac{1}{48} \tk^3 \text{Log}[\abs{1-\tk}] \biggl] \nonumber\,,
\end{align}
\begin{displaymath}
|L_H(k)|^2=\frac{A_H^2\,k_D^{3}}{512\,\pi^4}
\left\{\begin{array}{ll}
-\frac{4}{9}-\frac{4 \tk^2}{15}+\frac{\tk^3}{3} \quad
& {\rm for}  \quad 0 \le \tk \le 1 \\
\frac{4}{9}+\frac{16}{45 \tk^3}-\frac{4}{3 \tk}+\frac{4 \tk^2}{15}-\frac{\tk^3}{9}
& {\rm for}  \quad 1 \le \tk \le 2 
\end{array} 
\label{scalarE_n4}
\right. \,,
\end{displaymath}
\begin{displaymath}
|\sigma_B(k)|^2=\frac{A_H^2\,k_D^{3}}{1152\,\pi^4}
\left\{\begin{array}{ll}
\frac{253}{192}-\frac{9}{64 \tk^4}-\frac{9}{128 \tk^3}+\frac{29}{64 \tk^2}+\frac{55}{256 \tk}-\frac{159 \tk}{128}-\frac{19 \tk^2}{64}+\frac{413 \tk^3}{1536} \\
+\frac{\tk^3 \pi ^2}{96}-\frac{9 \text{Log}[1-\tk]}{64 \tk^5}+\frac{\text{Log}[1-\tk]}{2 \tk^3}-\frac{11 \text{Log}[1-\tk]}{16 \tk}+\frac{1}{2} \tk \text{Log}[1-\tk] \\
-\frac{11}{64} \tk^3 \text{Log}[1-\tk]+\frac{1}{8} \tk^3 \text{Log}[1-\tk] \text{Log}[\tk]-\frac{1}{16} \tk^3 \text{Log}[\tk]^2 \\
-\frac{1}{8} \tk^3 \text{PolyLog}\left[2,\frac{-1+\tk}{\tk}\right] \quad
& {\rm for}  \quad 0 \le \tk \le 1 \\
\frac{253}{192}-\frac{9}{64 \tk^4}-\frac{9}{128 \tk^3}+\frac{29}{64 \tk^2}+\frac{55}{256 \tk}-\frac{159 \tk}{128}-\frac{19 \tk^2}{64}+\frac{413 \tk^3}{1536} \\
-\frac{9 \text{Log}[-1+\tk]}{64 \tk^5}+\frac{\text{Log}[-1+\tk]}{2 \tk^3}-\frac{11 \text{Log}[-1+\tk]}{16 \tk}+\frac{1}{2} \tk \text{Log}[-1+\tk] \\
-\frac{11}{64} \tk^3 \text{Log}[-1+\tk]-\frac{1}{16} \tk^3 \text{Log}[-1+\tk] \text{Log}\left[\frac{1}{\tk}\right] \\
+\frac{1}{16} \tk^3 \text{PolyLog}\left[2,\frac{1}{\tk}\right]-\frac{1}{16} \tk^3 \text{PolyLog}\left[2,\frac{-1+\tk}{\tk}\right]
& {\rm for}  \quad 1 \le \tk \le 2 
\end{array} 
\label{scalarE_n4}
\right. \,,
\end{displaymath}
\begin{displaymath}
|\sigma_H(k)|^2=\frac{A_H^2\,k_D^{3}}{512\,\pi^4}
\left\{\begin{array}{ll}
-\frac{4}{27}+\frac{\tk}{18}+\frac{4 \tk^2}{135} \quad
& {\rm for}  \quad 0 \le \tk \le 1 \\
\frac{4}{27}+\frac{8}{135 \tk^3}-\frac{8}{27 \tk}+\frac{\tk}{18}-\frac{4 \tk^2}{135}
& {\rm for}  \quad 1 \le \tk \le 2 
\end{array} 
\label{scalarE_n4}
\right. \,.
\end{displaymath}

\subsubsection{$n_B,\,n_H=-1$}
\begin{displaymath}
|\rho_B(k)|^2=\frac{A_B^2\,k_D\,k_*^2}{512\,\pi^4}
\left\{\begin{array}{ll}
4-5 \tk+\frac{4 \tk^2}{3}+\frac{\tk^3}{4} \quad
& {\rm for}  \quad 0 \le \tk \le 1 \\
-4+\frac{8}{3 \tk}+3 \tk-\frac{4 \tk^2}{3}+\frac{\tk^3}{4}
& {\rm for}  \quad 1 \le \tk \le 2 \\	
\end{array} 
\label{scalarNE_n-1}
\right. \,,
\end{displaymath}
\begin{displaymath}
|\rho_H(k)|^2=\frac{A_H^2\,k_D\,k_*^2}{512\,\pi^4}
\left\{\begin{array}{ll}
-\frac{3}{2}+\frac{3 \tk}{4}+\frac{\tk \pi ^2}{12}+\frac{\text{Log}[1-\tk]}{2 \tk}-\frac{1}{2} \tk \text{Log}[1-\tk]-\frac{1}{2} \tk \text{Log}[1-\tk] \text{Log}\left[\frac{1}{\tk}\right] \notag\\
+\frac{1}{2} \tk \text{Log}[1-\tk] \text{Log}[\tk]-\frac{1}{2} \tk \text{Log}[\tk]^2-\tk \text{PolyLog}\left[2,\frac{-1+\tk}{\tk}\right] \quad
& {\rm for}  \quad 0 \le \tk \le 1 \\
-\frac{3}{2}+\frac{3 \tk}{4}+\frac{\text{Log}[-1+\tk]}{2 \tk}-\frac{1}{2} \tk \text{Log}[-1+\tk]-\frac{1}{2} \tk \text{Log}[-1+\tk] \text{Log}\left[\frac{1}{\tk}\right] \notag\\
+\frac{1}{2} \tk \text{PolyLog}\left[2,\frac{1}{\tk}\right]-\frac{1}{2} \tk \text{PolyLog}\left[2,\frac{-1+\tk}{\tk}\right]
& {\rm for}  \quad 1 \le \tk \le 2 \\	
\end{array} 
\label{scalarE_n-1}
\right. \,,
\end{displaymath}
\begin{displaymath}
|L_B(k)|^2=\frac{A_H^2\,k_D\,k_*^2}{512\,\pi^4}
\left\{\begin{array}{ll}
\frac{44}{15}-2 \tk-\frac{4 \tk^2}{105} \quad
& {\rm for}  \quad 0 \le \tk \le 1 \\
-\frac{44}{15}-\frac{64}{105 \tk^5}+\frac{16}{15 \tk^3}+\frac{8}{3 \tk}+\frac{2 \tk}{3}+\frac{4 \tk^2}{105}
& {\rm for}  \quad 1 \le \tk \le 2 
\end{array} 
\label{scalarE_n4}
\right. \,,
\end{displaymath}
\begin{displaymath}
|L_H(k)|^2=\frac{A_H^2\,k_*^2\,k_D}{512\,\pi^4}
\left\{\begin{array}{ll}
-\frac{1}{2}-\frac{1}{2 \tk^2}-\frac{1}{4 \tk}+\frac{3 \tk}{8}-\frac{\text{Log}[1-\tk]}{2 \tk^3}+\frac{\text{Log}[1-\tk]}{\tk}-\frac{1}{2} \tk \text{Log}[1-\tk] \quad
& {\rm for}  \quad 0 \le \tk \le 1 \\
-\frac{1}{2}-\frac{1}{2 \tk^2}-\frac{1}{4 \tk}+\frac{3 \tk}{8}-\frac{\text{Log}[-1+\tk]}{2 \tk^3}+\frac{\text{Log}[-1+\tk]}{\tk}-\frac{1}{2} \tk \text{Log}[-1+\tk]
& {\rm for}  \quad 1 \le \tk \le 2 
\end{array} 
\label{scalarE_n4}
\right. \,,
\end{displaymath}
\begin{displaymath}
|\sigma_B(k)|^2=\frac{A_H^2\,k_D\,k_*^2}{1152\,\pi^4}
\left\{\begin{array}{ll}
\frac{28}{5}-5 \tk+\frac{68 \tk^2}{105}+\frac{\tk^3}{16} \quad
& {\rm for}  \quad 0 \le \tk \le 1 \\
-\frac{28}{5}-\frac{48}{35 \tk^5}+\frac{16}{5 \tk^3}+\frac{8}{3 \tk}+3 \tk-\frac{68 \tk^2}{105}+\frac{\tk^3}{16}
& {\rm for}  \quad 1 \le \tk \le 2 
\end{array} 
\label{scalarE_n4}
\right. \,,
\end{displaymath}
\begin{displaymath}
|\sigma_H(k)|^2=\frac{A_H^2\,k_D\,k_*^2}{512\,\pi^4}
\left\{\begin{array}{ll}
-\frac{1}{4}-\frac{1}{12 k^2}-\frac{1}{24 k}+\frac{7 k}{48}+\frac{k \pi ^2}{108}+\frac{5}{72} k \text{Log}\left[\frac{1}{1-k}\right]-\frac{\text{Log}[1-k]}{12 k^3} \\
+\frac{2 \text{Log}[1-k]}{9 k}-\frac{5}{72} k \text{Log}[1-k]-\frac{1}{9} k \text{Log}[1-k] \text{Log}\left[\frac{1}{k}\right]-\frac{1}{9} k \text{Log}[k]^2 \\
-\frac{1}{9} k \text{PolyLog}\left[2,\frac{-1+k}{k}\right] \quad
& {\rm for}  \quad 0 \le \tk \le 1 \\
-\frac{1}{4}-\frac{1}{12 \tk^2}-\frac{1}{24 \tk}+\frac{7 \tk}{48}+\frac{5}{72} \tk \text{Log}\left[\frac{1}{-1+\tk}\right]-\frac{\text{Log}[-1+\tk]}{12 \tk^3} \\
+\frac{2 \text{Log}[-1+\tk]}{9 \tk}-\frac{5}{72} \tk \text{Log}[-1+\tk]-\frac{1}{18} \tk \text{Log}[-1+\tk] \text{Log}\left[\frac{1}{\tk}\right] \\
+\frac{1}{18} \tk \text{PolyLog}\left[2,\frac{1}{\tk}\right]-\frac{1}{18} \tk \text{PolyLog}\left[2,\frac{-1+\tk}{\tk}\right] \qquad
& {\rm for}  \quad 1 \le \tk \le 2 
\end{array} 
\label{scalarE_n4}
\right. \,.
\end{displaymath}

\subsubsection{$n_B,\,n_H=-3/2$}
\begin{displaymath}
|\rho_B(k)|^2=\frac{A_B^2\,k_*^3}{512\,\pi^4}
\left\{\begin{array}{ll}
\frac{232}{45 \sqrt{1-\tk}}+\frac{88}{15 \tk}-\frac{88}{15 \sqrt{1-\tk} \tk}+\frac{4 \tk}{3}-\frac{32 \tk}{45 \sqrt{1-\tk}}+\frac{64 \tk^2}{45 \sqrt{1-\tk}}+\frac{\tk^3}{9} \notag\\
-2 \pi +8 \text{Log}\left[1+\sqrt{1-\tk}\right]-4 \text{Log}[\tk] \quad
& {\rm for}  \quad 0 \le \tk \le 1 \\
-\frac{232}{45 \sqrt{-1+\tk}}+\frac{88}{15 \tk}+\frac{88}{15 \sqrt{-1+\tk} \tk}+\frac{4 \tk}{3}+\frac{32 \tk}{45 \sqrt{-1+\tk}}-\frac{64 \tk^2}{45 \sqrt{-1+\tk}} \notag\\
+\frac{\tk^3}{9}-4 \text{ArcTan}\left[\frac{1}{\sqrt{-1+\tk}}\right]+4 \text{ArcTan}\left[\sqrt{-1+\tk}\right]
& {\rm for}  \quad 1 \le \tk \le 2 \\	
\end{array} 
\label{scalarNE_n-3/2}
\right. \,,
\end{displaymath}
\begin{displaymath}
|\rho_H(k)|^2=\frac{A_H^2\,k_*^3}{512\,\pi^4}
\left\{\begin{array}{ll}
\frac{20}{3 \sqrt{1-\tk}}+\frac{4}{3 \tk}-\frac{4}{3 \sqrt{1-\tk} \tk}+2 \tk-\frac{16 \tk}{3 \sqrt{1-\tk}}-\pi -4 \text{Log}\left[1+\sqrt{1-\tk}\right]+2 \text{Log}[\tk] \quad
& {\rm for}  \quad 0 \le \tk \le 1 \\
-\frac{16}{3} \sqrt{-1+\tk}+\frac{4}{3 \tk}+\frac{4 \sqrt{-1+\tk}}{3 \tk}+2 \tk-2 \text{ArcTan}\left[\sqrt{\frac{1}{-1+\tk}}\right]+2 \text{ArcTan}\left[\sqrt{-1+\tk}\right]
& {\rm for}  \quad 1 \le \tk \le 2 \\	
\end{array} 
\label{scalarE_n-3/2}
\right. \,,
\end{displaymath}
\begin{displaymath}
|L_B(k)|^2=\frac{A_H^2\,k_*^3}{512\,\pi^4}
\left\{\begin{array}{ll}
\frac{10616}{1755 \sqrt{1-\tk}}-\frac{2048}{2925 \tk^5}+\frac{2048}{2925 \sqrt{1-\tk} \tk^5}-\frac{1024}{2925 \sqrt{1-\tk} \tk^4}+\frac{128}{135 \tk^3} \\
-\frac{9088}{8775 \sqrt{1-\tk} \tk^3}+\frac{3776}{8775 \sqrt{1-\tk} \tk^2}+\frac{88}{15 \tk}-\frac{10136}{1755 \sqrt{1-\tk} \tk}+\frac{32 \tk}{1755 \sqrt{1-\tk}} \\
-\frac{64 \tk^2}{1755 \sqrt{1-\tk}}-\frac{22 \pi }{15}+\frac{88}{15} \text{Log}\left[1+\sqrt{1-\tk}\right]-\frac{44 \text{Log}[\tk]}{15} \quad
& {\rm for}  \quad 0 \le \tk \le 1 \\
-\frac{10616}{1755 \sqrt{-1+\tk}}-\frac{2048}{2925 \tk^5}-\frac{2048}{2925 \sqrt{-1+\tk} \tk^5}+\frac{1024}{2925 \sqrt{-1+\tk} \tk^4}+\frac{128}{135 \tk^3} \\
+\frac{9088}{8775 \sqrt{-1+\tk} \tk^3}-\frac{3776}{8775 \sqrt{-1+\tk} \tk^2}+\frac{88}{15 \tk}+\frac{10136}{1755 \sqrt{-1+\tk} \tk}-\frac{32 \tk}{1755 \sqrt{-1+\tk}} \\
+\frac{64 \tk^2}{1755 \sqrt{-1+\tk}}-\frac{44}{15} \text{ArcTan}\left[\frac{1}{\sqrt{-1+\tk}}\right]+\frac{44}{15} \text{ArcTan}\left[\sqrt{-1+\tk}\right]
& {\rm for}  \quad 1 \le \tk \le 2 
\end{array} 
\label{scalarE_n4}
\right. \,,
\end{displaymath}
\begin{displaymath}
|L_H(k)|^2=\frac{A_H^2\,k_*^3}{512\,\pi^4}
\left\{\begin{array}{ll}
\frac{24}{7 \sqrt{1-\tk}}-\frac{128}{63 \tk^3}+\frac{128}{63 \sqrt{1-\tk} \tk^3}-\frac{64}{63 \sqrt{1-\tk} \tk^2}+\frac{8}{3 \tk}-\frac{184}{63 \sqrt{1-\tk} \tk}-\frac{32 \tk}{21 \sqrt{1-\tk}} \\
-\frac{2 \pi }{3}-\frac{8}{3} \text{Log}\left[1+\sqrt{1-\tk}\right]+\frac{4 \text{Log}[\tk]}{3} \quad
& {\rm for}  \quad 0 \le \tk \le 1 \\
-\frac{208}{21 \sqrt{-1+\tk}}-\frac{32 \sqrt{-1+\tk}}{3}-\frac{128}{63 \tk^3}+\frac{128}{63 \sqrt{-1+\tk} \tk^3}-\frac{64}{63 \sqrt{-1+\tk} \tk^2}+\frac{8}{3 \tk} \\
-\frac{16}{63 \sqrt{-1+\tk} \tk}+\frac{8 \sqrt{-1+\tk}}{3 \tk}+\frac{64 \tk}{7 \sqrt{-1+\tk}}-4 \text{ArcTan}\left[\sqrt{\frac{1}{-1+\tk}}\right] \\
+\frac{8}{3} \text{ArcTan}\left[\frac{1}{\sqrt{-1+\tk}}\right]+\frac{4}{3} \text{ArcTan}\left[\sqrt{-1+\tk}\right]
& {\rm for}  \quad 1 \le \tk \le 2 
\end{array} 
\label{scalarE_n4}
\right. \,,
\end{displaymath}
\begin{displaymath}
|\sigma_B(k)|^2=\frac{A_H^2\,k_*^3}{1152\,\pi^4}
\left\{\begin{array}{ll}
\frac{328}{39 \sqrt{1-\tk}}-\frac{512}{325 \tk^5}+\frac{512}{325 \sqrt{1-\tk} \tk^5}-\frac{256}{325 \sqrt{1-\tk} \tk^4}+\frac{128}{45 \tk^3}-\frac{8896}{2925 \sqrt{1-\tk} \tk^3} \\
+\frac{3872}{2925 \sqrt{1-\tk} \tk^2}+\frac{124}{15 \tk}-\frac{4664}{585 \sqrt{1-\tk} \tk}+\frac{4 \tk}{3}-\frac{32 \tk}{65 \sqrt{1-\tk}}+\frac{64 \tk^2}{65 \sqrt{1-\tk}}+\frac{\tk^3}{36} \\
-\frac{14 \pi }{5}+\frac{56}{5} \text{Log}\left[1+\sqrt{1-\tk}\right]-\frac{28 \text{Log}[\tk]}{5} \quad
& {\rm for}  \quad 0 \le \tk \le 1 \\
-\frac{32}{65} \sqrt{-1+\tk}-\frac{512}{325 \tk^5}+\frac{512 \sqrt{-1+\tk}}{325 \tk^5}+\frac{256 \sqrt{-1+\tk}}{325 \tk^4}+\frac{128}{45 \tk^3}-\frac{6592 \sqrt{-1+\tk}}{2925 \tk^3} \\
-\frac{544 \sqrt{-1+\tk}}{585 \tk^2}+\frac{124}{15 \tk}-\frac{1736 \sqrt{-1+\tk}}{195 \tk}+\frac{4 \tk}{3}-\frac{64}{65} \sqrt{-1+\tk} \tk+\frac{\tk^3}{36} \\
-\frac{28}{5} \text{ArcTan}\left[\sqrt{\frac{1}{-1+\tk}}\right]+\frac{28}{5} \text{ArcTan}\left[\sqrt{-1+\tk}\right]
& {\rm for}  \quad 1 \le \tk \le 2 
\end{array} 
\label{scalarE_n4}
\right. \,,
\end{displaymath}
\begin{displaymath}
|\sigma_H(k)|^2=\frac{A_H^2\,k_*^3}{512\,\pi^4}
\left\{\begin{array}{ll}
\frac{160 \sqrt{1-\tk}}{189}-\frac{64}{189 \tk^3}+\frac{64 \sqrt{1-\tk}}{189 \tk^3}+\frac{32 \sqrt{1-\tk}}{189 \tk^2}+\frac{16}{27 \tk}-\frac{88 \sqrt{1-\tk}}{189 \tk}+\frac{2 \tk}{9} \\
-\frac{2 \pi }{9}-\frac{8}{9} \text{Log}\left[1+\sqrt{1-\tk}\right]+\frac{4 \text{Log}[\tk]}{9} \quad
& {\rm for}  \quad 0 \le \tk \le 1 \\
\frac{248}{189 \sqrt{-1+\tk}}-\frac{64}{189 \tk^3}+\frac{64}{189 \sqrt{-1+\tk} \tk^3}-\frac{32}{189 \sqrt{-1+\tk} \tk^2}+\frac{16}{27 \tk}-\frac{40}{63 \sqrt{-1+k} \tk}+\frac{2 \tk}{9} \\
-\frac{160 \tk}{189 \sqrt{-1+\tk}}-\frac{4}{9} \text{ArcTan}\left[\sqrt{\frac{1}{-1+\tk}}\right]+\frac{4}{9} \text{ArcTan}\left[\sqrt{-1+\tk}\right] \quad
& {\rm for}  \quad 1 \le \tk \le 2 
\end{array} 
\label{scalarE_n4}
\right. \,.
\end{displaymath}

{\bf Correlators for vector perturbations}
\setcounter{subsubsection}{0}\\
Our exact results for $|\Pi_B^{(V)}(k)|^2$, $|\Pi_H^{(V)}(k)|^2$ and $X^{(V)}(k)$ 
are given for selected values of $n_B$ and $n_H$.

\subsubsection{$n_B,\,n_H=4$}
\be
|\Pi_B^{(V)}(k)|^2=\frac{A_B^2\,k_D^{11}}{256\,\pi^4\,k_*^8}\biggl[\frac{28}{165}-\frac{5 \tk}{12}+\frac{8 \tk^2}{15}-\frac{5 \tk^3}{12}+\frac{4 \tk^4}{21}-\frac{41 \tk^5}{960}+\frac{\tk^7}{640}-\frac{\tk^{11}}{118272}\biggl] \nonumber\,,
\label{vectorNE_n4}
\ee
\begin{displaymath}
|\Pi_H^{(V)}(k)|^2=\frac{A_H^2\,k_D^{11}}{256\,\pi^4\,k_*^8}
\left\{\begin{array}{ll}
\frac{4}{33}-\frac{\tk}{4}+\frac{4 \tk^2}{15}-\frac{\tk^3}{6}+\frac{12 \tk^4}{245}-\frac{4 \tk^6}{1575}+\frac{\tk^{11}}{53900} \quad
& {\rm for}  \quad 0 \le \tk \le 1 \\
-\frac{4}{33}-\frac{4}{2205 \tk^3}+\frac{2}{35 \tk}+\frac{23 \tk}{100}-\frac{4 \tk^2}{15}+\frac{\tk^3}{6}-\frac{12 \tk^4}{245}+\frac{4 \tk^6}{1575}-\frac{\tk^{11}}{161700}
& {\rm for}
\quad 1 \le \tk  \le 2
\end{array} 
\label{vectorE_n4}
\right. \,,
\end{displaymath}
\begin{displaymath}
X^{(V)}(k)=\frac{A_B\,A_H\,k_D^{11}}{512\,\pi^4\,k_*^8}
\left\{\begin{array}{ll}
\frac{4 \tk}{75}-\frac{\tk^2}{6}+\frac{23 \tk^3}{105}-\frac{7 \tk^4}{48}+\frac{2 \tk^5}{45}-\frac{\tk^6}{600}-\frac{\tk^7}{693}+\frac{\tk^{11}}{450450} \quad
& {\rm for}  \quad 0 \le \tk \le 1 \\
\frac{72}{385}+\frac{16}{17325 \tk^4}-\frac{8}{585 \tk^2}-\frac{8 \tk}{15}+\frac{13 \tk^2}{18}-\frac{58 \tk^3}{105}+\frac{79 \tk^4}{336} \\
-\frac{2 \tk^5}{45}-\frac{\tk^6}{600}+\frac{\tk^7}{693}-\frac{\tk^{11}}{450450}
& {\rm for}
\quad 1 \le \tk  \le 2
\end{array} 
\label{vectorM_n4}
\right. \,.
\end{displaymath}

\subsubsection{$n_B,\,n_H=3$}
\begin{displaymath}
|\Pi_B^{(V)}(k)|^2=\frac{A_B^2\,k_D^9}{256\,\pi^4\,k_*^6}
\left\{\begin{array}{ll}
\frac{28}{135}-\frac{5 \tk}{12}+\frac{296 \tk^2}{735}-\frac{2 \tk^3}{9}+\frac{92 \tk^4}{1575}-\frac{32 \tk^6}{10395}+\frac{2 \tk^9}{11025} \quad
& {\rm for}  \quad 0 \le \tk \le 1 \\
-\frac{28}{135}-\frac{32}{24255 \tk^5}+\frac{4}{945 \tk^3}+\frac{44}{525 \tk}+\frac{23 \tk}{60}-\frac{296 \tk^2}{735} \notag\\
+\frac{2 \tk^3}{9}-\frac{92 \tk^4}{1575}+\frac{32 \tk^6}{10395}-\frac{2 \tk^9}{33075}
& {\rm for}
\quad 1 \le \tk  \le 2
\end{array} 
\label{vectorNE_n3}
\right. \,,
\end{displaymath}
\be
|\Pi_H^{(V)}(k)|^2=\frac{A_H^2\,k_D^9}{256\,\pi^4\,k_*^6}\biggl[\frac{4}{27}-\frac{\tk}{4}+\frac{4 \tk^2}{21}-\frac{5 \tk^3}{72}+\frac{\tk^5}{192}-\frac{\tk^9}{24192}\biggr] \nonumber\,,
\label{vectorE_n3}
\ee
\begin{displaymath}
X^{(V)}(k)=\frac{A_B\,A_H\,k_D^9}{512\,\pi^4\,k_*^6}
\left\{\begin{array}{ll}
\frac{\tk}{15}-\frac{\tk^2}{6}+\frac{10 \tk^3}{63}-\frac{\tk^4}{16}+\frac{\tk^5}{315}+\frac{\tk^6}{360}+\frac{k^9}{54054} \quad
& {\rm for}  \quad 0 \le \tk \le 1 \\
-\frac{32}{135}-\frac{16}{4095 \tk^4}+\frac{8}{231 \tk^2}+\frac{8 \tk}{15}-\frac{23 \tk^2}{42}+\frac{2 \tk^3}{7} \\
-\frac{\tk^4}{16}-\frac{\tk^5}{315}+\frac{\tk^6}{360}-\frac{\tk^9}{54054}
& {\rm for}
\quad 1 \le \tk  \le 2
\end{array} 
\label{vectorM_n3}
\right. \,.
\end{displaymath}

\subsubsection{$n_B,\,n_H=2$}
\be
|\Pi_B^{(V)}(k)|^2=\frac{A_B^2\,k_D^7}{256\,\pi^4\,k_*^4}\biggl[\frac{4}{15}-\frac{5 \tk}{12}+\frac{4 \tk^2}{15}-\frac{\tk^3}{12}+\frac{7 \tk^5}{960}-\frac{\tk^7}{1920}\biggl] \nonumber\,,
\label{vectorNE_n2}
\ee
\begin{displaymath}
|\Pi_H^{(V)}(k)|^2=\frac{A_H^2\,k_D^7}{256\,\pi^4\,k_*^4}
\left\{\begin{array}{ll}
\frac{4}{21}-\frac{\tk}{4}+\frac{8 \tk^2}{75}-\frac{4 \tk^4}{315}+\frac{\tk^7}{1050} \quad
& {\rm for}  \quad 0 \le \tk \le 1 \\
-\frac{4}{21}-\frac{4}{525 \tk^3}+\frac{2}{15 \tk}+\frac{7 \tk}{36}-\frac{8 \tk^2}{75}+\frac{4 \tk^4}{315}-\frac{\tk^7}{3150} 
& {\rm for}  \quad 1 \le \tk \le 2 
\end{array} 
\label{vectorE_n2}
\right. \,,
\end{displaymath}
\begin{displaymath}
X^{(V)}(k)=\frac{A_B\,A_H\,k_D^7}{512\,\pi^4\,k_*^4}
\left\{\begin{array}{ll}
\frac{4 \tk}{45}-\frac{\tk^2}{6}+\frac{2 \tk^3}{21}-\frac{\tk^4}{144}-\frac{2 \tk^5}{315}+\frac{2 \tk^7}{10395} \quad
& {\rm for}  \quad 0 \le \tk \le 1 \\
\frac{32}{105}+\frac{16}{3465 \tk^4}-\frac{8}{189 \tk^2}-\frac{8 \tk}{15}+\frac{11 \tk^2}{30}-\frac{2 \tk^3}{21}-\frac{\tk^4}{144}+\frac{2 \tk^5}{315}-\frac{2 \tk^7}{10395}
& {\rm for}
\quad 1 \le \tk  \le 2
\end{array} 
\label{vectorM_n2}
\right. \,.
\end{displaymath}

\subsubsection{$n_B,\,n_H=1$}
\begin{displaymath}
|\Pi_B^{(V)}(k)|^2=\frac{A_B^2\,k_D^5}{256\,\pi^4\,k_*^2}
\left\{\begin{array}{ll}
\frac{28}{75}-\frac{5 \tk}{12}+\frac{4 \tk^2}{35}-\frac{8 \tk^4}{315}+\frac{\tk^5}{50} \quad
& {\rm for}  \quad 0 \le \tk \le 1 \\
-\frac{28}{75}-\frac{32}{1575 \tk^5}+\frac{4}{105 \tk^3}+\frac{4}{15 \tk}+\frac{\tk}{4}-\frac{4 \tk^2}{35}+\frac{8 \tk^4}{315}-\frac{\tk^5}{150} 
& {\rm for}  \quad 1 \le \tk \le 2 \\	
\end{array} 
\label{vectorNE_n1}
\right. \,,
\end{displaymath}
\be
|\Pi_H^{(V)}(k)|^2=\frac{A_H^2\,k_D^5}{256\,\pi^4\,k_*^2}\biggl[\frac{4}{15}-\frac{\tk}{4}+\frac{\tk^3}{24}-\frac{\tk^5}{320}\biggl] \nonumber\,,
\label{vectorE_n1}
\ee
\begin{displaymath}
X^{(V)}(k)=\frac{A_B\,A_H\,k_D^5}{512\,\pi^4\,k_*^2}
\left\{\begin{array}{ll}
\frac{2 \tk}{15}-\frac{\tk^2}{6}+\frac{2 \tk^3}{105}+\frac{\tk^4}{48}+\frac{\tk^5}{315} \quad
& {\rm for}  \quad 0 \le \tk \le 1 \\
-\frac{8}{15}-\frac{16}{315 \tk^4}+\frac{8}{35 \tk^2}+\frac{8 \tk}{15}-\frac{\tk^2}{6}-\frac{2 \tk^3}{105}+\frac{\tk^4}{48}-\frac{\tk^5}{315}
& {\rm for}
\quad 1 \le \tk  \le 2
\end{array} 
\label{vectorM_n1}
\right. \,.
\end{displaymath}

\subsubsection{$n_B,\,n_H=0$}
\begin{align}
|\Pi_B^{(V)}(k)|^2=\frac{A_B^2\,k_D^3}{256\,\pi^4}\biggl[&\frac{53}{96}+\frac{1}{32 \tk^4}+\frac{1}{64 \tk^3}-\frac{1}{32 \tk^2}-\frac{5}{384 \tk}-\frac{29 \tk}{64}-\frac{5 \tk^2}{96}+\frac{55 \tk^3}{768}+\frac{\log[\abs{1-\tk}]}{32 \tk^5}-\frac{\log[\abs{1-\tk}]}{24 \tk^3}  \nonumber\\
&-\frac{\log[\abs{1-\tk}]}{16 \tk}+\frac{1}{8} \tk \log[\abs{1-\tk}]-\frac{5}{96} \tk^3 \log[\abs{1-\tk}]\biggr] \nonumber\,,
\label{vectorNE_n0}
\end{align}
\begin{displaymath}
|\Pi_H^{(V)}(k)|^2=\frac{A_H^2\,k_D^3}{256\,\pi^4}
\left\{\begin{array}{ll}
\frac{4}{9}-\frac{\tk}{4}-\frac{4 \tk^2}{15}+\frac{\tk^3}{6} \quad
& {\rm for}  \quad 0 \le \tk \le 1 \\	
-\frac{4}{9}-\frac{4}{45 \tk^3}+\frac{2}{3 \tk}-\frac{\tk}{4}+\frac{4 \tk^2}{15}-\frac{\tk^3}{18}
& {\rm for}
\quad 1 \le \tk  \le 2
\end{array} 
\label{vectorE_n0}
\right. \,,
\end{displaymath}
\begin{displaymath}
X^{(V)}(k)=\frac{A_B\,A_H\,k_D^3}{512\,\pi^4}
\left\{\begin{array}{ll}
-\frac{23}{280}+\frac{1}{14 \tk^3}+\frac{1}{28 \tk^2}-\frac{37}{210 \tk}+\frac{8 \tk}{21}-\frac{17 \tk^2}{140}-\frac{914 \tk^3}{11025} \\
+\frac{1}{6} \text{Log}[1-\tk]+\frac{\text{Log}[1-\tk]}{14 \tk^4}-\frac{\text{Log}[1-\tk]}{5 \tk^2}-\frac{4}{105} \tk^3 \text{Log}[1-\tk]+\frac{8}{105} \tk^3 \text{Log}[\tk] \quad
& {\rm for}  \quad 0 \le \tk \le 1 \\
\frac{2033}{2520}+\frac{16}{245 \tk^4}+\frac{1}{14 \tk^3}-\frac{199}{700 \tk^2}-\frac{37}{210 \tk}-\frac{44 \tk}{105}-\frac{17 \tk^2}{140} \\
+\frac{914 \tk^3}{11025}+\frac{1}{6} \text{Log}[-1+\tk]+\frac{\text{Log}[-1+\tk]}{14 \tk^4}-\frac{\text{Log}[-1+\tk]}{5 \tk^2}-\frac{4}{105} \tk^3 \text{Log}[-1+\tk]
& {\rm for}
\quad 1 \le \tk  \le 2
\end{array} 
\label{vectorM_n0}
\right. \,.
\end{displaymath}

\subsubsection{$n_B,\,n_H=-1$}
\begin{displaymath}
|\Pi_B^{(V)}(k)|^2=\frac{A_B^2\,k_D\,k_*}{256\,\pi^4}
\left\{\begin{array}{ll}
\frac{28}{15}-\frac{7 \tk}{4}+\frac{16 \tk^2}{105} \quad
& {\rm for}  \quad 0 \le \tk \le 1 \\
-\frac{28}{15}+\frac{32}{105 \tk^5}-\frac{4}{15 \tk^3}+\frac{4}{3 \tk}+\frac{11 \tk}{12}-\frac{16 \tk^2}{105} \quad
& {\rm for}  \quad 1 \le \tk \le 2 \\	
\end{array} 
\label{vectorNE_n-1}
\right. \,,
\end{displaymath}
\begin{displaymath}
|\Pi_H^{(V)}(k)|^2=\frac{A_H^2\,k_D\,k_*}{256\,\pi^4}
\left\{\begin{array}{ll}
\frac{7}{8}+\frac{1}{8 \tk^2}+\frac{1}{16 \tk}-\frac{15 \tk}{32}-\frac{\tk \pi ^2}{24}+\frac{\text{Log}[1-\tk]}{8 \tk^3}-\frac{\text{Log}[1-\tk]}{2 \tk}+\frac{3}{8} \tk \text{Log}[1-\tk] \\
+\frac{1}{4} \tk \text{Log}[1-\tk] \text{Log}\left[\frac{1}{\tk}\right]-\frac{1}{4} \tk \text{Log}[1-\tk] \text{Log}[\tk]+\frac{1}{4} \tk \text{Log}[\tk]^2 \notag\\
+\frac{1}{2} \tk \text{PolyLog}\left[2,\frac{-1+\tk}{\tk}\right] \quad
& {\rm for}  \quad 0 \le \tk \le 1 \\
\frac{7}{8}+\frac{1}{8 \tk^2}+\frac{1}{16 \tk}-\frac{15 \tk}{32}+\frac{\log[-1+\tk]}{8 \tk^3}-\frac{\log[-1+\tk]}{2 \tk}+\frac{3}{8} \tk \log[-1+\tk] \\
+\frac{1}{4} \tk \log[-1+\tk] \log\left[\frac{1}{\tk}\right]-\frac{1}{4} \tk \text{PolyLog}\left[2,\frac{1}{\tk}\right]+\frac{1}{4} \tk \text{PolyLog}\left[2,\frac{-1+\tk}{\tk}\right] \quad
& {\rm for}  \quad 1 \le \tk \le 2 \\	
\end{array} 
\label{vectorE_n-1}
\right. \,,
\end{displaymath}
\begin{displaymath}
X^{(V)}(k)=\frac{A_B\,A_H\,k_D\,k_*}{512\,\pi^4}
\left\{\begin{array}{ll}
\frac{17}{120}-\frac{1}{10 \tk^3}-\frac{1}{20 \tk^2}+\frac{3}{10 \tk}-\frac{28 \tk}{225}-\frac{\tk^2}{12}-\frac{1}{2} \text{Log}[1-\tk]-\frac{\text{Log}[1-\tk]}{10 \tk^4} \\
+\frac{\text{Log}[1-\tk]}{3 \tk^2}+\frac{4}{15} \tk \text{Log}[1-\tk]-\frac{8}{15} \tk \text{Log}[\tk] \quad
& {\rm for}  \quad 0 \le \tk \le 1 \\
\frac{17}{120}+\frac{16}{25 \tk^4}-\frac{1}{10 \tk^3}-\frac{169}{180 \tk^2}+\frac{3}{10 \tk}+\frac{28 \tk}{225}-\frac{\tk^2}{12}-\frac{1}{2} \text{Log}[-1+\tk] \\
-\frac{\text{Log}[-1+\tk]}{10 \tk^4}+\frac{\text{Log}[-1+\tk]}{3 \tk^2}+\frac{4}{15} \tk \text{Log}[-1+\tk]
& {\rm for}
\quad 1 \le \tk  \le 2
\end{array} 
\label{vectorM_n-1}
\right. \,.
\end{displaymath}

\subsubsection{$n_B,\,n_H=-3/2$}
\begin{displaymath}
|\Pi_B^{(V)}(k)|^2=\frac{A_B^2\,k_*^3}{256\,\pi^4}
\left\{\begin{array}{ll}
\frac{4936}{1755 \sqrt{1-\tk}}+\frac{1024}{2925 \tk^5}-\frac{1024}{2925 \sqrt{1-\tk} \tk^5}+\frac{512}{2925 \sqrt{1-\tk} \tk^4}-\frac{32}{135 \tk^3}+\frac{2464}{8775 \sqrt{1-\tk} \tk^3} \\
-\frac{848}{8775 \sqrt{1-\tk} \tk^2}+\frac{44}{15 \tk}-\frac{5176}{1755 \sqrt{1-\tk} \tk}+\frac{\tk}{3}-\frac{224 \tk}{1755 \sqrt{1-\tk}}+\frac{448 \tk^2}{1755 \sqrt{1-\tk}}-\frac{14 \pi }{15} \\
+\frac{56}{15} \text{Log}\left[1+\sqrt{1-\tk}\right]-\frac{28 \text{Log}[\tk]}{15} \quad
& {\rm for}  \quad 0 \le \tk \le 1 \\
-\frac{4936}{1755 \sqrt{-1+\tk}}+\frac{1024}{2925 \tk^5}+\frac{1024}{2925 \sqrt{-1+\tk} \tk^5}-\frac{512}{2925 \sqrt{-1+\tk} \tk^4}-\frac{32}{135 \tk^3}-\frac{2464}{8775 \sqrt{-1+\tk} \tk^3} \\
+\frac{848}{8775 \sqrt{-1+\tk} \tk^2}+\frac{44}{15 \tk}+\frac{5176}{1755 \sqrt{-1+\tk} \tk}+\frac{\tk}{3}+\frac{224 \tk}{1755 \sqrt{-1+\tk}}-\frac{448 \tk^2}{1755 \sqrt{-1+\tk}} \\
-\frac{28}{15} \text{ArcTan}\left[\sqrt{\frac{1}{-1+\tk}}\right]+\frac{28}{15} \text{ArcTan}\left[\sqrt{-1+\tk}\right] \quad
& {\rm for}  \quad 1 \le \tk \le 2 \\	
\end{array} 
\label{vectorNE_n-3/2}
\right. \,,
\end{displaymath}
\begin{displaymath}
|\Pi_H^{(V)}(k)|^2=\frac{A_H^2\,k_*^3}{256\,\pi^4}
\left\{\begin{array}{ll}
-\frac{64 \sqrt{1- \tk}}{21}+\frac{32}{63  \tk^3}-\frac{32 \sqrt{1- \tk}}{63  \tk^3}-\frac{16 \sqrt{1- \tk}}{63  \tk^2}-\frac{4}{3  \tk}+\frac{8 \sqrt{1- \tk}}{7  \tk}- \tk \\
+\frac{2 \pi }{3}+\frac{8}{3} \text{Log}\left[1+\sqrt{1- \tk}\right]-\frac{4 \text{Log}[ \tk]}{3} \quad
& {\rm for}  \quad 0 \le \tk \le 1 \\
\frac{64 \sqrt{-1+\tk}}{21}+\frac{32}{63 \tk^3}+\frac{32 \sqrt{-1+\tk}}{63 \tk^3}+\frac{16 \sqrt{-1+\tk}}{63 \tk^2}-\frac{4}{3 \tk}-\frac{8 \sqrt{-1+\tk}}{7 \tk}-\tk \\
+\frac{4}{3} \text{ArcTan}\left[\sqrt{\frac{1}{-1+\tk}}\right]-\frac{4}{3} \text{ArcTan}\left[\sqrt{-1+\tk}\right] \quad
& {\rm for}  \quad 1 \le \tk \le 2 \\	
\end{array} 
\label{vectorE_n-3/2}
\right. \,,
\end{displaymath}
\begin{displaymath}
X^{(V)}(k)=\frac{A_B\,A_H\,k_*^3}{512\,\pi^4}
\left\{\begin{array}{ll}
-\frac{16}{9}+\frac{6464}{3465 \sqrt{1-\tk}}+\frac{1024}{3465 \tk^4}-\frac{1024}{3465 \sqrt{1-\tk} \tk^4}+\frac{512}{3465 \sqrt{1-\tk} \tk^3} \\
+\frac{64}{105 \tk^2}-\frac{1984}{3465 \sqrt{1-\tk} \tk^2}+\frac{32}{99 \sqrt{1-\tk} \tk}-\frac{4768 \tk}{3465 \sqrt{1-\tk}}-\frac{64 \tk^2}{693 \sqrt{1-\tk}}+\frac{\pi }{3} \quad
& {\rm for}  \quad 0 \le \tk \le 1 \\
-\frac{16}{9}-\frac{1028}{1155 \sqrt{-1+\tk}}+\frac{1024}{3465 \tk^4}+\frac{3904}{3465 \sqrt{-1+\tk} \tk^4}-\frac{1952}{3465 \sqrt{-1+\tk} \tk^3} \\
+\frac{64}{105 \tk^2}-\frac{6296}{3465 \sqrt{-1+\tk} \tk^2}+\frac{76}{99 \sqrt{-1+\tk} \tk}+\frac{1696 \tk}{1155 \sqrt{-1+\tk}}-\frac{64 \tk^2}{693 \sqrt{-1+\tk}} \\
+\frac{2}{3} \text{ArcTan}\left[\frac{1}{\sqrt{-1+\tk}}\right]-\frac{2}{3} \text{ArcTan}\left[\sqrt{-1+\tk}\right] \quad
& {\rm for}  \quad 1 \le \tk \le 2 \\	
\end{array} 
\label{vectorM_n-3/2}
\right. \,.
\end{displaymath}

{\bf Correlators for tensor perturbations}
\\Our exact results for $|\Pi_B^{(T)}(k)|^2$, $|\Pi_H^{(T)}(k)|^2$ and $X^{(T)}(k)$ are given for selected values of $n_B$ and $n_H$.
\setcounter{subsubsection}{0}

\subsubsection{$n_B,\,n_H=4$}
\be
|\Pi_B^{(T)}(k)|^2=\frac{A_B^2\,k_D^{11}}{256\,\pi^4\,k_*^8}\biggl[\frac{56}{165}-\frac{7 \tk}{6}+\frac{88 \tk^2}{45}-\frac{41 \tk^3}{24}+\frac{16 \tk^4}{21}-\frac{61 \tk^5}{480}-\frac{3 \tk^7}{640}+\frac{109 \tk^{11}}{709632}\biggl] \nonumber\,,
\label{tensorNE_n4}
\ee
\begin{displaymath}
|\Pi_H^{(T)}(k)|^2=\frac{A_H^2\,k_D^{11}}{1024\,\pi^4\,k_*^8}
\left\{\begin{array}{ll}
-\frac{8}{33}+\tk-\frac{8 \tk^2}{5}+\frac{4 \tk^3}{3}-\frac{24 \tk^4}{49}+\frac{8 \tk^6}{225}-\frac{6 \tk^{11}}{13475} \quad
& {\rm for}  \quad 0 \le \tk \le 1 \\
\frac{8}{33}-\frac{16}{2205 \tk^3}-\frac{23 \tk}{25}+\frac{8 \tk^2}{5}-\frac{4 \tk^3}{3}+\frac{24 \tk^4}{49}-\frac{8 \tk^6}{225}+\frac{2 \tk^{11}}{13475}
& {\rm for}  \quad 1 \le \tk \le 2 
\end{array} 
\label{tensorE_n4}
\right. \,,
\end{displaymath}
\begin{displaymath}
X^{(T)}(k)=\frac{A_B\,A_H\,k_D^{11}}{256\,\pi^4\,k_*^8}
\left\{\begin{array}{ll}
\frac{16 \tk}{75}-\frac{\tk^2}{3}+\frac{8 \tk^3}{21}-\frac{13 \tk^4}{48}+\frac{8 \tk^5}{63}-\frac{43 \tk^6}{1200}+\frac{16 \tk^7}{3465}-\frac{68 \tk^{11}}{225225} \quad
& {\rm for}  \quad 0 \le \tk \le 1 \\
\frac{48}{77}-\frac{16}{17325 \tk^4}-\frac{16}{819 \tk^2}-\frac{32 \tk}{15}+\frac{161 \tk^2}{45}-\frac{64 \tk^3}{21}+\frac{421 \tk^4}{336} \notag\\
-\frac{8 \tk^5}{63}-\frac{43 \tk^6}{1200}-\frac{16 \tk^7}{3465}+\frac{68 \tk^{11}}{225225}
& {\rm for}
\quad 1 \le \tk  \le 2
\end{array} 
\label{tensorM_n4}
\right. \,.
\end{displaymath}

\subsubsection{$n_B,\,n_H=3$}
\begin{displaymath}
|\Pi_B^{(T)}(k)|^2=\frac{A_B^2\,k_D^9}{256\,\pi^4\,k_*^6}
\left\{\begin{array}{ll}
\frac{56}{135}-\frac{7 \tk}{6}+\frac{1112 \tk^2}{735}-\frac{127 \tk^3}{144}+\frac{296 \tk^4}{1575}+\frac{104 \tk^6}{10395}-\frac{29 \tk^9}{11025} \quad
& {\rm for}  \quad 0 \le \tk \le 1 \\
-\frac{56}{135}+\frac{16}{24255 \tk^5}+\frac{8}{945 \tk^3}+\frac{32}{525 \tk}+\frac{37 \tk}{30}-\frac{1112 \tk^2}{735}+\frac{43 \tk^3}{48} \notag\\
-\frac{296 \tk^4}{1575}-\frac{104 \tk^6}{10395}+\frac{29 \tk^9}{33075}
& {\rm for}
\quad 1 \le \tk  \le 2
\end{array} 
\label{tensorNE_n3}
\right. \,,
\end{displaymath}
\be
|\Pi_H^{(T)}(k)|^2=\frac{A_H^2\,k_D^9}{1024\,\pi^4\,k_*^6}\biggl[-\frac{8}{27}+\tk-\frac{8 \tk^2}{7}+\frac{5 \tk^3}{9}-\frac{\tk^5}{16}+\frac{5 \tk^9}{6048}\biggr] \nonumber\,,
\label{tensorE_n3}
\ee
\begin{displaymath}
X^{(T)}(k)=\frac{A_B\,A_H\,k_D^9}{256\,\pi^4\,k_*^6}
\left\{\begin{array}{ll}
\frac{4 k}{15}-\frac{\tk^2}{3}+\frac{104 \tk^3}{315}-\frac{3 \tk^4}{16}+\frac{4 \tk^5}{63}-\frac{7 \tk^6}{720}-\frac{46 \tk^9}{27027} \quad
& {\rm for}  \quad 0 \le \tk \le 1 \\
-\frac{16}{27}+\frac{16}{4095 \tk^4}+\frac{16}{1155 \tk^2}+\frac{32 \tk}{15}-\frac{55 \tk^2}{21}+\frac{152 \tk^3}{105} \notag\\
-\frac{3 \tk^4}{16}-\frac{4 \tk^5}{63}-\frac{7 \tk^6}{720}+\frac{46 \tk^9}{27027}
& {\rm for}
\quad 1 \le \tk  \le 2
\end{array} 
\label{tensorM_n3}
\right. \,.
\end{displaymath}

\subsubsection{$n_B,\,n_H=2$}
\be
|\Pi_B^{(T)}(k)|^2=\frac{A_B^2\,k_D^7}{256\,\pi^4\,k_*^4}\biggl[\frac{8}{15}-\frac{7 \tk}{6}+\frac{16 \tk^2}{15}-\frac{7 \tk^3}{24}-\frac{13 \tk^5}{480}+\frac{11 \tk^7}{1920}\biggl] \nonumber\,,
\label{tensorNE_n2}
\ee
\begin{displaymath}
|\Pi_H^{(T)}(k)|^2=\frac{A_H^2\,k_D^7}{1024\,\pi^4\,k_*^4}
\left\{\begin{array}{ll}
-\frac{8}{21}+\tk-\frac{16 \tk^2}{25}+\frac{8 \tk^4}{63}-\frac{8 \tk^7}{525} \quad
& {\rm for}  \quad 0 \le \tk \le 1 \\
\frac{8}{21}-\frac{16}{525 \tk^3}-\frac{7 \tk}{9}+\frac{16 \tk^2}{25}-\frac{8 \tk^4}{63}+\frac{8 \tk^7}{1575}
& {\rm for}  \quad 1 \le \tk \le 2 
\end{array} 
\label{tensorE_n2}
\right. \,,
\end{displaymath}
\begin{displaymath}
X^{(T)}(k)=\frac{A_B\,A_H\,k_D^7}{256\,\pi^4\,k_*^4}
\left\{\begin{array}{ll}
\frac{16 \tk}{45}-\frac{\tk^2}{3}+\frac{32 \tk^3}{105}-\frac{19 \tk^4}{144}+\frac{8 \tk^5}{315}-\frac{16 \tk^7}{1485} \quad
& {\rm for}  \quad 0 \le \tk \le 1 \\
\frac{16}{15}-\frac{16}{3465 \tk^4}-\frac{64}{945 \tk^2}-\frac{32 \tk}{15}+\frac{9 \tk^2}{5}-\frac{32 \tk^3}{105}-\frac{19 \tk^4}{144}-\frac{8 \tk^5}{315}+\frac{16 \tk^7}{1485}
& {\rm for}
\quad 1 \le \tk  \le 2
\end{array} 
\label{tensorM_n2}
\right. \,.
\end{displaymath}

\subsubsection{$n_B,\,n_H=1$}
\begin{displaymath}
|\Pi_B^{(T)}(k)|^2=\frac{A_B^2\,k_D^5}{256\,\pi^4\,k_*^2}
\left\{\begin{array}{ll}
\frac{56}{75}-\frac{7 \tk}{6}+\frac{64 \tk^2}{105}+\frac{\tk^3}{16}+\frac{8 \tk^4}{63}-\frac{4 \tk^5}{25} \quad
& {\rm for}  \quad 0 \le \tk \le 1 \\
 -\frac{56}{75}+\frac{16}{1575 \tk^5}+\frac{8}{105 \tk^3}+\frac{3 \tk}{2}-\frac{64 \tk^2}{105}+\frac{\tk^3}{16}-\frac{8 \tk^4}{63}+\frac{4 \tk^5}{75}
& {\rm for}  \quad 1 \le \tk \le 2 \\	
\end{array} 
\label{tensorNE_n1}
\right. \,,
\end{displaymath}
\be
|\Pi_H^{(T)}(k)|^2=\frac{A_H^2\,k_D^5}{1024\,\pi^4\,k_*^2}\biggl[-\frac{8}{15}+\tk-\frac{\tk^3}{3}+\frac{3 \tk^5}{80}\biggl] \nonumber\,,
\label{tensorE_n1}
\ee
\begin{displaymath}
X^{(T)}(k)=\frac{A_B\,A_H\,k_D^5}{256\,\pi^4\,k_*^2}
\left\{\begin{array}{ll}
\frac{8 \tk}{15}-\frac{\tk^2}{3}+\frac{8 \tk^3}{21}-\frac{5 \tk^4}{48}-\frac{4 \tk^5}{45} \quad
& {\rm for}  \quad 0 \le \tk \le 1 \\
-\frac{16}{15}+\frac{16}{315 \tk^4}+\frac{32 \tk}{15}-\frac{\tk^2}{3}-\frac{8 \tk^3}{21}-\frac{5 \tk^4}{48}+\frac{4 \tk^5}{45}
& {\rm for}
\quad 1 \le \tk  \le 2
\end{array} 
\label{tensorM_n1}
\right. \,.
\end{displaymath}

\subsubsection{$n_B,\,n_H=0$}
\begin{displaymath}
|\Pi_B^{(T)}(k)|^2=\frac{A_B^2\,k_D^3}{256\,\pi^4}
\left\{\begin{array}{ll}
\frac{293}{192}-\frac{1}{64 \tk^4}-\frac{1}{128 \tk^3}-\frac{17}{192 \tk^2}-\frac{35}{768 \tk}-\frac{397 \tk}{384}-\frac{17 \tk^2}{192}+\frac{181 \tk^3}{1536} \\
+\frac{\tk^3 \pi ^2}{96}-\frac{\text{Log}[1-\tk]}{64 \tk^5}-\frac{\text{Log}[1-\tk]}{12 \tk^3}+\frac{5 \text{Log}[1-\tk]}{16 \tk}-\frac{1}{4} \tk \text{Log}[1-\tk] \\
+\frac{7}{192} \tk^3 \text{Log}[1-\tk]+\frac{1}{8} \tk^3 \text{Log}[1-\tk] \text{Log}[\tk]-\frac{1}{16} \tk^3 \text{Log}[\tk]^2 \\
-\frac{1}{8} \tk^3 \text{PolyLog}\left[2,\frac{-1+\tk}{\tk}\right] \quad
& {\rm for}  \quad 0 \le \tk \le 1 \\	
\frac{293}{192}-\frac{1}{64 \tk^4}-\frac{1}{128 \tk^3}-\frac{17}{192 \tk^2}-\frac{35}{768 \tk}-\frac{397 \tk}{384}-\frac{17 \tk^2}{192}+\frac{181 \tk^3}{1536} \\
-\frac{\text{Log}[-1+\tk]}{64 \tk^5}-\frac{\text{Log}[-1+\tk]}{12 \tk^3}+\frac{5 \text{Log}[-1+\tk]}{16 \tk}-\frac{1}{4} \tk \text{Log}[-1+\tk] \\
+\frac{7}{192} \tk^3 \text{Log}[-1+\tk]-\frac{1}{16} \tk^3 \text{Log}[-1+\tk] \text{Log}\left[\frac{1}{\tk}\right] \\
+\frac{1}{16} \tk^3 \text{PolyLog}\left[2,\frac{1}{\tk}\right]-\frac{1}{16} \tk^3 \text{PolyLog}\left[2,\frac{-1+\tk}{\tk}\right]
& {\rm for}
\quad 1 \le \tk  \le 2
\end{array} 
\label{tensorNE_n0}
\right. \,,
\end{displaymath}
\begin{displaymath}
|\Pi_H^{(T)}(k)|^2=\frac{A_H^2\,k_D^3}{1024\,\pi^4}
\left\{\begin{array}{ll}
-\frac{8}{9}+\tk+\frac{8 \tk^2}{5}-\frac{4 \tk^3}{3} \quad
& {\rm for}  \quad 0 \le \tk \le 1 \\	
\frac{8}{9}-\frac{16}{45 \tk^3}+\tk-\frac{8 \tk^2}{5}+\frac{4 \tk^3}{9}
& {\rm for}
\quad 1 \le \tk  \le 2
\end{array} 
\label{tensorE_n0}
\right. \,,
\end{displaymath}
\begin{displaymath}
X^{(T)}(k)=\frac{A_B\,A_H\,k_D^3}{256\,\pi^4}
\left\{\begin{array}{ll}
\frac{9}{280}-\frac{1}{14 \tk^3}-\frac{1}{28 \tk^2}+\frac{8}{105 \tk}+\frac{263 \tk}{210}-\frac{199 \tk^2}{840}-\frac{1928 \tk^3}{11025} \notag\\
+\frac{1}{6} \text{Log}[1-\tk]-\frac{\text{Log}[1-\tk]}{14 \tk^4}+\frac{\text{Log}[1-\tk]}{10 \tk^2}-\frac{1}{2} \tk^2 \text{Log}[1-\tk] \notag\\
+\frac{32}{105} \tk^3 \text{Log}[1-\tk]-\frac{64}{105} \tk^3 \text{Log}[k] \quad
& {\rm for}  \quad 0 \le \tk \le 1 \\
\frac{9041}{2520}-\frac{16}{245 \tk^4}-\frac{1}{14 \tk^3}-\frac{473}{700 \tk^2}+\frac{8}{105 \tk}-\frac{409 \tk}{210}-\frac{199 \tk^2}{840} \notag\\
+\frac{1928 \tk^3}{11025}+\frac{1}{6} \text{Log}[-1+\tk]-\frac{\text{Log}[-1+\tk]}{14 \tk^4}+\frac{\text{Log}[-1+\tk]}{10 \tk^2} \notag\\
-\frac{1}{2} \tk^2 \text{Log}[-1+\tk]+\frac{32}{105} \tk^3 \text{Log}[-1+\tk]
& {\rm for}
\quad 1 \le \tk  \le 2
\end{array} 
\label{tensorM_n0}
\right. \,.
\end{displaymath}

\subsubsection{$n_B,\,n_H=-1$}
\begin{displaymath}
|\Pi_B^{(T)}(k)|^2=\frac{A_B^2\,k_D\,k_*}{256\,\pi^4}
\left\{\begin{array}{ll}
\frac{56}{15}-\frac{5 \tk}{2}-\frac{8 \tk^2}{105}+\frac{\tk^3}{16} \quad
& {\rm for}  \quad 0 \le \tk \le 1 \\
-\frac{56}{15}-\frac{16}{105 \tk^5}-\frac{8}{15 \tk^3}+\frac{16}{3 \tk}+\frac{\tk}{6}+\frac{8 \tk^2}{105}+\frac{\tk^3}{16}
& {\rm for}  \quad 1 \le \tk \le 2 \\	
\end{array} 
\label{tensorNE_n-1}
\right. \,,
\end{displaymath}
\begin{displaymath}
|\Pi_H^{(T)}(k)|^2=\frac{A_H^2\,k_D\,k_*}{1024\,\pi^4}
\left\{\begin{array}{ll}
-\frac{5}{2}+\frac{1}{2 \tk^2}+\frac{1}{4 \tk}+\frac{9 \tk}{8}+\frac{\tk \pi ^2}{6}+\frac{\text{Log}[1-\tk]}{2 \tk^3}-\frac{1}{2} \tk \text{Log}[1-\tk] \notag\\
+2 \tk \text{Log}[1-\tk] \text{Log}[\tk]-\tk \text{Log}[\tk]^2-2 \tk \text{PolyLog}\left[2,\frac{-1+\tk}{\tk}\right] \quad
& {\rm for}  \quad 0 \le \tk \le 1 \\
-\frac{5}{2}+\frac{1}{2 \tk^2}+\frac{1}{4 \tk}+\frac{9 \tk}{8}+\frac{\text{Log}[-1+\tk]}{2 \tk^3}-\frac{1}{2} \tk \text{Log}[-1+\tk] \notag\\
-\tk \text{Log}[-1+\tk] \text{Log}\left[\frac{1}{\tk}\right]+\tk \text{PolyLog}\left[2,\frac{1}{\tk}\right]-\tk \text{PolyLog}\left[2,\frac{-1+\tk}{\tk}\right]
& {\rm for}  \quad 1 \le \tk \le 2 \\	
\end{array} 
\label{tensorE_n-1}
\right. \,,
\end{displaymath}
\begin{displaymath}
X^{(T)}(k)=\frac{A_B\,A_H\,k_D\,k_*}{256\,\pi^4}
\left\{\begin{array}{ll}
\frac{53}{120}+\frac{1}{10 \tk^3}+\frac{1}{20 \tk^2}+\frac{13}{15 \tk}+\frac{481 \tk}{450}-\frac{7 \tk^2}{24}-\frac{5}{2} \text{Log}[1-\tk] \notag\\
+\frac{\text{Log}[1-\tk]}{10 \tk^4}+\frac{5 \text{Log}[1-\tk]}{6 \tk^2}+\frac{16}{15} \tk \text{Log}[1-\tk]+\frac{1}{2} \tk^2 \text{Log}[1-\tk]-\frac{32}{15} \tk \text{Log}[\tk] \quad
& {\rm for}  \quad 0 \le \tk \le 1 \\
\frac{53}{120}-\frac{16}{25 \tk^4}+\frac{1}{10 \tk^3}+\frac{329}{180 \tk^2}+\frac{13}{15 \tk}-\frac{31 \tk}{450}-\frac{7 \tk^2}{24}-\frac{5}{2} \text{Log}[-1+\tk] \notag\\
+\frac{\text{Log}[-1+\tk]}{10 \tk^4}+\frac{5 \text{Log}[-1+\tk]}{6 \tk^2}+\frac{16}{15} \tk \text{Log}[-1+\tk]+\frac{1}{2} \tk^2 \text{Log}[-1+\tk]
& {\rm for}
\quad 1 \le \tk  \le 2
\end{array} 
\label{tensorM_n-1}
\right. \,.
\end{displaymath}

\subsubsection{$n_B,\,n_H=-3/2$}
\begin{displaymath}
|\Pi_B^{(T)}(k)|^2=\frac{A_B^2\,k_*^3}{256\,\pi^4}
\left\{\begin{array}{ll}
\frac{16304}{1755 \sqrt{1-\tk}}-\frac{512}{2925 \tk^5}+\frac{512}{2925 \sqrt{1-\tk} \tk^5}-\frac{256}{2925 \sqrt{1-\tk} \tk^4}-\frac{64}{135 \tk^3}+\frac{3968}{8775 \sqrt{1-\tk} \tk^3} \\
-\frac{2176}{8775 \sqrt{1-\tk} \tk^2}+\frac{28}{3 \tk}-\frac{16496}{1755 \sqrt{1-\tk} \tk}-\frac{2 \tk}{3}+\frac{64 \tk}{351 \sqrt{1-\tk}}-\frac{128 \tk^2}{351 \sqrt{1-\tk}}+\frac{\tk^3}{36}-\frac{28 \pi }{15} \\
+\frac{112}{15} \text{Log}\left[1+\sqrt{1-\tk}\right]-\frac{56 \text{Log}[\tk]}{15} \quad
& {\rm for}  \quad 0 \le \tk \le 1 \\
-\frac{16304}{1755 \sqrt{-1+\tk}}-\frac{512}{2925 \tk^5}-\frac{512}{2925 \sqrt{-1+\tk} \tk^5}+\frac{256}{2925 \sqrt{-1+\tk} \tk^4}-\frac{64}{135 \tk^3}-\frac{3968}{8775 \sqrt{-1+\tk} \tk^3} \\
+\frac{2176}{8775 \sqrt{-1+\tk} \tk^2}+\frac{28}{3 \tk}+\frac{16496}{1755 \sqrt{-1+\tk} \tk}-\frac{2 \tk}{3}-\frac{64 \tk}{351 \sqrt{-1+\tk}}+\frac{128 \tk^2}{351 \sqrt{-1+\tk}}+\frac{\tk^3}{36} \\
-\frac{56}{15} \text{ArcTan}\left[\sqrt{\frac{1}{-1+\tk}}\right]+\frac{56}{15} \text{ArcTan}\left[\sqrt{-1+\tk}\right]
& {\rm for}  \quad 1 \le \tk \le 2 \\	
\end{array} 
\label{tensorNE_n-3/2}
\right. \,,
\end{displaymath}
\begin{displaymath}
|\Pi_H^{(T)}(k)|^2=\frac{A_H^2\,k_*^3}{1024\,\pi^4}
\left\{\begin{array}{ll}
\frac{208}{21 \sqrt{1-\tk}}+\frac{128}{63 \tk^3}-\frac{128}{63 \sqrt{1-\tk} \tk^3}+\frac{64}{63 \sqrt{1-\tk} \tk^2}+\frac{16}{63 \sqrt{1-\tk} \tk}+4 \tk \notag\\
-\frac{64 \tk}{7 \sqrt{1-\tk}}-\frac{4 \pi }{3}-\frac{16}{3} \text{Log}\left[1+\sqrt{1-\tk}\right]+\frac{8 \text{Log}[\tk]}{3} \quad
& {\rm for}  \quad 0 \le \tk \le 1 \\
\frac{208}{21 \sqrt{-1+\tk}}+\frac{128}{63 \tk^3}-\frac{128}{63 \sqrt{-1+\tk} \tk^3}+\frac{64}{63 \sqrt{-1+\tk} \tk^2}+\frac{16}{63 \sqrt{-1+\tk} \tk}+4 \tk \notag\\
-\frac{64 \tk}{7 \sqrt{-1+\tk}}-\frac{8}{3} \text{ArcTan}\left[\frac{1}{\sqrt{-1+\tk}}\right]+\frac{8}{3} \text{ArcTan}\left[\sqrt{-1+\tk}\right]
& {\rm for}  \quad 1 \le \tk \le 2 \\	
\end{array} 
\label{tensorE_n-3/2}
\right. \,,
\end{displaymath}
\begin{displaymath}
X^{(T)}(k)=\frac{A_B\,A_H\,k_*^3}{256\,\pi^4}
\left\{\begin{array}{ll}
-\frac{88}{9}+\frac{5248 \sqrt{1-\tk}}{693}-\frac{1024}{3465 \tk^4}+\frac{1024 \sqrt{1-\tk}}{3465 \tk^4}+\frac{512 \sqrt{1-\tk}}{3465 \tk^3}+\frac{128}{21 \tk^2}-\frac{2304 \sqrt{1-\tk}}{385 \tk^2} \notag\\
-\frac{2048 \sqrt{1-\tk}}{693 \tk}+\frac{640}{693} \sqrt{1-\tk} \tk+\frac{2 \tk^2}{3}+\frac{7 \pi }{3} \quad
& {\rm for}  \quad 0 \le \tk \le 1 \\
-\frac{88}{9}-\frac{4100}{693 \sqrt{-1+\tk}}-\frac{1024}{3465 \tk^4}-\frac{3904}{3465 \sqrt{-1+\tk} \tk^4}+\frac{1952}{3465 \sqrt{-1+\tk} \tk^3} \notag\\
+\frac{128}{21 \tk^2}-\frac{2152}{3465 \sqrt{-1+\tk} \tk^2}+\frac{1564}{3465 \sqrt{-1+\tk} \tk}+\frac{5248 \tk}{693 \sqrt{-1+\tk}}+\frac{2 \tk^2}{3} \notag\\
-\frac{640 \tk^2}{693 \sqrt{-1+\tk}}+\frac{14}{3} \text{ArcTan}\left[\frac{1}{\sqrt{-1+\tk}}\right]-\frac{14}{3} \text{ArcTan}\left[\sqrt{-1+\tk}\right]
& {\rm for}  \quad 1 \le \tk \le 2 \\	
\end{array} 
\label{tensorE_n-3/2}
\right. \,.
\end{displaymath}

\end{widetext}


\end{document}